\documentclass[12pt]{elsarticle}
\usepackage{graphicx,epsfig,color,booktabs}
\usepackage{amssymb,amsmath,a4wide,url}
\usepackage{multirow}
\usepackage{rotating}

\journal{Physics Reports}

\begin{document}

\begin{frontmatter}
\title{Vital nodes identification in complex networks}
\author{{Linyuan L\"u}$^{b,a,*}$, Duanbing Chen$^{a,c}$, Xiao-Long Ren$^{d}$, Qian-Ming Zhang$^{c}$, Yi-Cheng Zhang$^{**}$, Tao Zhou$^{c,***}$\\
\textnormal{$^{a}$ Institute of Fundamental and Frontier Sciences and Big Data Research Center, University of Electronic Science and Technology of China, Chengdu 610054, PR China\\
$^{b}$ Alibaba Research Center for Complexity Sciences, Hangzhou Normal University, Hangzhou, 310036, PR China\\
$^{c}$ CompleX Lab, Web Sciences Center, University of Electronic Science and Technology of China, Chengdu, 611731, PR China\\
$^{d}$ Department of Humanities, Social and Political Sciences, ETH Zurich, Zurich, CH-8092, Switzerland\\
$^{*}$ linyuan.lv@gmail.com
$^{**}$ yi-cheng.zhang@unifr.ch
$^{***}$ zhutou@ustc.edu}}
\begin{abstract}
  \textbf{Real networks exhibit heterogeneous nature with nodes playing far different roles in structure and function. To identify vital nodes is thus very significant, allowing us to control the outbreak of epidemics, to conduct advertisements for e-commercial products, to predict popular scientific publications, and so on. The vital nodes identification attracts increasing attentions from both computer science and physical societies, with algorithms ranging from simply counting the immediate neighbors to complicated machine learning and message passing approaches. In this review, we clarify the concepts and metrics, classify the problems and methods, as well as review the important progresses and describe the state of the art. Furthermore, we provide extensive empirical analyses to compare well-known methods on disparate real networks, and highlight the future directions. In despite of the emphasis on physics-rooted approaches, the unification of the language and comparison with cross-domain methods would trigger interdisciplinary solutions in the near future.}
\end{abstract}
\end{frontmatter}

\newpage
\tableofcontents
\pagenumbering{arabic}
\newpage

\section{Introduction}\label{Chapter1}

The last decade has witnessed a great change where the research on networks, being of limited interests mainly from mathematical society under the name \emph{graph theory}, have received a huge amount of attention from many branches of sciences \cite{Newman2010BOOK,Chen2012BOOK}. Barab\'asi \cite{Barabasi2013PTRSA37120120375} argued that we have seen the emergence of \emph{network science}, which is an attempt to understand networks emerging in nature, technology and society using a unified set of tools and principles. Recently, the focus of network science has been shifting from discovering macroscopic statistical regularities (e.g., small-world \cite{Watts1998Nature393440}, scale-free \cite{Barabasi1999Science286509} and assortative mixing \cite{Newman2002PRL89208701}) to unfolding mecroscopic structural organization (e.g., communities \cite{Newman2004PRE69026113} and motifs \cite{Alon2007NatRev8450}), and further to uncovering the explicit roles played by such microscopic elements as individual nodes \cite{Pei2013JSMP12002} and links \cite{Csermely2006Book}.

The scale-free property \cite{Barabasi1999Science286509,Caldarelli2007BOOK,Barabasi2009Science325412} indicates that the roles of different nodes in the structure and function of a network may be largely different. Indeed, to identify vital nodes associated with some certain structural or functional objectives is very significant, which allows us to better control the outbreak of epidemics \cite{Pastor-Satorras2002PRE65036104,Cohen2003PRL91247901}, conduct successful advertisements for e-commercial products \cite{Leskovec2007ACM15,Lu2012PhysRep5191}, prevent catastrophic outages in power grids or the Internet \cite{Motter2002PRE66065102,Motter2004PRL93098701,Albert2004PRE69025103}, optimize the use of limited resources to facilitate information propagation \cite{Chen2013Book}, discover drug target candidates and essential proteins \cite{Csermely2013PT138333}, maintain the connectivity or design strategies for connectivity breakdowns in communication networks \cite{Albert2000Nature406378,Cohen2001PRL863682,Resende2006BOOK}, identify the best player from the records of professional sport competitions \cite{Radicchi2011PLoSONE6e17249}, and predict successful scientists as well as popular scientific publications based on co-authorship and citation networks \cite{Radicchi2009PRE80056103,Zhou2012NJP14033033,Ding2011JASIST62236} (to name just a few, see more examples in Chapter \ref{Chapter10}).

However, to identify vital nodes is not a trivial task. Firstly, criteria of vital nodes are diverse. Sometimes it needs for the nodes whose initial immunization will best protect the whole population in an epidemic spreading, sometimes it requires the nodes whose damage will lead to the widest cascading failures, and so on. Therefore, to find a universal index that best quantifies nodes' importance in every situation is not possible. Even for an explicitly given objective function, a method's performance may be highly different for different networks or under different parameters of the objective function. Secondly, the indices that require only local information of a node and the parameter-free indices are usually simpler and of lower computational complexity compared with indices based on global topological information or with many tunable parameters, but the accuracies of local and parameter-free indices are usually poor. Therefore, to find a nice tradeoff between local and global indices or between parameter-free and multi-parameter indices is a challenge. Thirdly, most known methods were essentially designed for identify individual vital nodes instead of a set of vital nodes, while the latter is more relevant to real applications since we often try to immunize or push advertisements to a group of people instead of only one person. However, putting the two most influential spreaders together does not generate a most influential set with two spreaders, because the two spreaders' influences may be largely overlapped. In fact, many heuristic algorithms with ideas directly borrowed from the identification of individual vital nodes do not perform well in identifying a set of vital nodes. Therefore, to identify a set of vital nodes emerges as a significant yet very difficult challenge recently. Lastly, to design efficient and effective methods for some new types of networks, such as spatial, temporal and multilayer networks, is a novel task in this research domain.

Thanks for both the challenging and significance, the vital nodes identification attracts increasing attentions recently. Our motivation for writing this review is fourfold. Firstly, it lacks a systematic review in this direction, and thus here we would like to clarify the concepts and metrics, classify the problems and methods, as well as review the important progresses and describe the state of the art. Secondly, we intend to make extensive empirical comparisons with well-known methods on disparate real networks under different objective functions, so that we can go one step towards the comprehensive understanding of the similarities and dissimilarities in different methods, as well as the applicabilities of different methods. Thirdly, although this review emphasizes on physics-rooted approaches and omits purely machine learning methods, we carefully choose the language that can be easily accepted by both computer scientists and physicists. We hope this effort will further trigger new perspectives and interdisciplinary solutions. Fourthly, we would like to highlight some open challenges for future studies in this domain.

This review is organized as follows. First of all, we will introduce methods in identifying individual vital nodes. Specifically speaking, in Chapter \ref{Chapter2} we will introduce centrality indices only based on structural information. In Chapter \ref{Chapter3} we will apply iterative refinement processes to rank nodes' importance. These processes are usually driven by some Markovian dynamics, such as the random walk. In Chapter \ref{Chapter4}, we will discuss how to quantify the nodes' importance by looking at the affects from the removal of one node or a group of nodes. All the above methods do not consider the features of dynamical processes involved in the objective function, while in Chapter \ref{Chapter5}, we will introduce the methods taking into consideration both the specific dynamical rules and the parameters in the objective dynamical processes. Then in Chapter \ref{Chapter6}, we will turn from the identification of individual vital nodes to the identification of a set of vital nodes, emphasizing the physics-rooted methods such as the message passing theory and percolation model. We will focus on two specific types of networks, weighted networks and bipartite networks, respectively in Chapter \ref{Chapter7} and Chapter \ref{Chapter8}. In Chapter \ref{Chapter9}, we will show extensive empirical analyses on representative methods and present their advantages, disadvantages and applicabilities subject to different networks and objective functions. We will exhibit some interesting and relevant applications of vital nodes identification algorithms in Chapter \ref{Chapter10}. Lastly, we conclude this review with the outlook of this field in Chapter \ref{Chapter11}.

\section{Structural centralities}\label{Chapter2}

The influence of a node is largely affected and reflected by the topological structure of the network it belongs to. In fact, the majority of known methods in identifying vital nodes only make use of the structural information, which allows the wide applications independent to the specific dynamical processes under consideration. The concept \emph{centrality} was just proposed to answer the question that how to characterize a node's importance according to the structure \cite{Freeman1977Sociometry4035,Bonacich1987AJS1170,Borgatti2005SNet2755,Borgatti2006SNet28466}. Generally speaking, a centrality measure assigns a real value to each node in the network, where the values produced are expected to provide a ranking of nodes subject to their importance. In this chapter, we will introduce the so-called \emph{structural centralities} that can be obtained based solely on structural information, however, one should notice that the centralities introduced in the next chapter are also obtained by using structural information. Since in Chapter \ref{Chapter3}, dynamical processes (e.g., random walk) and iterative refinement methods are utilized to explore the structural properties, we classify the related algorithms into another class named iterative refinement methods.

Due to the wide meanings of \emph{importance} from different aspects, many methods have been proposed. A node's influence is highly correlated to its capacity to impact the behaviors of its surrounding neighbors. For example, an influential user on twitter.com has the potential to spread news or views to more audiences directly. Therefore, a straightforward and efficient algorithm is to directly count the number of a node's immediate neighbors, resulting in the degree centrality. Chen \emph{et al.}~\cite{Chen2012PhysicaA1777} proposed an improved version to the degree centrality, called LocalRank algorithm, which takes into consideration of the information contained in the fourth-order neighbors of each node. These two algorithms are all based on the number of links among neighborhoods, while it is known that the local interconnectedness plays a negative role in the information spreading process~\cite{Eguiluz2002PRL108701,Petermann2004PRE66116}. Thus, ClusterRank~\cite{Chen2013PLoSONE8e77455} was proposed by considering both the number of immediate neighbors and the clustering coefficient~\cite{Watts1998Nature393440} of a node. In general, with the same number of neighbors, the larger the clustering coefficient of a node is, the smaller its influence is. Recently, Kitsak \emph{et al.}~\cite{Kitsak2010NatPhys6888} argued that the location of a node (whether in the central place) is more important than its degree, and they applied the $k$-core decomposition~\cite{Dorogovtsev2006PRL40601,Carmi2007PNAS10411150} that iteratively decomposes the network according to the residual degree of the nodes. The highest core order, corresponding to the smallest core a node belongs to, is then defined as this node's coreness, which is considered as a more accurate index in quantifying a node's influence in spreading dynamics~\cite{Kitsak2010NatPhys6888}. The famous H-index~\cite{Hirsch2005PNAS10216569} was also employed to quantify the influences of users in social networks~\cite{Korn2009PA2221}. A very interesting result is that the degree centrality, H-index and coreness can be considered as the initial, intermediate and steady states of a sequence driven by an discrete operator~\cite{Lu2016NatCom710168}.

The above-mentioned centrality measures are essentially based on the neighborhood of a node, while from the viewpoint of information dissemination, the node who has the potential to spread the information faster and vaster is more vital, which should be largely affected by the paths of propagation. Both \emph{eccentricity} centrality~\cite{Hage1995SNet57} and \emph{closeness} centrality~\cite{Sabidussi1966Psychometrika581,Freeman1979SNet215} think that the shorter the distance a node from all other nodes, the faster the information disseminated. However, eccentricity centrality only considers the maximum distance among all the shortest paths to the other nodes, which is very sensitive to the existence of a few unusual paths. In comparison, closeness centrality eliminates the disturbance through summarizing the distances between the target node and all other nodes. The \emph{betweenness} centrality of a node is defined as the ratio of the shortest paths which pass through the target node among all the shortest paths in the network~\cite{Freeman1977Sociometry4035}. Generally speaking, the node with the smallest closeness centrality has the best vision of the information flow, while the node with the largest betweenness centrality has the strongest control over the information flow. As a matter of fact, information is not necessary to propagate along the shortest paths. Katz centrality~\cite{Katz1953Psychometrika1839} considers all the paths in the network and assigns less weights on the longer paths. Similar to the Katz centrality, the subgraph centrality~\cite{Estrada2005PRE71} counts the number of the closed paths and gives less weights on the longer paths. The information index~\cite{Stephenson1989SNet1} also assumes that the information will lose during every hop in the network and therefore the longer the path, the more the loss. Accordingly, it computes the influence of a node by measuring the information contained in all possible paths from the target node to all the others.

As discussed before, this chapter roughly classifies structural centralities into neighborhood-based and path-based centralities, and then introduces the most representative ones.

    \subsection{Neighborhood-based centralities}

        \subsubsection{Degree centrality}

        In an undirected simple network $G(V,E)$ with $V$ and $E$ being the set of nodes and set of links respectively, the degree of a node $v_{i}$, denoted as $k_{i}$, is defined as the number of directly connected neighbors of $v_{i}$. Mathematically, $k_{i}=\sum_{j} a_{ij}$, where $A=\{a_{ij}\}$ is the adjacency matrix, that is, $a_{ij}=1$ if $v_i$ and $v_j$ are connected and 0 otherwise. Degree centrality is the simplest index to identify nodes' influences: the more connections a node has, the greater the influence of the node gets.

        To compare the influences of nodes in different networks, the normalized degree centrality is defined as
        \begin{equation}\label{Eq_NormlDegree}
          DC(i)=\frac{k_{i}}{n-1},
        \end{equation}
        where $n=|V|$ is the number of nodes in $G$ and $n-1$ is the largest possible degree. Notice that, the above normalization is indeed just for convenience, that is to say, nodes in different networks are generally not comparable even using the normalized degree centrality since the organization, function and density of those networks are different.

        The simplicity and low computational complexity of degree centrality give it a wide range of applications. Sometimes, degree centrality performs surprisingly good. For example, in the study of network vulnerability, the degree-targeted attack can destroy the scale-free networks and exponential networks very effectively, comparing with some selecting methods based on more complicated centralities like betweenness centrality, closeness centrality and eigenvector centrality~\cite{Iyer2013PLoSONEe59613}. In addition, when the spreading rate is very small, degree centrality is a much better index to identify the spreading influences of nodes than eigenvector centrality and some other well-known centralities~\cite{Klemm2012SciRep2292,Liu2016SciRep621380}.

        In a directed network $D(V,E)$, every link is associated with a direction, and then we should respectively consider a node's out-degree and in-degree. For example, thinking of twitter.com where a directed link exists from node $v_j$ to node $v_i$ if $v_j$ follows $v_i$, then, the in-degree of a node $v_i$ (i.e., the number of nodes having a directed link pointing to $v_i$) reflects $v_i$'s popularity while the out-degree of $v_i$ (i.e., the number of links from $v_i$ to other nodes) to some extent represents $v_i$'s social activity~\cite{Wasserman1994BOOK}. In weighted networks, the degree centrality are usually replaced by the strength, defined as the sum of weights of the associated links. More details will be presented in Chapter \ref{Chapter7}.

        \subsubsection{LocalRank}

        Degree centrality could be less accurate in the evaluation of the influences of nodes since it makes use of very limited information ~\cite{Chen2012PhysicaA1777,Kitsak2010NatPhys6888}. As an extension of the degree centrality, Chen \emph{et al.}~\cite{Chen2012PhysicaA1777} proposed an effective local-information-based algorithm, LocalRank, which fully considers the information contained in the fourth-order neighbors of each node. The LocalRank score of node $v_{i}$ is defined as
        \begin{equation}\label{Eq_LocalRank}
          LR(i) = \sum\limits_{j \in \Gamma_i} {Q(j)},
        \end{equation}
        \begin{equation}
          Q(j){\rm{ = }}\sum\limits_{k \in \Gamma_j} {R(k)},
        \end{equation}
        where $\Gamma_i$ is the set of the nearest neighbors of $v_{i}$ and $R(k)$ is the number of the nearest and the next nearest neighbors of $v_{k}$. LocalRank algorithm has much lower time complexity than typical path-based centralities. In fact, the computational complexity of LocalRank algorithm is $O(n\langle k\rangle ^2)$ that grows almost linearly with the scale of the networks. The LocalRank algorithm can also be extended weighted networks~\cite{Gao2013PhysicaA3925490}.

        \subsubsection{ClusterRank}

        The local clustering (also being referred to as local interconnectedness in some literatures) usually plays a negative role in the spreading processes~\cite{Eguiluz2002PRL108701,Petermann2004PRE66116}, as well as in the growth of an evolving network~\cite{Chen2013PLoSONE8e77455}. Unlike the degree centrality and LocalRank algorithm, ClusterRank not only considers the number of the nearest neighbors, but also takes into account the interactions among them~\cite{Chen2013PLoSONE8e77455}. ClusterRank is defined in directed networks, in which a link from $v_{i}$ to $v_{j}$, denoted as $(i\rightarrow j)$, implies that the information or disease will spread from $v_{i}$ to $v_{j}$. The ClusterRank score of a node $v_{i}$ is defined as
        \begin{equation}\label{Eq_ClusterRank}
          CR(i) = f({c_i})\sum\limits_{j \in \Gamma_i} {(k_j^{{\rm{out}}} + 1)},
        \end{equation}
        where $f({c_i})$ is a function of the clustering coefficient $c_{i}$ of the node $v_{i}$ in the directed network $D$, which is defined as~\cite{Chen2013PLoSONE8e77455}:
        \begin{equation}\label{Eq_ClusteringCoeff}
          c_{i}=\frac{|\{(j\rightarrow k)|j,k \in \Gamma_i^{out}\}|}{k_{i}^{out}(k_{i}^{out}-1)},
        \end{equation}
        where $k_{i}^{out}$ is the out-degree of $v_{i}$ and $\Gamma_i^{out}$ is the set of the nearest out-neighbors of $v_{i}$ (i.e., the nodes being directly pointed by $v_i$). Since the local clustering plays a negative role, $f({c_i})$ should be negatively correlated with $c_{i}$. Chen \emph{et al.}~\cite{Chen2013PLoSONE8e77455} adopted a decreasingly exponential function $f(c_{i})=10^{-c_{i}}$ as well as some other forms of $f({c_i})$, such as $\alpha^{-c_{i}}$ or $c_{i}^{\alpha}$ with a new parameter $\alpha$, which achieves similar results.

        From a related viewpoint, for the case of message spreading in a network with many communities, the information will rapidly diffuse in the local area once it reaches a community. The node who connects to many communities will have the potential to accelerate the information spreading in the global scope. Accordingly, Zhao \emph{et al.}~\cite{Zhao2014CJC43} proposed an index to compute the influence of a node by counting the number of communities the node connects with. Similarly, nodes across structural holes~\cite{Burt2004AJS110349} between groups are more likely to express valuable ideas and have higher influences, and thus Su \emph{et al.}~\cite{Su2015APS64020101} and Han \emph{et al.}~\cite{Han2015APS64058902} proposed ranking algorithms that takes into account structural holes.

        \subsubsection{Coreness}

        Degree centrality only takes into account the number of the nearest neighbors and asserts that the nodes with the same degree have the same influence in the network. Kitsak \emph{et al.}~\cite{Kitsak2010NatPhys6888} argued that the location of a node is more significant than its immediate neighbors in evaluating its spreading influence. That is to say, if a node is located in the core part of the network, the influence of the node will be higher than the one which is located in the periphery. Therefore, Kitsak \emph{et al.}~\cite{Kitsak2010NatPhys6888} proposed coreness as a better indicator for a node's spreading influence, which can be obtained by using the $k$-core (also called $k$-shell) decomposition~\cite{Dorogovtsev2006PRL40601} in networks.

\begin{figure}
          \centering
          \includegraphics[width=9cm]{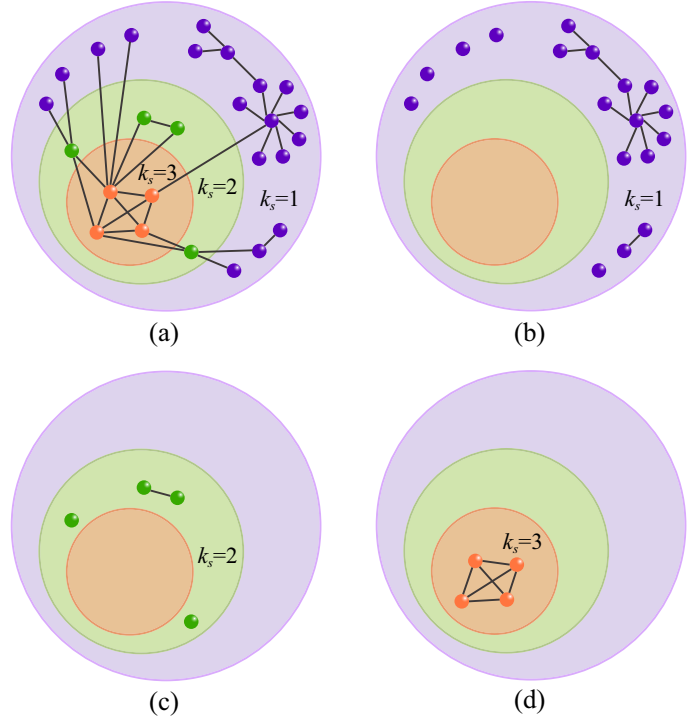}\\
          \caption{Schematic diagram of the $k$-core decomposition. After \cite{Kitsak2010NatPhys6888}.}\label{Figure1}
        \end{figure}

        Given an undirected simple network $G$, initially, the coreness $c_i$ of every isolated node $v_i$ (i.e., $k_i=0$) is defined as $c_i=0$ and these nodes are removed before the $k$-core decomposition. Then in the first step of $k$-core decomposition, all the nodes with degree $k=1$ will be removed. This will cause a reduction of the degree values to the remaining nodes. Continually remove all the nodes whose residual degree $k\le{1}$, until all the remaining nodes' residual degrees $k>1$. All the removed nodes in the first step of the decomposition form the $1$-shell and their coreness $k_s$ are all equal to $1$. In the second step, all the remaining nodes whose degrees $k=2$ will be removed in the first place. Then iteratively remove all the nodes whose residual degrees $k\le{2}$ until all the remaining nodes' whose residual degrees $k>2$. The removed nodes in the second step of the decomposition form the $2$-shell and their coreness $k_s$ are two. The decomposition process will continue until all the nodes are removed. At last, the coreness of a node $v_{i}$ equals its corresponding shell layer. Fig.~\ref{Figure1} is a simple schematic diagram of the $k$-core decomposition. Apparently, a node with a larger coreness means that the node is located in a more central position and is possibly more important in the network.

        As an index for nodes' influences, coreness can be applied to large-scale network easily for its low computational complexity, and thus has found applications in many real networks~\cite{Kitsak2010NatPhys6888,Zeng2013PLA1031,Pei2014SciRep5547}. However, there is still some space to improve the coreness. Firstly, coreness cannot be used in some classical modeled networks, such as Barabasi–Albert (BA) networks~\cite{Barabasi1999Science286509} and tree-like networks, where the coreness values of all nodes are the very small and indistinguishable. Secondly, coreness is highly coarse-grained, resulting in many indistinguishable nodes with same coreness. Thirdly, coreness only considers the residual degree of each node, however, nodes with the same $k$-shell layer may have different connecting patterns to the nodes outside this layer and if nodes in a high-order $k$-shell mostly connect to other nodes in the same shell or even higher-order shells, they are possibly less influencial~\cite{Liu2015SciRep59602}. To solve these problems, Zeng \emph{et al.}~\cite{Zeng2013PLA1031} proposed a mixed degree decomposition procedure in which both the number of links connecting to the remaining nodes, denoted as the residual degree $k_{i}^{r}$, and the number of links connecting to the removed nodes, denoted as the exhausted degree $k_{i}^{e}$, are considered. The mixed degree of a node $v_i$ is defined as
        \begin{equation}
        k_{i}^{mix}=k_{i}^{r}+\alpha k_{i}^{e},
        \end{equation}
        where $\alpha$ is a tunable parameter. Liu \emph{et al.}~\cite{Liu2013PhysicaA4154} proposed an improved method to distinguish the influences of the nodes in the same layer by measuring the sum of the shortest distances from the target node to all the nodes in the network core, i.e., the node set with the highest coreness. Hu \emph{et al.}~\cite{Hu2013APS62140101} proposed a new model by combining $k$-core and community properties of networks. Luo \emph{et al.}~\cite{Luo2016arXiv160107700} suggested that the weak ties and strong ties should be considered separately in the $k$-core decomposition. Min \emph{et al.} \cite{Min2015PLoSONE10e0136831} proposed an algorithm based on surveys on human behavior and social mechanisms. Pei \emph{et al.} \cite{Pei2014SciRep5547} found that the important nodes are consistently located in the $k$-core across dissimilar social platforms. Borge-Holthoefer and Moreno \cite{Borge-Holthoefer2012PRE85026116} studied the $k$-core decomposition in rumor dynamics. Liu \emph{et al.}~\cite{Liu2015SciRep513172} proposed a novel and effective method that firstly remove redundant links and then apply the routine $k$-core decomposition.

        \subsubsection{H-index}

\begin{figure}
          \centering
          \includegraphics[width=13cm]{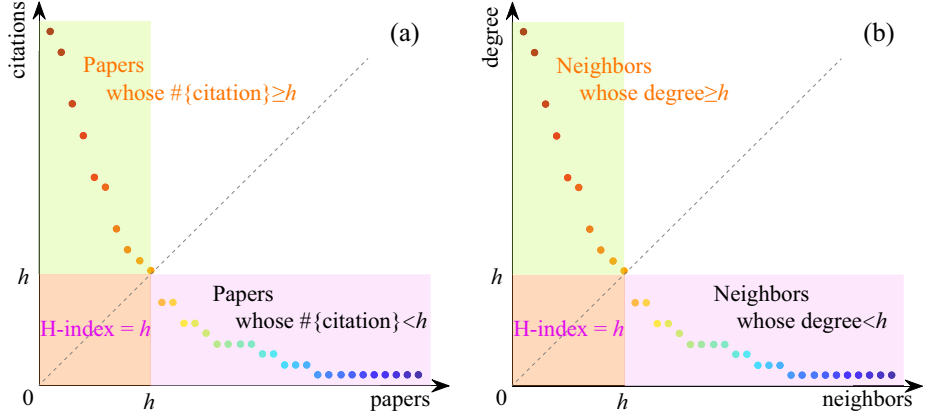}\\
          \caption{Explanation of calculating the H-index of (a) a scholar, and (b) a user in a social network. Decreasingly ranking (a) all the papers according to their citations or (b) all the neighbors of the target user according to their degrees, the H-index equals the side length of the largest square which starts from the origin (0,0). After \cite{Lu2016NatCom710168}.}\label{Figure2}
        \end{figure}

        The iterative $k$-core decomposition process requires global topological information of the network, which conditions its application to very large-scale dynamical networks~\cite{Lu2016NatCom710168}. Unlike coreness, H-index (also known as Hirsch index~\cite{Hirsch2005PNAS10216569}) is a local centrality in which every node only needs a few pieces of information, i.e., the degrees of its neighbors. H-index was originally used to measure the academic impacts of researchers or journals based on their publications and citations~\cite{Hirsch2005PNAS10216569,Braun2006Scientometrics169,Hirsch2007PNAS19193}, which is defined as the maximum value $h$ such that there exists at least $h$ papers, each with citations no less than $h$~\cite{Hirsch2005PNAS10216569,Lu2016NatCom710168}. Recently, H-index was extended to quantify the influence of users in social networks. The H-index of a user $v_{i}$ is defined as the largest $h$ satisfies that $v_{i}$ has at least $h$ neighbors for each with a degree no less than $h$~\cite{Korn2009PA2221}. Fig.~\ref{Figure2} is an insightful explanation of calculating the H-index of a scholar and a user in a social network.

        Mathematically, we can define an operator $H$ on a finite number of real variables $\{x_1,x_2,\cdots,x_m\}$ that returns the maximum integer $h$ such that among $\{x_1,x_2,\cdots,x_m\}$ there are at least $h$ elements whose values are no less than $h$. Accordingly, the H-index of a node $v_{i}$ in a social network can be written as
        \begin{equation}\label{Eq_HIndex}
          h_{i} = H(k_{j_1},k_{j_2},\cdots,k_{j_{k_i}}),
        \end{equation}
        where $k_{j_1},k_{j_2},\cdots,k_{j_{k_i}}$ is the sequence of degree values of $v_i$'s neighbours. The zero-order H-index of $v_{i}$ is defined as $h_{i}^{(0)}=k_{i}$ and the $n$-order H-index is iteratively defined as~\cite{Lu2016NatCom710168}
        \begin{equation}\label{Eq_nHIndex}
          h_{i}^{(n)} = H(k_{j_1}^{(n-1)},k_{j_2}^{(n-1)},\cdots,k_{j_{k_i}}^{(n-1)}).
        \end{equation}
        Thus, the classical H-index of $v_{i}$ equals the first-order H-index, i.e., $h_{i}=h_{i}^{(1)}$. L\"u \emph{et al.}~\cite{Lu2016NatCom710168} proved that after finite steps, the H-indices of every node $v_i$, say $h_i^{(0)},h_i^{(1)},h_i^{(2),\cdots}$ will converge to its coreness, namely
        \begin{equation}
        c_i=h_i^{\infty}.
        \end{equation}
        Therefore, degree, classical H-index, and coreness are the initial state, intermediate state, and steady state driven by the operator $H$, and all the other H-indices (in different orders) can also be used to measure nodes' importance~\cite{Lu2016NatCom710168}. The H-index can be easily extended to directed and weighted networks, in which the degree of a node is replaced by its in-degree, out-degree or node strength. Thereby the corresponding $H_{in}$-index, $H_{out}$-index, and weighted H-index, as well as in-coreness, out-coreness, and weighted coreness can also be defined~\cite{Lu2016NatCom710168}.

    \subsection{Path-based centralities}

        \subsubsection{Eccentricity}

\begin{figure}
          \centering
          \includegraphics[width=5cm]{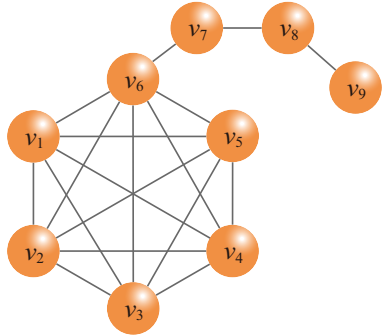}\\
          \caption{Illustration of an example network, where the most influential node should be $v_6$, but according to the eccentricity index, $v_7$ is the most influential one with ECC(7)=2.}\label{Figure3}
        \end{figure}

        In a connected network, define $d_{ij}$ as the shortest path length between node $v_{i}$ and $v_{j}$. It is believed that the shorter the distance a node $v_{i}$ from all other nodes is, the more centric the node is. Thus, the eccentricity of a node $v_{i}$ is defined as the maximum distance among all the shortest paths to the other nodes~\cite{Hage1995SNet57}:
        \begin{equation}\label{Eq_Eccentricity}
          ECC(i)=\max_{v_j\neq v_i}\{d_{ij}\},
        \end{equation}
        where $v_j$ runs over all nodes other than $v_i$. The node with a smaller eccentricity is assumed to be more influential. To compare eccentricities in different networks, the normalized eccentricity of $v_i$ is defined as
        \begin{equation}
        ECC'(i)=\frac{ECC(i)-ECC_{\min}}{ECC_{\max}-ECC_{\min}},
        \end{equation}
        where $ECC_{\max}$ and $ECC_{\min}$ are the largest and smallest eccentricities in the network, respectively. Notice that, the maximum distance can be affected by a few unusually long paths, then the eccentricity may fail to reflect the importance of nodes (see an example in Fig.~\ref{Figure3}).

        \subsubsection{Closeness centrality}

        In comparison, closeness centrality eliminates the disturbance through summarizing all the distances between the target node and all other nodes. For a connected network, the closeness centrality of a node $v_{i}$ is defined as the inverse of the mean geodesic distance from $v_i$ to all other nodes~\cite{Sabidussi1966Psychometrika581,Freeman1979SNet215}:
        \begin{equation}\label{Eq_Closeness}
          CC(i)=\frac{n-1}{\sum_{j\ne i}d_{ij}}.
        \end{equation}
        Obviously, the larger the closeness is, the more central the node is. Closeness centrality can also be understood as the inverse of the average propagation length of information in the network. In general, the node with the highest closeness value has the best vision of the information flow. Unfortunately, the original definition has a major shortcoming: when the network is not connected (in directed networks, the network has to be strongly connected), there exists some node pairs with $d_{ij}=\infty$. Thus, a very popular approach is to calculate the closeness centrality in terms of the inverse of the harmonic mean distances between the nodes:
        \begin{equation}
          CC(i)=\frac{1}{n-1} \sum_{j\ne i} \frac{1}{d_{ij}},
        \end{equation}
        in which $1/\infty = 0$. The closeness centrality of a node reflects how efficiently it exchanges information with others. Inspired by this, the network efficiency~\cite{Latora2001PRL87198701} is defined as the average efficiency of the nodes among the networks $G$:
        \begin{equation}
          Eff(G)=\frac{1}{n(n-1)} \sum_{i=1}^{n} \sum_{j=1,j\ne i}^{n} \frac{1}{d_{ij}}.
        \end{equation}

        \subsubsection{Katz centrality}

        Unlike closeness centrality that considers only the shortest path lengths between the node pairs, Katz centrality computes influences of nodes by considering all the paths in the network. Katz centrality holds that a shorter path plays a more important role. Supposing the length of a path between node $v_{i}$ and $v_{j}$ is $d_{ij}=p$, the contribution of the path to the importance of node $v_{i}$ and $v_{j}$ is $s^{p}$, where $s\in (0,1)$ is a tunable parameter. Obviously, the contribution of the longer paths will be largely suppressed if $s$ is small. Assuming that $l_{ij}^{p}$ is the number of paths between $v_i$ and $v_j$ with length $p$, we have $(l_{ij}^p)=A^p$ where $A$ is the adjacency matrix of the network. Therefore, the interaction matrix which depicts the relationship between any pair of nodes is obtained as~\cite{Katz1953Psychometrika1839}:
        \begin{equation}\label{Eq_Katz}
          K =(k_{ij})= sA + {s^2}{A^2} +  \cdots  + {s^p}{A^p} +  \cdots  = {\left( {I - sA} \right)^{ - 1}} - I,
        \end{equation}
        where $I$ is the identity matrix. In order to ensure the establishment of the right side of the Eq.~(\ref{Eq_Katz}), $s$ must be less than the reciprocal of the largest eigenvalue of $A$. The value of $k_{ij}$ is also known as Katz similarity between node $v_i$ and $v_j$ in network science. Katz centrality of a node $v_i$ is defined as~\cite{Katz1953Psychometrika1839}:
        \begin{equation}\label{Eq_KatzCentrality}
          Katz(i) = \sum_{j\neq i} {{k_{ij}}}.
        \end{equation}
        Katz centrality has a graceful mathematical structure. However, the high computational complexity of Katz centrality makes it difficult to be used in large-scale networks.

        \subsubsection{Information index}

        Information centrality index (also known as S-Z centrality index) measures the importance of nodes based on the information contained in all possible paths between pairs of nodes in the network~\cite{Stephenson1989SNet1}. The definition of ``information" comes from the theory of statistical estimation. It is supposed that there exists noise in every link, causing losing during every single transition of the information. The longer the path is, the more the loss is. Mathematically, the total amount of the information transmitted on a path equals the reciprocal of the path length. The quantity of the information that can be transmitted between a pair of nodes $v_i$ and $v_j$ equals the summation of the information transmitted through every possible path between them, denoted as $q_{ij}$. Information index considers all possible paths in the connected network, but does not require to enumerate them. In fact, $q_{ij}$ has been proved to be equivalent to the electrical conductance in an electrical network~\cite{Altmann1993SNet1,Poulin2000SNet187}. Based on the electrical network theory, the total amount of the information between $v_i$ and $v_j$ can be obtained as
        \begin{equation}
          {q_{ij}} = {\left( {{r_{ii}} + {r_{jj}} - 2{r_{ij}}} \right)^{ - 1}},
        \end{equation}
        with $r_{ij}$ being an element from the matrix
        \begin{equation}
         R=({r_{ij}}) = {(D - A + F)^{ - 1}},
        \end{equation}
        where $D$ is an $n$-dimensional diagonal matrix whose elements are the degrees of the corresponding nodes, and $F$ is an $n$-dimensional matrix whose elements all equal $1$. The information index of $v_{i}$ is then defined as the harmonic average of $q_{ij}$~\cite{Poulin2000SNet187}:
        \begin{equation}\label{Eq_InfoIndex}
          INF\left( i \right) = {\left[ {\frac{1}{n}\mathop \sum \limits_j \frac{1}{{{q_{ij}}}}} \right]^{ - 1}}.
        \end{equation}
        The information index of $v_{i}$ can also be calculated through arithmetic average~\cite{Altmann1993SNet1}. It is a variation of closeness centrality from mathematical perspective, with a different way to measure the contribution of each path. This index can be easily extended to weighted networks~\cite{Stephenson1989SNet1}.

        \subsubsection{Betweenness centrality}

        The betweenness centrality was firstly proposed by Bavelas in 1948~\cite{Bavelas1948HumOrgan16}, and then restated by Shimbel~\cite{Shimbel1953BullMathBiophys501} and Shaw~\cite{Shaw1954JPsy139} in the view of a node's potential power in controlling the information flow in a network. In 1977, Freeman~\cite{Freeman1977Sociometry4035} generalized the graph theoretical notion of the betweenness and extended it to both connected and unconnected networks, showing it the way we are using today. Generally, there exists more than one shortest paths starting from node $v_s$ and ending at $v_t$. The controllability of information flow of $v_i$ can be computed by counting all of the shortest paths passing through $v_i$. Thus, the betweenness centrality of node $v_{i}$ can be defined as
        \begin{equation}\label{Eq_Betweenness}
          BC\left( i \right) = \mathop \sum \limits_{i \ne s,i \ne t,s \ne t} \frac{{g_{st}^i}}{{{g_{st}}}},
        \end{equation}
        where $g_{st}$ is the number of the shortest paths between $v_{s}$ and $v_{t}$ and $g_{st}^i$ is the number of the paths which pass through $v_i$ among all the shortest paths between $v_{s}$ and $v_{t}$. Let's consider two extreme cases in a star network with $n$ nodes. Obviously, a leaf node isn't on any shortest path and thus its betweenness centrality equals $0$ according to Eq.~(\ref{Eq_Betweenness}), while for the central node in the star, in another extreme situation, its betweenness centrality equals $(n-1)(n-2)/2$, which is the maximum possible value of betweenness centrality. Thus, the normalized betweenness centrality of node $v_i$ in an undirected network is
        \begin{equation}\label{Eq_Betweenness2}
          BC'\left( i \right) = \mathop \frac{2}{{\left( {n - 1} \right)\left( {n - 2} \right)}} \sum \limits_{i \ne s,i \ne t,s \ne t} \frac{{g_{st}^i}}{{{g_{st}}}}.
        \end{equation}
        For the convenience of the calculation, researchers also adopte an approximately normalized form of betweenness~\cite{Newman2010BOOK}, say
         \begin{equation}\label{Eq_Betweenness3}
          BC'\left( i \right) = \mathop \frac{1}{n^2} \sum \limits_{i \ne s,i \ne t,s \ne t} \frac{{g_{st}^i}}{{{g_{st}}}}.
        \end{equation}

        Goh \emph{et al.}~\cite{Goh2001PRL278701} studied the transport of data packets in scale-free networks in which all the packets passed through the shortest paths. If there are more than one shortest paths between a given pair of nodes, the data packet will encounter branching points and will be equally distributed to these paths. In fact, the betweenness centrality of node $v_i$ is equivalent to the load at $v_i$ when every node sends one data packet to every other node~\cite{Goh2001PRL278701}, with the interference and delay of the data packet being neglected. Goh \emph{et al.}~\cite{Goh2001PRL278701} found that the distribution of the betweenness centralities follows a power law, which is valid for both undirected and directed scale-free networks.

        Everett and Borgatti~\cite{Everett1999JMathSociol181} proposed a group betweenness centrality to measure the centrality of a group of nodes, in terms of the shortest paths which pass through at least one of the nodes in this group. Kolaczyk \emph{et al.}~\cite{Kolaczyk2009SNet190} further discussed the relationship of group betweenness centrality and co-betweenness centrality, which computes the centrality of a group of nodes in terms of the shortest paths that pass through all the nodes in the group.

        However, many factors, such as loading balancing and fault tolerance, may lead to some compromised strategies in which data packets are not always delivered through the shortest path in the real commercial transport network~\cite{Dolev2010JACM5725,Tang2011PRE84026116}. It has also been shown that the selection of the shortest paths between all pairs of nodes may lead to traffic congestion problem~\cite{Yan2006PRE73046108}. Freeman \emph{et al.}~\cite{Freeman1991SNet141} proposed an algorithm called flow betweenness centrality, which considers all the paths between a given pair of nodes. Mathematically, the flow betweenness centrality of $v_i$ is defined as
        \begin{equation}\label{Eq_FBC}
          FBC(i)= \sum \limits_{s \neq t \neq i} \frac{\tilde g_{st}^i}{{\tilde g}_{st}},
        \end{equation}
        where ${\tilde g}_{st}$ is the maximal flow starting from $v_s$ and ending at $v_t$ and $\tilde g_{st}^i$ is the amount of flow that starts from $v_s$ and ends at $v_t$ passing through $v_i$. In the max-flow problem, a \emph{s-t cut} is a partition which divides nodes $v_s$ and $v_t$ into two disconnected components. The cut set capacity is the summation of the capacities of the individual links composing the set. The famous \emph{min-cut, max-flow} theorem proved that the maximum flow from $v_s$ to $v_t$ exactly equals the minimum cut capacity~\cite{Freeman1991SNet141,Ford2015Book}.

        Communicability betweenness centrality also takes all the paths into consideration by introducing a scaling so that the longer paths carry less contributions~\cite{Estrada2009PhysicaA764}. If $W_{st}^{\left( p \right)}$ is the number of the paths connecting node $v_s$ and $v_t$ with length $p$, the communicability between them can be defined as
        \begin{equation}
          G_{st} = \mathop \sum \limits_{p=0}^\infty \frac{1}{p!}W_{st}^{\left( p \right)} = \sum \limits_{p = 1}^\infty  \frac{{{{\left( {{A^p}} \right)}_{st}}}}{{p!}} = {\left( {{{\rm{e}}^A}} \right)_{st}},
        \end{equation}
        where $A$ is the adjacency matrix. Then, the communicability betweenness centrality of node $v_i$ in an undirected network is
        \begin{equation}\label{Eq_CBC}
          CBC(i) = \frac{2}{{\left( {n - 1} \right)\left( {n - 2} \right)}} \sum \limits_{i \ne s,i \ne t,s \ne t} \frac{{G_{st}^i}}{{{G_{st}}}},
        \end{equation}
        where $G_{st}^i$ is the corresponding communicability between $v_s$ and $v_t$ which involves node $v_i$. The lower and upper bounds of communicability betweenness centrality were discussed in~\cite{Estrada2009PhysicaA764}.

        Random-walk betweenness centrality is another famous betweenness variant that counts all paths in networks and gives more weights to shorter paths~\cite{Newman2005SNet39}. Just as its name implies, this algorithm counts the expected times a node is traversed in a random walk between an arbitrary pair of nodes. The random-walk betweenness centrality of node $v_i$ can be written as
        \begin{equation}\label{Eq_RWBC}
          RWBC\left( i \right) = \frac{{\mathop \sum \nolimits_{s \ne t} I_{st}^i}}{{n\left( {n - 1} \right)/2}},
        \end{equation}
        where $I_{st}^i$ is the number of walks traversed $v_i$ starting from $v_s$ and ending at $v_t$. To calculate $I_{st}^i$, we firstly construct the Laplacian matrix of the network as
        \begin{equation}
        L=D-A,
        \end{equation}
        and then compute the Moore-Penrose pseudoinverse~\cite{Moors1920BAMS26394,Penrose1955PCPS51406} of the $L$, say $T=L^+$. Finally, we can get
        \begin{equation}
        I_{st}^i = \frac{1}{2}\mathop \sum \nolimits_j {a_{ij}}\left| {T_{is} - T_{it} - T_{js} + T_{jt}} \right|.
        \end{equation}
        More detailed computational procedure as well as the difference between betweenness centrality and random-walk betweenness centrality can be found in Refs.~\cite{Newman2005SNet39,Zhou2006CPL232327,Fouss2007IEEE19355}.

        \subsubsection{Subgraph centrality}

        Subgraph centrality of a node $v_i$ is defined as a weighted sum of the numbers of all closed paths starting from and ending at $v_i$~\cite{Estrada2005PRE71}. Similar to closeness centrality and information index, a path with shorter length will make a greater contribution to the importance of revelent nodes. This rule is inspired by the observation of motifs~\cite{Milo2002Science298824} in real-world networks. The number of closed paths of length $p$ starting from and ending at $v_i$ can be obtained by the $i$th diagonal element of the $p$th power of the adjacency matrix, i.e.,  $(A^p)_{ii}$. In fact, for $p=1$, $(A^1)_{ii}=0$, and for $p=2$, $(A^2)_{ii}=k_i$. The subgraph centrality of node $v_{i}$ is then defined as
        \begin{equation}\label{Eq_Subgraph}
          SC\left( i \right) = \mathop \sum \limits_{p = 1}^\infty  \frac{(A^p)_{ii}}{{p!}}.
        \end{equation}
        If the network is a simple network with $n$ nodes, the subgraph centrality of node $v_{i}$ can be calculated as~\cite{Estrada2005PRE71}
        \begin{equation}\label{Eq_Subgraph2}
          SC\left( i \right) = \mathop \sum \limits_{p = 1}^\infty  \frac{(A^p)_{ii}}{{p!}} = \mathop \sum \limits_{j = 1}^n {\left( {\xi _j^i} \right)^2}{{\rm{e}}^{{\lambda _j}}},
        \end{equation}
        where ${\lambda _j}\left( {j = 1,2, \ldots ,n} \right)$ is the eigenvalue of $A$, ${\overrightarrow{\xi} _j}$ is the corresponding eigenvector of ${\lambda _j}$, and $\xi _j^i$ is the $i$th entity of ${\overrightarrow{\xi} _j}$. However, if the network is a simple and connected network with $n>1$, the subgraph centrality of node $v_{i}$ satisfies the inequality~\cite{Estrada2005PRE71}
        \begin{equation}\label{Eq_Subgraph3}
          SC\left( i \right)  \le \frac{1}{n}\left( {{e^{n - 1}} + \frac{{n - 1}}{e}} \right).
        \end{equation}
        The equality holds if and only if the network is complete. Subgraph centrality has good performance on finding more important nodes, and can also be used to detect motifs in networks~\cite{Estrada2005PRE71}.

\section{Iterative refinement centralities}\label{Chapter3}

The influence of a node is not only determined by the number of its neighbors (as what we have discussed in Chapter \ref{Chapter2}), but also determined by the influence of its neighbors, known as the \emph{mutual enhancement effect}~\cite{Wittenbaum1999JPSP77967}. In this chapter, we will select some typical iterative refinement centralities in which every node gets support from its neighbors. Among these algorithms, eigenvector centrality~\cite{Bonacich1972JMS113} and cumulative nomination algorithm~\cite{Poulin2000SNet187} are designed in undirected networks, while PageRank~\cite{Brin1998CNIS107}, HITs~\cite{Kleinberg1999JACM46604} and their variants are mainly used in directed networks. PageRank was originally used to rank web pages and was the core algorithm of Google search engine. To solve the dangling node problem\footnote{PageRank is actually a random walk process on directed network. If the network is strongly connected, the probability of the walker on each node can reach a steady state. However, if there exists some nodes without outgoing links, named dangling nodes, the walker will be trapped at these nodes, leading to the failture of PageRank.}, PageRank introduces a random jumping factor which is a tunable parameter whose optimal value depends on both the network structure and the objective function. LeaderRank \cite{Lu2011PLoSONE6e21202} gives a simple but effective solution by adding a ground node which connects to every node through bidirectional links. The ground node, associated with $2n$ links, make the network strongly connected and eliminate all the dangling nodes. Then, the steady distribution of the visiting probabilities on all nodes for a random walk is used to quantify nodes' importance. As nodes may play different roles in directed networks, HITs algorithm evaluates each node from two aspect: authority and hub~\cite{Kleinberg1999JACM46604}. In a directed network, the authority score of a node equals the summation of the hub scores of all the nodes that point to this node while the hub score of a node equals the summation of the authority scores of all the nodes being pointed of by this node.

    \subsection{Eigenvector centrality}
    Eigenvector centrality supposes that the influence of a node is not only determined by the number of its neighbors, but also determined by the influence of each neighbor~\cite{Bonacich1972JMS113,Bonacich1987AJS1170}. The centrality of a node is proportional to the summation of the centralities of the nodes to which it is connected~\cite{Bonacich2007SNet555}. The importance of a node $v_i$, denoted by $x_i$, is
    \begin{equation}\label{Eq_EC}
      x_i = c\mathop \sum \limits_{j = 1}^n {a_{ij}}{x_j},
    \end{equation}
    which can be written in the matrix form as
    \begin{equation}\label{Eq_EC2}
      \overrightarrow{x} = cA\overrightarrow{x},
    \end{equation}
    where $c$ is a proportionality constant. Generally, $c= 1 \left.\right/ \lambda$ in which $\lambda$ is the largest eigenvalue of $A$. Eigenvector centrality can be efficiently computed by power iteration method~\cite{Hotelling1936Psychometrika27}. In the beginning of power iteration, the score of each node is initialized as $1$. Each node then shares out its score evenly to its connected neighbors and receives new values in every iteration round. This process repeats until the values of nodes reach the steady state. From the viewpoint of this iteration method, PageRank algorithm is a variant of the eigenvector centrality.

    Eigenvector centrality score much prefers to concentrate in a few nodes under common conditions, making it hard to distinguish among the nodes. Martin \emph{et al.}~\cite{Martin2014PRE52808} proposed a modified eigenvector centrality, called nonbacktracking centrality, based on the leading eigenvector of the Hashimoto or nonbacktracking matrix~\cite{Krzakala2013PNAS20935} of the undirected networks. The main idea of nonbacktracking centrality is that: when calculating the centrality score of node $v_i$, the values of $v_i$'s neighbors in the summation will no longer consider the effect of $v_i$, which is similar to the cavity network method that will be introduced in Chapter~\ref{Chapter6}. In directed networks, however, it is common that many nodes only have out-degree, leading to zero status after the first round of the power iteration. To solve this problem, Bonacich \emph{et al.}~\cite{Bonacich2001SNet23191} proposed a variant of eigenvector centrality, called alpha centrality:
    \begin{equation}\label{Eq_AlphaCentrality}
      \overrightarrow{x} = \alpha A\overrightarrow{x} + \overrightarrow{e},
    \end{equation}
    where $\overrightarrow{e}$ is a vector of the exogenous sources of status and $\alpha$ is a parameter which reflects the relative importance of endogenous versus exogenous factors. In Ref.~\cite{Bonacich2001SNet23191}, $\overrightarrow{e}$ was assumed to be a vector of ones, resulting in essentially the same solution with Katz centrality.

    \subsection{Cumulative nomination}

    Eigenvector centrality may not always be an ideal method for its slow rate of convergence and sometimes trapping into an endless loop. Cumulative nomination scheme supposes that more central individuals will be nominated more often in social networks~\cite{Poulin2000SNet187}, and considers the nomination values of each node and its neighbors'. Let $\tilde p_i(t-1)$ denotes the accumulated number of nominations of node $v_i$ after $t-1$ iterations. At the beginning, the nomination value of each node is initially set as one. Then every node gets nominations from all of its neighbors and the updated value equals its original value plus the summation of the neighbors' values in each iteration. The cumulative nomination of node $v_i$ after $t$ iterations is
    \begin{equation}\label{Eq_CumulativeNomination}
      \tilde p_i(t) = \tilde p_i(t-1) + \mathop \sum \limits_j {a_{ij}}\tilde p_j(t-1).
    \end{equation}
    The normalized cumulative nomination of $v_i$ can be computed as
    \begin{equation}\label{Eq_CumulativeNomination2}
      p_i(t) = \frac{{\tilde p_i(t)}}{{\mathop \sum \nolimits_j \tilde p_j(t)}}.
    \end{equation}
    The iteration of the nomination will stops if the normalized cumulative nominations of all nodes reach the steady state. Cumulative nomination has the similar formula with alpha centrality. While the difference is that the vector $\overrightarrow{e}$ in alpha centrality is a fixed vector, however, the corresponding element in cumulative nomination, say $\tilde p_i^t$, equals the value of the node in the last iteration, which improves the rate of convergence.

    \subsection{PageRank}

    PageRank algorithm~\cite{Brin1998CNIS107} is a famous variant of eigenvector centrality and is used to rank websites in Google search engine and other commercial scenarios~\cite{Langville2011BOOK}. The traditional keyword-based websites ranking algorithms are vulnerable to malicious attack which improves the influence of a website by increasing the density of irrelevant keywords. PageRank distinguishes the importance of different websites by random walking on the network constructed from the relationships of web pages. Similar to eigenvector centrality, PageRank supposes that the importance of a web page is determined by both the quantity and the quality of the pages linked to it. Initially, each node (i.e., page) gets one unit PR value. Then every node evenly distributes the PR value to its neighbors along its outgoing links. Mathematically, the PR value of node $v_i$ at $t$ step is
    \begin{equation}\label{Eq_PageRank}
      PR{_i}\left( t \right) = \mathop \sum \limits_{j = 1}^n {a_{ji}}\frac{P{R_j}\left( {t - 1} \right)}{k_j^{\rm{out}}},
    \end{equation}
    where $n$ is the total number of nodes in the network, and $k_j^{\rm{out}}$ is the out-degree of node $v_j$. The above iteration will stops if the PR values of all nodes reach the steady state. A major drawback of the above random walk process is that the PR value of a dangling node (the node with zero out-degree) cannot be redistributed, and then Eq. (\ref{Eq_PageRank}) cannot guarantee the convergence (there are some other cases where Eq. (\ref{Eq_PageRank}) will not converge~\cite{Chen2012BOOK}). To solve these problems, a random jumping factor was introduced by assuming that the web surfer will browse the web pages along the links with probability $s$, and leave the current page and open a random page with probability $1-s$. Accordingly, Eq. (\ref{Eq_PageRank}) is modified as
    \begin{equation}\label{Eq_PageRank2}
      P{R_i}\left( t \right) = s\mathop \sum \limits_{j = 1}^n {a_{ji}}\frac{{P{R_j}\left( {t - 1} \right)}}{{k_j^{out}}} + ({\rm{1 - }}s)\frac{{\rm{1}}}{n}.
    \end{equation}
    Eq. (\ref{Eq_PageRank2}) will return to Eq. (\ref{Eq_PageRank}) when $s=1$. The random jumping probability $s$ is usually set around $0.85$, but indeed is should be tested in different scenarios~\cite{Brin1998CNIS107}.

    PageRank has been utilized far beyond its initial intent and design~\cite{Gleich2015SIAM57321}. It has been applied to rank a broad range of objects via their network structure: to rank images~\cite{Jing2008ACM307} and books~\cite{Meng2009WebS22}, to rank genes~\cite{Morrison2005BMC6233} and proteins~\cite{Freschi2007IEEE42} in Biology and Bioinformatics, to rank molecules~\cite{Mooney2012JCC33853} in Chemistry, to rank brain regions~\cite{Zuo2012CC221682} and neurons~\cite{Crofts2011IntMath7233} in Neuroscience, to rank host names~\cite{Arasu2002WWW107}, Lonux kernels~\cite{Chepelianskii2010arXiv10035455} and programming interfaces~\cite{Kim2013ACM4193} in complex information systems, to rank leaders~\cite{Java2006WWW22,Weng2010ICWSWDM261,Lu2011PLoSONE6e21202} in social networks, to rank scientists~\cite{Liu2005IPM411462,Ding2009JASIST602229}, papers~\cite{Chen2007JInf18,Ma2008IPM44800,Sayyadi2009SDM533}, and journals~\cite{Bollen2006Scientometrics69669,Butler2008NatureNews4516,West2010CRL71236} in Bibliometrics, to rank players~\cite{Radicchi2011PLoSONE6e17249} and teams~\cite{Govan2008SAS2008} in sports, and so on. Ghoshal and Barab{\'a}si~\cite{Ghoshal2011NatComm2394} studied the emergence of super-stable nodes in perturbed networks when evaluated the spreading ablities of the nodes with PageRank. They investigated the rankings for different topological properties and found that PageRank is sensitive to pertubations in random networks while is stable in scale-free networks.

    \subsection{LeaderRank}

    The random jumping probability for each node is the same in PageRank, implying that a web surfer has the same probability to leave from an informative web page and from a trivial page, which doesn't conform to the actual situation. In addition, how to determine the parameter $s$ to achieve the best ranking depends on specific scenarios. LeaderRank gives a simple yet effective solution via adding a ground node which connects to all other nodes through $n$ bidirectional links~\cite{Lu2011PLoSONE6e21202}. Thus, the network is strongly connected and consists of $n+1$ nodes and $m+2n$ directed links. Initially, each node except the ground node is assigned one unit score. Every node distributes its score to its neighbors equally along the outgoing links. The value of node $v_i$ at time step $t$ stage is
    \begin{equation}\label{Eq_LeaderRank}
      LR_{i}\left( {\rm{t}} \right) = \mathop \sum \limits_{j = 1}^{n + 1} a_{ji}\frac{L{R_j}\left( {t - 1} \right)}{k_j^{\rm{out}}}.
    \end{equation}
    After the scores of all nodes reach the steady state, the value of the ground node will be evenly distributed to all other nodes. The final score of node $v_i$ is
    \begin{equation}\label{Eq_LeaderRank2}
      LR_{i} = L{R_i}\left( {{t_c}} \right) + \frac{{L{R_g}\left( {{t_c}} \right)}}{n}.
    \end{equation}

    This adaptive and parameter-free algorithm has very good performance for online social networks. Extensive experiments show that LeaderRank converges faster since the network is strongly connected with the diameter being only $2$~\cite{Lu2011PLoSONE6e21202,Li2014PhysicaA40447}. LeaderRank outperforms PageRank in terms of ranking effectiveness, as well as robustness against manipulations and noisy data~\cite{Lu2011PLoSONE6e21202}. The same idea, i.e., adding a ground node, was also proved to be effective to solve the accuracy-diversity dilemma in recommender systems~\cite{Zhou2013PLoSONE8e70094}. Li \emph{et al.}~\cite{Li2014PhysicaA40447} further improved the LeaderRank algorithm by allowing popular nodes get more values from the ground node in the random walks. For a node $v_i$ whose in-degree is $k_i^{in}$, the weight of the link from the ground node to $v_i$ is $w_{gi}= (k_i^{in})^\alpha$ where $\alpha$ is a free parameter. The weights of other links remain unchanged. Thus, the score of $v_i$ at time step $t$ for the improved LeaderRank~\cite{Li2014PhysicaA40447} is
    \begin{equation}\label{Eq_LeaderRankImproved}
      L{R_i}\left( {\rm{t}} \right) = \mathop \sum \limits_{j = 1}^{n + 1} w_{ji}\frac{L{R_j}\left( {t - 1} \right)}{\sum\nolimits_l^{n + 1} w_{jl} }.
    \end{equation}

    \subsection{HITs}

    HITs algorithm considers two roles of each node in the network, namely authority and hub~\cite{Kleinberg1999JACM46604}. In the World Wide Web, the authoritative websites are always reliable and provides original information of specific topics while the hub websites are those who link to many related authorities. Hubs and authorities exhibit a mutually reinforcing relationship: a good hub points to many authorities, and a good authority is pointed at by many hubs~\cite{Kleinberg1999JACM46604,Lempel2000ComputNet33387}. In a directed network, the authority score of a node equals the summation of the hub scores of all the nodes that point to this node while the hub score of a node equals the summation of the authority scores of all the nodes being pointed at by this node. In a network of $n$ nodes, denote the authority score and hub score of node $v_i$ at time $t$ by $a_i(t)$ and $h_i(t)$, respectively. In the begining, the hub scores of all nodes are assigned as $1$. Mathematically, the authority and hub values of node $v_i$ at time $t$ are
    \begin{equation}\label{Eq_HITs}
      {a'}_i(t) = \mathop \sum \limits_{j = 1}^n {a_{ji}}h_j(t - 1), \qquad
      {h'}_i(t) = \mathop \sum \limits_{j = 1}^n {a_{ij}}{a'}_j(t).
    \end{equation}
    After each iteration, the scores of each node should be normalized as
    \begin{equation}\label{Eq_HITs2}
      {a}_i(t) = \frac{{a'}_i(t)}{{\left\| {{a'}(t)} \right\|}}, \qquad
      {h}_i(t) = \frac{{h'}_i(t)}{{\left\| {{h'}(t)} \right\|}},
    \end{equation}
    where $\left\| x \right\| = \sqrt {\sum\nolimits_i {{(x_i)^2}} } $ is the norm of $\overrightarrow{x}$.

    The iteration will cease if the normalized scores of all nodes reach the steady state. The matrix form of HITs algorithm and its further discussion can be found in Refs.~\cite{Newman2010BOOK,Chen2012BOOK}. HITs algorithm is convergent, and the final authority weights (hub weights) are the coordinates of the normalized greatest eigenvector of $A^TA$ (of $AA^T$) in which $A^TA$ is the co-citation matrix~\cite{Small1973JASIS24265} while $AA^T$ is the bibliographic matrix~\cite{Kessler1963AMDOC1410} in the field of bibliometrics.

    Chakrabarti \emph{et al.} improved the HITs algorithm by analyzing both text and links of the WWW~\cite{Chakrabarti1998CNIS3065}. To determine the authority for a given topic or query, HITs algorithm exhibits the topic drift phenomenon in the hypertext link structure which includes tightly interconnected clusters~\cite{Borodin2001ACM415}. Ng \emph{et al.}~\cite{Ng2001ACM258} studied the stability of ranking under small perturbations using HITs algorithm. Benzi \emph{et al.}~\cite{Benzi2013LAA4382447} mapped the directed network onto a bipartite undirected network and computed the hub and authority rankings using matrix functions. HITs algorithm and its variants are also widely used in many bipartite networks (see the detailed introduction to BiHITS in Chapter \ref{Chapter8})~\cite{Deng2009ACMSIGKDD239}.

   \subsection{SALSA}

\begin{figure}
       \centering
       \includegraphics[width=8cm]{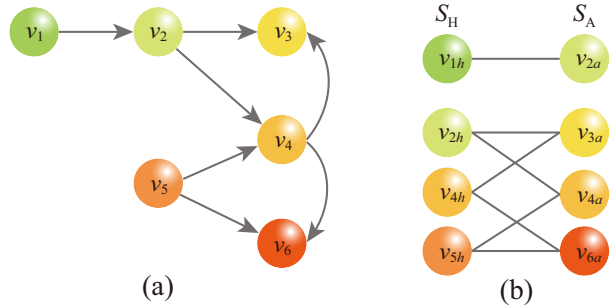}
       \caption{An example of building a bipartite undirected network from a directed network, where in the bipartite representation, the isolated nodes are removed. After~\cite{Lempel2000ComputNet33387}.}\label{Figure4}
     \end{figure}

    Short for \emph{stochastic approach for link structure analysis}~\cite{Lempel2000ComputNet33387}, SALSA is a famous variant of HITs algorithm, which builds upon the stochastic properties of random walks on directed networks. 
    The first step of SALSA is mapping the directed network onto a bipartite undirected network. The nodes with non-zero out-degree constitute the set of hubs (i.e., $S_H$), while the nodes with non-zero in-degree constitute the set of authorities (i.e., $S_A$). If hub node $v_{i_h}$ points to the authority node $v_{i_a}$ in the original directed network, then the two nodes will be connected in the bipartite network. Fig.~\ref{Figure4} gives an example of this mapping process. It is assumed that each step of the random walk on the original network consists of two adjacent edges in the corresponding bipartite network, and the two edges will inevitably start off from different sides. Every path of length $2$ represents a traversal of one link in the proper direction (passing from the hub-side to the authority-side), and a retreat along a link ~\cite{Lempel2000ComputNet33387}. For example, the path $\{(v_{2h},v_{3a}),(v_{3a},v_{4h})\}$ in Fig.~\ref{Figure4} represents a traversal of the directed link from $v_2$ to $v_3$ and a retreat along the link from $v_4$ to $v_3$. In the initial stage, the values of the nodes in hub-side are assigned to be $1$, denoting by $\overrightarrow{h}(0)$. Thus the values of nodes in hub-side at time $t$ can be computed as

     \begin{equation}\label{Eq_SALSA}
       \overrightarrow{h}(t) = \tilde H{\overrightarrow{h}(t-1)},
     \end{equation}
where
     \begin{equation}
       {(\tilde H)_{ij}} = {\tilde h_{ij}} = \mathop \sum \limits_{i,j \in {S_H},\alpha  \in {S_A}} \frac{{{a_{i\alpha }}}}{{{k_j}}} \cdot \frac{{{a_{j\alpha }}}}{{{k_\alpha }}}.
     \end{equation}

Similarly, the values of authorities can be obtained by the random walk starting from authority-side:
     \begin{equation}\label{Eq_SALSA2}
       \overrightarrow{a}(t) = \tilde A{\overrightarrow{a}(t - 1)},
     \end{equation}
where
     \begin{equation}
       {(\tilde A)_{\alpha \beta }} = {\tilde a_{\alpha \beta }} = \mathop \sum \limits_{\alpha ,\beta  \in {S_A},{\rm{ }}i \in {S_H}} \frac{{{a_{i\alpha }}}}{{{k_\beta }}}\frac{{{a_{i\beta }}}}{{{k_i}}}.
     \end{equation}

Notice that, the two random walks that used to calculate the scores of hubs and authorities are independent to each other, which is much different with HITs where the two values are mutual reinforced. However, both SALSA and HITs employ the same meta-algorithm, and SALSA is equivalent to a weighted in-degree analysis of the link-structure of the network, making it computationally more efficient than HITs ~\cite{Lempel2000ComputNet33387}.

\section{Node operation}\label{Chapter4}

Scale-free networks are stable against random failures but vulnerable under intentional attack~\cite{Albert2000Nature406378, Cohen2001PRL863682}, which implies that some nodes are more important on preserving the network connectivity. According to~\cite{Albert2000Nature406378, Cohen2001PRL863682}, a node (or a set of nodes) is important if its removal would largely shrink the giant component. Finding a set of nodes by node removal and contraction methods coincides with the idea of determining a system's core in system science~\cite{Li2004SysE2213}. The core of a system is defined as a set of nodes whose importance can be simply quantified by the number of connected components showing up after their removal~\cite{Wang1993JSEE41}. Nevertheless, different connected components might have different amounts of nodes and different functions. Thus, only considering the number of components is not able to distinguish the roles of different nodes located in different positions. In contrast, some path-based centralities are sophisticated on such issue, like closeness centrality and betweenness centrality (see details in Section~\ref{Section5_1}). For example, Li \emph{et al.}~\cite{Li2004SysE2213} developed a shortest-distance-based method by quantifying the damage to network connectivity.

Following the same node removal strategy, some research focused on the damage to network's stability or robustness. In specific, one node (or a set of nodes) is considered to be important if its removal would largely reduce the network's stability or robustness, or make the network more vulnerable. Many path-based methods have been proposed to measure the vulnerability. For example, Bao \emph{et al.}~\cite{Bao2012SysEE34168} proposed that a robust network should have more disjoint paths (two paths are disjoint if they have no common intermediary nodes \cite{Sun2004BOOK}) among the nodes. Dangalchev~\cite{Dangalchev2006PhysicaA365556} evaluated the average closeness centrality after the removal of nodes, named residual closeness, to measure the vulnerability of network. Besides, Chen \emph{et al.}~\cite{Chen2004JCIC25129} adopted the number of spanning tree of networks to measure the reliability of communication networks, which assumes a more reliable network tends to have more spanning trees. The number of spanning trees is apparently related to the number of distinct paths between nodes. From this point of view, these methods are all path-based. Besides, the removal of nodes would also impact dynamical processes on networks~\cite{Restrepo2006PRL97094102}. Therefore, the importance of node can also be measured by the change in the largest eigenvalue of the network adjacency matrix~\cite{Restrepo2006PRL97094102}.

Another kind of operation on nodes is to contract a node and its neighbors into a new node, named node contraction method. After contracting one node, if the entire network becomes more agglomerate, this node is considered to be more important~\cite{Tan2006SysETP1179}. For instance, the star network will be contracted to one single node after contracting the central node. In the following, we classify all these methods into four categories basing on different kinds of operations on nodes and different objective functions under consideration.

    \subsection{Connectivity-sensitive method}
    As we described above, we can measure the damage to network connectivity after nodes removal from three aspects, i.e., the size of giant component, the number of connected components, and the shortest distance between nodes. The first two are very intuitive. As to the shortest distance, directly calculating the change of average shortest distance of networks is not exact enough. In particular, the loss of the network connectivity comes from three situations~\cite{Li2004SysE2213}: (i) the disconnections between the set of removed nodes and the remaining nodes; (ii) the disconnections between the removed nodes; (iii) the disconnections between the remaining nodes after node removal. The first two are considered as direct loss of the network connectivity, while the third one is indirect loss. A basic assumption is that the direct loss and short-distance-connections are more destructive than the indirect loss and long-distance-connections.

    The loss of connectivity between two nodes can be quantified by the reciprocal of distance. After the removal of one node, say $v_i$, suppose some disjunct pairs of nodes appear and are denoted by set $E$ (including both direct and indrect losses). The importance of $v_i$ can be defined as~\cite{Li2004SysE2213}:
        \begin{equation}\label{EqDelNodeShortPath}
          DSP(i) = \sum_{(j,k)\in E} \frac{1}{d_{jk}},
        \end{equation}
    where $d_{jk}$ is the distance between node $v_j$ and $v_k$ before the removal of $v_i$. When $j=i$ or $k=i$, $DSP(i)$ is the direct loss; when $j\neq k\neq i$, it is the indirect loss. This formula is simple but usually not effective if only one node is removed, since a large-scale real-world network is not likely to be broke into pieces by removing only one node. In this case, the indirect loss is $0$, and DSP degenerates to the sum of the inverse of distances between the removed node to all other nodes, equivalents to the closeness centrality.

    \subsection{Stability-sensitive method}
    Although the removal of one node is not likely to crush real-world networks, it would indeed impact the network stability or vulnerability. A strong evidence can be found in communication networks, where removing a vital node may not lead to the failure of transferring messages, but is very likely to delay the transmission, and even leads to information congestion. In other words, the transmittability would be damaged and may lead to unstable transmission. From the aspect of network structure, there are some metrics for the stability or vulnerability of networks. Therein, the shortest distances between nodes are the mostly applied. For instance, closeness centrality can be considered as a measurement of propagation length of information in communication networks. If the closeness of a network (i.e., the summation of closenesses of all nodes) is high, the transmission through this network would be much efficient. This measurement was employed to measure the network vulnerability by Dangalchev~\cite{Dangalchev2006PhysicaA365556}. If we go one step further, this achievement can be utilized to measure the importance of nodes. Namely, a node is considered to be more important if its removal would make the network more vulnerable. In the following, we present some related methods of measuring network vulnerability based on node removal strategy.

    The residual closeness centrality is a variant of the closeness centrality proposed by Latora and Marchiori~\cite{Latora2001PRL87198701}. Dangalchev introduced an exponential function $f(d_{jk})=2^{d_{jk}}$ to redefine the closeness of node $v_i$, as
        \begin{equation}
          C(i) = \sum_{k\neq i}\frac{1}{2^{d_{ik}}}.    \nonumber
        \end{equation}
        Then the closeness of $G$ is defined as
        \begin{equation}\label{EqDelNodeResidual}
          RCC = \sum_{i}C(i) = \sum_{i}\sum_{k\neq i}\frac{1}{2^{d_{ik}}}.
        \end{equation}
    For a star graph with $n$ nodes $RCC=\frac{n-1}{2}+(n-1)(\frac{1}{2}+\frac{n-2}{4})=\frac{(n-1)(n+2)}{4}$, while for a chain with $n$ nodes $RCC=2n-4+\frac{1}{2^{n-2}}$. After removing a node, the vulnerability of network would change, which can be captured by $RCC(i)$, i.e., the closeness of the remaining network $G\backslash \{v_i\}$ after removing $v_i$. Then the vulnerability of network can be obtained by the residual closeness $R=\mathrm{min}_i\{RCC(i)\}$.

    In accordance with this idea, namely adopting the change of vulnerability of network after node removal to measure the importance of nodes, some research also quantified the vulnerability from other aspects. For example, Rao \emph{et al.}~\cite{Rao2009CompE3514} deemed that information would not propagate along longer paths until no shortest path is valid. Thus they argued that the vulnerability can be measured by the number of the the shortest paths between nodes. To quantify the vulnerability, they compared the connectivity between nodes in real-world networks with that in the corresponding fully connected networks. The difference is measured by average equivalent shortest paths, which is calculated by $y_{ij}=x_{ij}/\mu$. $x_{ij}$ is the number of the shortest paths between $v_i$ and $v_j$ in real-world networks; the length of the shortest path is $d_{ij}$ and $\mu$ is the number of the shortest pathes with length not bigger than $d_{ij}$ between two nodes in the corresponding fully connected network. Then the vulnerability of the network is calculated by the average value of $y_{ij}$ of the whole network.

    Besides the shortest paths, Bao \emph{et al.}~\cite{Bao2012SysEE34168} claimed that the number of disjoint paths is also able to measure the vulnerability of networks. Two paths are disjoint if they do not share any nodes. Thus, differing from the method based on the shortest paths, this method considers that the diversity of passageways between nodes is more important to measure the vulnerability of the accessibility between nodes. Similar to the method based on equivalent paths, this method also employes the fully connected networks as benchmark. The vulnerability between nodes $v_i$ and $v_j$ is defined as $V(v_i,v_j)=P_{ij}/P_{ij}^{\mathrm{full}}$, $i\neq j$, where $P_{ij}$ is the number of disjoint paths between $v_i$ and $v_j$, and $P_{ij}^{\mathrm{full}}$ is the number of disjoint paths in the corresponding fully connected networks. Then the vulnerability of the network can be calculated by $V(G)=\sum{P_{ij}}/\sum{P_{ij}^{\mathrm{full}}}$.

    In addition, Chen \emph{et al.}~\cite{Chen2004JCIC25129} assumed that a node is more important if its removal would lead to a subgraph with a lower number of spanning trees. This correlation is ingenious since a well connected graph usually has multiple spanning trees, while a disconnected graph has none. Namely the stability of a network is related to the number of spanning trees. A spanning tree of an undirected graph $G$ is a subgraph that is a tree which includes all of the nodes of $G$. The number of spanning trees can be calculated by employing the Laplacian matrix $L=D-A$. Jungnickel~\cite{Jungnickel2005BOOK} proved that the number of spanning trees $t_0$ can be calculated by $t_0=|M_{pq}|$, where $M_{pq}$ is the minor corresponding to an arbitrary entry $l_{pq}$ in $L$. Accordingly, after removing $v_i$, the number of spanning trees of the remaining network can be obtained, denoted by $t_{G\backslash \{v_i\}}$. Then the importance of the node can be calculated through~\cite{Chen2004JCIC25129}
        \begin{equation}\label{EqDelNodeSpanT}
          DST(i) = 1- \frac{t_{G\backslash \{v_i\}}}{t_0}.
        \end{equation}

    Compared with the connectivity sensitive methods, the methods depicted in this subsection are more sensitive against the removal of only one node. But when the network is very fragile, even randomly removing only one node would break the network into pieces, the advantage of the stability sensitive methods would not be obvious any more. Especially, the method based on spanning tree would fail if the node removal leads to disconnected components.

    \subsection{Eigenvalue-based method}

    Many dynamical processes on networks are determined by the largest eigenvalue, denoted by $\lambda$, of the network adjacency matrix $A$~\cite{Grassberger1983MathBio63157,Cohen2000PRL854626,Vazquez2003PRE67015101,Restrepo2006Chaos16015107}. Thus the relative change of $\lambda$ after removing one node (or a set of nodes) could reflect the impact of this node (or the set of nodes) on the dynamical process. Specifically, the larger change caused by the node removal, the more important the removed node is. This principle was utilized by Restrepo \emph{et al.}~\cite{Restrepo2006PRL97094102} to measure the dynamical importance of nodes (and edges). Denote by $\overrightarrow{u}$ and $\overrightarrow{v}$ the right and left eigenvectors of $A$ respectively, then $A\overrightarrow{u}=\lambda \overrightarrow{u}$ and $\overrightarrow{v}^TA=\lambda \overrightarrow{v}^T$. The dynamical importance of node $v_k$ is proportional to the amount $-\Delta\lambda_k$ by which $\lambda$ decreases upon the removal of $v_k$, written as
        \begin{equation}\label{EqDynaImportance}
          I_k=-\frac{\Delta\lambda_k}{\lambda}.
        \end{equation}
    Due to the complexity of computation, a perturbative analysis was introduced~\cite{Restrepo2006PRL97094102} in order to obtain an approximation $\hat{I}$ to $I$ (see a similar perturbative analysis in~\cite{Lu2015PNAS1122325}). Define $A+\Delta A$ as the adjacency matrix after removing a node. Correspondingly, $\lambda+\Delta\lambda$ was defined as the largest eigenvalue of $A+\Delta A$, and $\overrightarrow{u}+\Delta \overrightarrow{u}$ is the corresponding right eigenvector. Upon the removal of node $v_k$, the perturbation matrix $\Delta A$ is given by $(\Delta A)_{lm}=-A_{lm}(\delta_{lk}+\delta_{mk})$, and $\Delta u_k=-u_k$, where $\delta_{lk}$ equals 1 if $l=k$, and 0 otherwise. Since the left and right eigenvectors have zero $k$th entry after the removal of node $v_k$, $\Delta \overrightarrow{u}$ can be set as $\Delta \overrightarrow{u}=\delta \overrightarrow{u}-u_k\hat{e}_k$, where $\hat{e}_k$ is the unit vector for the $k$ component, and $\delta \overrightarrow{u}$ is assumed to be small. Then, left multiplying $(A+\Delta A)(\overrightarrow{u}+\Delta \overrightarrow{u}) = (\lambda+\Delta\lambda)(\overrightarrow{u}+\Delta \overrightarrow{u})$ by $\overrightarrow{v}^T$ and neglecting second order terms $\overrightarrow{v}^T\Delta A\delta \overrightarrow{u}$ and $\Delta\lambda \overrightarrow{v}^T\delta \overrightarrow{u}$, they obtained $\Delta\lambda=(\overrightarrow{v}^T\Delta A\overrightarrow{u}-u_k\overrightarrow{v}^T\Delta A\hat{e}_k)/(\overrightarrow{v}^T\overrightarrow{u}-v_ku_k) = (-2\lambda u_kv_k-\lambda u_kv_k)/(v^Tu-v_ku_k)$. For a real-world network, $n$ is usually far larger than 1, then $u_kv_k\gg \overrightarrow{v}^T\overrightarrow{u}$ is a reasonable assumption. As a result, the approximation can be finally calculated through
        \begin{equation}
          \hat{I}_k=\frac{v_ku_k}{\overrightarrow{v}^T\overrightarrow{u}}.
        \end{equation}
    This method can be applied to both directed networks and weighted networks, and be also used to measure the dynamical importance of a link but with a different formula $\hat{I}_{ij}=A_{ij}v_iu_j/\lambda \overrightarrow{v}^T\overrightarrow{u}$.

    \subsection{Node contraction method}

    Node contraction was once applied in coarse-graining analyses of complex networks~\cite{Kim2004PRL93168701,Gfeller2007PRL99038701,Song2005Nature433392}, which contracts a node and its neighbor nodes into a new node. If a node is a very important core node, the entire network will be more agglomerate after contracting this node~\cite{Tan2006SysETP1179}. The key point of this method is to quantify the agglomeration degree of a network, which is determined by both the number of the nodes, i.e., $n$, and the average shortest distance, i.e., $\langle d\rangle$. If a network has both small $n$ and $\langle d\rangle$, it has high agglomeration degree. This is easy to be understood from sociology perspective: a social network is more agglomerate, if it has fewer people (small $n$) and the members could communicate with each other conveniently (small $\langle d\rangle$). Then the agglomeration degree of a network $G$ is defined as~\cite{Tan2006SysETP1179}:
        \begin{equation}\label{EqAggloDegree}
          \partial[G] = \frac{1}{n\cdot\langle d\rangle}=\frac{1}{n\cdot\frac{\sum_{i\neq j}d_{ij}}{n(n-1)}}=\frac{n-1}{\sum_{i\neq j}d_{ij}},
        \end{equation}
    where $d_{ij}$ is the distance between $v_i$ and $v_j$. When $n=1$, $\partial[G]$ is set to 1. Then $0<\partial[G]\leq1$. The importance of a node can be reflected by of change of $\partial[G]$ after contracting the node, which is defined as:
        \begin{eqnarray}\label{EqDelNodeContract}
          IMC(i) &=& 1- \frac{\partial[G]}{\partial[G_{\circledcirc v_i}]}      \nonumber \\
                 &=& \frac{n\cdot \langle d\rangle_G - (n-k_i)\cdot \langle d\rangle_{G_{\circledcirc v_i}}}{n\cdot \langle d\rangle_G},
        \end{eqnarray}
    where $\partial[G_{\circledcirc v_i}]$ is the agglomeration degree of the network after contracting node $v_i$, $\langle d\rangle_{G}$ denotes the average distance among all nodes of $G$, $k_i$ is the degree of node $v_i$. Apparently, $IMC(i)$ is jointly determined by the number of $v_i$'s neighbors and the location of $v_i$ in $G$. If $k_i$ is large, the contraction of $v_i$ would largely reduce the number of nodes in $G$, indicating that node with larger degree tends to be more important. Meanwhile, if $v_i$ is passed through by many shortest paths, the contraction of $v_i$ would largely shorten the average distance of $G$, leading to larger IMC of node $v_i$. Hence one can see that this metric embodies both the idea of degree centrality and betweenness centrality. Nevertheless, the node contraction method cannot be applied to large-scale networks, because the calculation of the average distance $\langle d\rangle_{G_{\circledcirc v_i}}$ ($i=1,2,\cdots,n$) for every node is very time consuming.

    Employing the node contraction method, Wang \emph{et al.}~\cite{Wang2011ProcediaE151600} introduced the impact of links to redefine the node importance. Firstly, a line graph $G^*$ of initial network $G$ is constructed, which represents the adjacencies between links of $G$ \cite{Harary1960RDCMDP9161,Whitney1932AmerJMath54150}. Then the node's importance in $G^*$ is obtained by rewriting Eq. (\ref{EqDelNodeContract}):
        \begin{equation}\label{EqAggloDegInvNet}
          IMC(i) = \alpha\cdot IMC_G(i) + \beta\cdot\sum_{j\in S}IMC_{G^*}(j),
        \end{equation}
    where $IMC_G(i)$ is $v_i$'s importance in $G$, and $IMC_{G^*}(j)$ is $v_j$'s importance in $G^*$. $S$ denotes the set of corresponding nodes of $v_i$ in $G^*$, i.e., the links that contain $v_i$ in the initial network G. In addition, the node contraction method has been applied weighted networks $G_w$ by redefining the agglomeration degree as \cite{Zhu2009SysEE311902}:
        \begin{equation}\label{EqAggloDegWNet}
          \partial[G_w] = \frac{1}{s\cdot\langle d\rangle_G},
        \end{equation}
    where $s=\sum_{i}s_i$, and $s_i=\sum_j{w_{ij}}$ is the strength of node $v_i$ (see more detail introduction about weighted networks in Chapter \ref{Chapter7}). Note that $\langle d\rangle_G$ is still the average distance among nodes in unweighted network corresponding to $G_w$.

\section{Dynamics-sensitive methods}\label{Chapter5}

One major reason to identify vital nodes is to find out the nodes playing critical roles in some specific dynamical processes. Therefore, the influences of a node or a set of nodes on some given dynamical processes are usually treated as criteria for vital nodes. For example, for an arbitrary node, setting this node as the infected seed and then the total number of ever infected nodes in a susceptible-infected-recovered (SIR) process \cite{Hethcote2000SIAM42599} is widely used as a metric quantifying the importance of this node (see details in Chapter \ref{Chapter9}). An ideal situation is that there is a golden structural centrality which gives correct rank of nodes subject to their influences on almost every dynamics. However, such fairy tale was proved to be wrong. In fact, even for a given dynamics, structural centralities perform far differently under different dynamical parameters. For example, again in the SIR process, the degree centrality can better identify influential spreaders when the spreading rate $\beta$ is very small while the eigenvector centrality performs better when $\beta$ is close to the epidemic threshold $\beta_c$ \cite{Klemm2012SciRep2292,Liu2016SciRep621380}. Based on extensive simulations, \v{S}iki\'c \emph{et al.} \cite{Sikic2013EPJB861} showed that for a given SIR process with two parameters (i.e., the spreading rate and recovering rate), the rank of nodes' influences largely depends on the parameters, that is to say, in principle any structural centrality cannot well estimate nodes' influences since it gives the same rank under different dynamical parameters.

According to the above argument, if we would like to uncover the roles of nodes in some network dynamics and we can estimate the related parameters in advance, then we are supposed to design better-performed methods to identify vital nodes than structural centralities by taking into account the features and parameters of the target dynamics. We thus call such kind of methods as \emph{dynamics-sensitive methods}. Notice that, in some methods like LeaderRank \cite{Lu2011PLoSONE6e21202}, we apply dynamics like random walk on the network, but we do not call LeaderRank as a dynamics-sensitive method since the random walk is used to rank nodes while itself is not the target dynamics. In fact, in the original paper for LeaderRank \cite{Lu2011PLoSONE6e21202}, a variant of SIR model \cite{Yang2007PLA364189} was considered as the target dynamics. In this section, we will classify known dynamics-sensitive methods into three categories and introduce the state-of-the-art progresses as well as discuss some open issues.

    \subsection{Path counting methods}\label{Section5_1}

    In principle, any path connecting nodes $i$ and $j$ can be used as a passageway to deliver $i$'s influences onto $j$'s state and vice versa. General speaking, the influence will decay as the increase of the path length with the decaying function which depends on the target dynamics. Therefore, it is very straightforward to estimate a node's influence by counting the number of paths originated from this node to all other nodes, where each path is assigned a weight related to its length. Such idea also embodies in some well-known structural centralities like Katz index \cite{Katz1953Psychometrika1839} and accessibility \cite{Travencolo2008PLA37389}. This subsection will introduce the combination of the above path counting idea with the specific features and parameters of the target dynamics. Usually, the dynamical features are used to design the decaying function of path weight on path length.

    Considering a network of $n$ nodes whose states are described by a time-dependent real vector $\overrightarrow{x}=(x_1,\cdots,x_n)$. For any discrete coupled linear dynamics
        \begin{equation}
           \overrightarrow{x}(t+1)=M \overrightarrow{x}(t),
        \label{TZ02}
        \end{equation}
    with $M$ an $n\times n$ real matrix, Klemm \emph{et al.} \cite{Klemm2012SciRep2292} proved that if the largest eigenvalue $\mu_{\max}$ of $M$ equals zero and non-degenerate, then the projection of $\overrightarrow{x}(0)$ on the left eigenvector of $M$ for $\mu_{\max}$ is all the system remembers at large times about the initial condition $\overrightarrow{x}(0)$. Denote $\overrightarrow{c}$ the left eigenvector for $\mu_{\max}$, then the $i$th entry $c_i$ quantifies the extent to which the initial condition at node $v_i$ affects the final state. $c_i$ is thus called \emph{dynamical influence} (DI) that measures $v_i$'s influence in the target dynamics $M$. The eigenvector $\overrightarrow{c}$ can be estimated by the power iteration method that applies higher and higher powers of $M$ to a uniform vector $\overrightarrow{w}^{(0)}=(1,1,\cdots,1)$, say
        \begin{equation}
          \overrightarrow{w}^{(l)}=(1,1,\cdots,1)M^l,
        \label{TZ01}
        \end{equation}
    where $l$ is a natural number. If the largest eigenvalue of $M$ is non-degenerate and larger in magnitude than the other eigenvalues, then $\overrightarrow{c}$ can be obtained in the limitation
        \begin{equation}
        \lim_{l\rightarrow \infty} \frac{\overrightarrow{w}^{(l)}}{\|\overrightarrow{w}^{(l)}\|} = \frac{\overrightarrow{c}}{\|\overrightarrow{c}\|}.
        \end{equation}
    Imaging $M$ as the adjacency matrix of the network, then according to Eq. (\ref{TZ01}), $w^{(l)}_i$ is the number of all possible walks of length $l$ originated from $v_i$. This calculation process shows off the idea of path counting behind the definition of DI.

    The method to obtain the dynamical influence is very general and can be applied in many representative dynamics \cite{Klemm2012SciRep2292}, such as the SIR model \cite{Hethcote2000SIAM42599}, voter model \cite{Sood2005PRL94178701}, Ising model \cite{Dorogovtsev2002PRE66016104}, Kuramoto model \cite{Acebron2005RMP77137}, and so on. Taking the discrete SIR model as an example, if at each time step, an infected node will infect each of its susceptible neighbors with probability $\beta$ and then relax to the recovered state in the next time step, the dynamics can be expressed as
        \begin{equation}
        \overrightarrow{x}(t+1)=-\overrightarrow{x}(t) + \beta A^T \overrightarrow{x}(t),
        \end{equation}
    where $x_i(t)$ is the probability of node $v_i$ to be infected at time step $t$ and $A$ is the adjacency matrix of the network. Consulting Eq. (\ref{TZ02}),
        \begin{equation}
        M=\beta A^T-I,
        \end{equation}
    with $I$ being the identity matrix. Denote $\alpha_{\max}$ the largest eigenvalue of $A$, only when $\beta=1/\alpha_{\max}$ (this is exactly the epidemic threshold for SIR model \cite{Castellano2010PRL105218701}), $\mu_{\max}=0$, and at this point, the dynamical influence $\overrightarrow{c}$ is the same to the right eigenvector of $A$ for $\alpha_{\max}$, namely the dynamical influence in this case equals the eigenvector centrality. Furthermore, Li \emph{et al.} \cite{Li2012CPL29048903} applied a similar method to study the SIS model by considering the effects of a perturbation on the equilibrium state of the dynamics. They provided complementary analysis on the suitability of DI, emphasizing on the importance of the existence of spectral gap and showing an example network (i.e., the homogeneous community network) where DI is invalid.

    Ide \emph{et al.} \cite{Ide2013PCS24227} considered the susceptible-infected-susceptible (SIS) model \cite{Hethcote2000SIAM42599}, where at each time step, an infected node will infect each of its susceptible neighbors with probability $\beta$ and then return to the susceptible state in the next time step with probability $\delta$. Denote $\overrightarrow{x}(t)$ the system state whose $i$th entry $x_i(t)$ is the probability that the $i$th node is in the infected state at time step $t$, then according to Eq. (\ref{TZ02}),
    \begin{equation}
    M=(1-\delta)I+\beta A.
    \end{equation}
    Since $\overrightarrow{x}(t)=M^t \overrightarrow{x}(0)$, the accumulative infection probability vector can be written as
    \begin{equation}
    \sum^{\infty}_{t=0} \overrightarrow{x}(t) = (I+M+M^2+\cdots+M^t+\cdots)\overrightarrow{x}(0)=(I-M)^{-1}\overrightarrow{x}(0),
    \label{TZ03}
    \end{equation}
    which represents nodes' influences in the long time limit. Denote $\alpha=\beta/\delta$ the effective infection rate, the Eq. (\ref{TZ03}) can be rewritten as
    \begin{equation}
    (I-M)^{-1}\overrightarrow{x}(0) = \frac{1}{\delta} (I-\alpha A)^{-1} \overrightarrow{x}(0) \propto (I-\alpha A)^{-1} \overrightarrow{e},
    \label{TZ04}
    \end{equation}
    where $\overrightarrow{e}=(1,1,\cdots,1)^T$. Note that, here we assume each node has the same probability to be initially infected and a node's influence is quantified by the frequency of its infected states during the whole spreading process. As $(I-\alpha A)^{-1}=I+\alpha A+\alpha^2 A^2+\cdots+\alpha^t A^t+\cdots$ and $(\alpha^t A^t)_{ij}$ represents the sum of probabilities that $i$ infects $j$ through paths with length $t$, Eq. (\ref{TZ04}) is a typical path counting method.

    Very interestingly, Eq. (\ref{TZ04}) has exactly the same form as the well-known alpha centrality \cite{Bonacich2001SNet23191}: $C_\alpha=(I-\alpha A)^{-1} \overrightarrow{e}$. The difference is that in the alpha centrality, the parameter $\alpha$ is a free parameter while in Eq. (\ref{TZ04}), it has explicit dynamical meaning $\alpha=\beta/\delta$. In the beginning of this Chapter, we argue that in the SIR model, the degree centrality can better identify influential spreaders when the spreading rate $\beta$ is very small while the eigenvector centrality performs better when $\beta$ is close to the epidemic threshold $\beta_c$ \cite{Klemm2012SciRep2292,Liu2016SciRep621380}. Indeed, this argument is also valid for SIS model and susceptible-infected (SI) model \cite{Liu2016SciRep621380,Ide2013PCS24227}. Looking at the alpha centrality or Eq. (\ref{TZ04}), when $\alpha$ is close to zero,
    \begin{equation}
    \lim_{\alpha \rightarrow 0}(I-\alpha A)^{-1} \overrightarrow{e} = \lim_{\alpha \rightarrow 0} (I+\alpha A+\alpha^2 A^2+\cdots+\alpha^t A^t+\cdots)\overrightarrow{e} \approx (I+\alpha A)\overrightarrow{e},
    \end{equation}
    indicating that the degree centrality is dominant in the alpha centrality when $\alpha \rightarrow 0$. When $\alpha$ is close to the critical point $1/\alpha_{\max}$,
    \begin{equation}
    \lim_{\alpha \rightarrow 1/\alpha_{\max}} (I-\alpha A)^{-1} \overrightarrow{e} = \lim_{\alpha \rightarrow 1/\alpha_{\max}} \sum^n_{i=1} \frac{1}{1-\alpha \lambda_i} \overrightarrow{v_i} \overrightarrow{v_i}^T \overrightarrow{e} \approx \overrightarrow{v_1} \overrightarrow{v_1}^T \overrightarrow{e},
    \label{TZ05}
    \end{equation}
    where $\overrightarrow{v_i}$ represents the eigenvector for eigenvalue $\lambda_i$, with $\lambda_1=\alpha_{\max}$ being the largest eigenvalue. Obviously, the approximation in Eq. (\ref{TZ05}) works only if the spectral gap exists, which is in accordance with the argument of Ref.~\cite{Li2012CPL29048903}. Eq. (\ref{TZ05}) suggests that the eigenvector centrality is dominant in the alpha centrality when $\alpha$ is close to the epidemic threshold. Is that a simple and nice explanation of the above argument? We think so.

    Bauer and Lizier \cite{Bauer2012EPL9968007} proposed a brave method that directly counts the number of possible infection walks of various lengths in SIS and SIR models. They define the impact of node $v_i$ as
    \begin{equation}
    I_i=\lim_{L\rightarrow \infty} \sum^L_{k=1} \sum_j q(i,j,k),
    \end{equation}
    where $j$ runs over all nodes including node $v_i$ and $q(i,j,k)$ is the probability that node $v_j$ is infected through a path of length $k$ given that the infections started at node $v_i$, with an assumption that all infection paths are independent to each other. In the SIS model with spreading rate $\beta$ and returning rate $\delta=1$, with the independent path assumption,
    \begin{equation}
    q(i,j,k)=1-(1-\beta^k)^{s^k_{ij}},
    \end{equation}
    where $s^k_{ij}$ is the number of different paths of length $k$ between $i$ and $j$, which equals $(A^k)_{ij}$. Therefore,
    \begin{equation}
    I^{SIS}_i=\lim_{L\rightarrow \infty} \sum^L_{k=1} \sum_j \left[ 1-(1-\beta^k)^{(A^k)_{ij}} \right].
    \end{equation}

    Note that, in the SIS model, a path can pass through a node by multiple times, so that it is a little bit different from the traditional definition of path but same to the definition of walk. In fact, Bauer and Lizier \cite{Bauer2012EPL9968007} called their method as a walk counting approach. While if we consider the SIR model, the corresponding definition is exactly the same to the traditional meaning of path, or the so-called self-avoiding walk. The SIR model is thus more complicated when applying Bauer and Lizier's method (see more details in \cite{Bauer2012EPL9968007}). Bauer and Lizier's method directly embodied the idea of path counting and their method performs very good in the simulations \cite{Bauer2012EPL9968007}, while the disadvantages lie in two aspects: the path independent assumption is too strong and the computational complexity is very high. In a word, their method leaves to us two challenges: how to eliminate the coherence among different paths and how to effectively and efficiently estimate the number of paths.

    Traffic dynamics is another typical dynamics in information networks and transportation networks \cite{Wang2007JKPS50134,Chen2012MPE732698}. In addition to the possibly heterogeneous packet generating rates and link bandwidths \cite{Tang2011PRE84026116}, the dynamical features of network traffic are mainly determined by the routing table, which lists the paths through which a packet generated by a source node $v_s$ can be delivered to its target node $v_t$ \cite{Yan2006PRE73046108}. Given the specific routing table, Dolev \emph{et al.} \cite{Dolev2010JACM5725} proposed a so-called routing betweenness centrality (RBC) to measure the importance of a node in a traffic dynamics. The RBC of an arbitrary node $v_i$ is defined as the expected number of packets that pass through $v_i$, say
    \begin{equation}
    RBC(i)=\sum_{v_s,v_t\in V} \delta_{s,t}(i)T(s,t),
    \end{equation}
    where $V$ is the set of nodes in the network, $\delta_{s,t}(i)$ is the probability that a packet generated by the source node $v_s$ and destined to leave the network at the target node $v_t$ will pass through the node $v$, and $T(s,t)$ is the number of packets sent from a source node $v_s$ to a target node $v_t$. With a given routing table (or routing rules), $\delta_{s,t}(i)$ can be written as
    \begin{equation}
    \delta_{s,t}(i)=\sum_{u\in Pred_{s,t}(i)} \delta_{s,t}(u)R(s,u,i,t),
    \end{equation}
    where $R$ is the routing table and $R(s,u,i,t)$ records the probability that $v_u$ will forward to $v_i$ a packet with source address $v_s$ and target address $v_t$, and $Pred_{s,t}(i)=\{u|R(s,u,i,t)>0\}$ is the set of all immediate predecessors of $v_i$ given the source address $v_s$ and target address $v_t$. The routing betweenness centrality is a general method to identify vital nodes in network traffic given the specific dynamical rules (i.e., the routing table $R$), which counts the number of paths passing through the target node, different from the path counting method for spreading dynamics that counts the number of paths originated from the target node.

    \subsection{Time-aware methods}

    In the above subsection, when talking about a node's influence or importance, we usually consider to which extent its initial state or a perturbation on it can impact the final state of the system, like the propagation coverage in spreading dynamics. Sometimes we would like to know a node's influence within finite time steps, which leads to the so-called time-aware methods.

    Consider the Boolean network dynamics \cite{Kauffman1969JTB22437} where at each time step, each node can be in one of two states $\{0,1\}$, which can be probabilistically or deterministically obtained from the states of its neighbors in the last time step. Though to be extremely simple, Boolean network dynamics have found wide applications in explaining and analyzing biological functions like gene regulation in living cells \cite{Shmulevich2002Bioinformatics18261,Bornholdt2005Science310449} and social phenomena like minority game in social networks \cite{Zhou2005PRE72046139,Chakraborti2015PhysRep5521}.

    Ghanbarnejad and Klemm \cite{Ghanbarnejad2012EPL9958006} considered a general Boolean dynamics
    \begin{equation}
    \overrightarrow{x}(t+1)=f(\overrightarrow{x}(t)),
    \end{equation}
    where the state vector $\overrightarrow{x}(t) \in \{0,1\}^N$ and $f$ is the Boolean function. The influence of an arbitrary node $v_i$ after $t$ time step is quantified by the change of system state resulted from a perturbation at $v_i$ (i.e., the flip of $v_i$'s state). Ghanbarnejad and Klemm define
    \begin{equation}
    H_i(t)=\{ \overrightarrow{x}\in \{0,1\}^n: f^t(\overrightarrow{x})\neq f^t(\overrightarrow{x}^{\updownarrow i}) \}
    \end{equation}
    as the set of initial conditions such that a perturbation at $v_i$ will cause the change of the system state after $t$ time steps, where $\overrightarrow{x}^{\updownarrow i}$ is the state vector that is different from $\overrightarrow{x}$ only with the $i$th entry. Assuming that all $2^n$ state vectors occur with the same probability, then the fraction of initial conditions allowing the spreading of $v_i$'s flip for at least $t$ time steps reads
    \begin{equation}
    h_i(t)=\frac{|H_i(t)|}{2^n},
    \end{equation}
    which is called the dynamical impact of node $v_i$ for $t$ steps.

    Given the Boolean function $f$, a challenging issue is to estimate a node $v_i$'s dynamical impact $h_i(t)$. Given a state vector $\overrightarrow{x}$, if a node $v_i$'s flip will change another node $v_j$'s state in the next step, we set $\partial^{(i)}f_j(\overrightarrow{x})=1$ where $f_j$ is the Boolean function on node $v_j$, otherwise it is zero, namely
    \begin{eqnarray}
    \partial^{(i)}f_j(\overrightarrow{x})=&\left\{ \begin{matrix}
       1~~~~~~~~~~f_j(\overrightarrow{x})\neq f_j(\overrightarrow{x}^{\updownarrow i}),  \\
       0~~~~~~~~~~~~~~~~~~\texttt{otherwise}.  \\
    \end{matrix}
    \right.
    \end{eqnarray}
    Again assuming all $2^n$ state vectors occur with the same probability, $v_i$'s immediate impact on $v_j$ can be defined as
    \begin{equation}
    m_{ij}(f)=\frac{1}{2^n} \sum_{x\in \{0,1\}^n} \partial^{(i)}f_j(\overrightarrow{x}).
    \end{equation}
    $M$ is called activity matrix, which is different from the adjacency matrix $A$ as it contains the dynamical features of $f$. Denote $p_j(t)$ the probability that node $v_j$'s state is changed at time $t$ resulted from an initial perturbation at node $v_i$, i.e., the probability that $\left[ f^t(\overrightarrow{x})\right]_j \neq \left[ f^t(\overrightarrow{x}^{\updownarrow i})\right]_j$. Note that, the impacts of a perturbation from different spreading paths can not be directly summed up in the Boolean dynamics, since two flips equals no flips. In despite of this fact, Ghanbarnejad and Klemm made a strong assumption that neglects the correlation in the spreading of a node's influence, that is to say
    \begin{equation}
    p_j(t)\propto \sum^n_{i=1} m_{ij}p_i(t-1).
    \end{equation}
    Therefore, the eigenvector of $M$ for the largest eigenvalue can be used to quantify nodes' influences in the long time limit. Although the two assumptions (i.e., the equal occurring probability of all states and the absence of correlations in influence spreading) seem to be too strong, Ghanbarnejad and Klemm tested the correlations between indices for node importance and $h(t)$ for different $t$ and found that the eigenvector centrality based on $M$ outperforms the eigenvector centrality based on $A$, supporting the advantage of dynamics-sensitive centralities. In addition, they showed that for small $t$, degree centrality is better than eigenvector centrality while for very large $t$, eigenvector centrality is much better. Note that, the Ghanbarnejad-Klemm method \cite{Ghanbarnejad2012EPL9958006} is not a really time-aware method since the eigenvector of $M$ contains no temporal information, however, their simulation results have clearly demonstrated the significance of temporal factor, similar to the contribution by \v{S}iki\'c \emph{et al.} \cite{Sikic2013EPJB861}, who have showed the relevance of dynamical parameters.

    Liu \emph{et al.} \cite{Liu2016SciRep621380} proposed a so-called dynamics-sensitive (DS) centrality to predict the outbreak size at a given time step $t$, which can be directly applied in quantifying the spreading influences of nodes. Using a different method from eigenvector centrality (see details in \cite{Liu2016SciRep621380}), Liu \emph{et al.} showed that the DS centrality of nodes at time step $t$ can be written as a vector
    \begin{equation}
    \overrightarrow{DS}(t)=\left[ (\beta/\delta)A+(\beta/\delta)^2A^2+\cdots+(\beta/\delta)^tA^t \right]\overrightarrow{e},
    \end{equation}
    where $A$ is the adjacency matrix, and $\beta$ and $\delta$ are the spreading rate and recovering rate in SIR model. Given the time step $t$, the DS centrality performs much better than the methods without temporal factor. Looking at Eq. (\ref{TZ04}) or the alpha centrality, though being rooted in different ideas, the DS centrality is indeed a temporal cutoff of alpha centrality as the absence of the identity matrix does not affect the ranking of nodes' influences.

\subsection{Others}
    As we have mentioned above, \v{S}iki\'c \emph{et al.} \cite{Sikic2013EPJB861} have pointed out the fact that the ranking of nodes' influences in SIR model varies considerably for different parameter sets $(\beta,\delta)$. Very straightforwardly, they define the so-called \emph{epidemic centrality} for an arbitrary node $v_i$ as
    \begin{equation}
    Z^i=\int^1_0d\beta\int^1_0d\delta X^i_{\beta,\delta},
    \end{equation}
    where $X^i_{p,q}$ measures the epidemic coverage (i.e., the fraction of nodes in state $R$ at the end of the spreading). Here we use $Z^i$ to respect the original paper and avoid the possible confusion with eigenvector centrality. Since some specific parameter spaces may be more interesting (e.g., surrounding the critical dynamical regime), \v{S}iki\'c \emph{et al.} argued that the averaging should be performed using some nonuniform weights, say
    \begin{equation}
    Z^i=\int^1_0d\beta\int^1_0d\delta w(\beta,\delta) X^i_{\beta,\delta}.
    \end{equation}
    \v{S}iki\'c \emph{et al.} \cite{Sikic2013EPJB861} raised a very important issue, but their solution is not a real solution since if we can accurately calculate the epidemic centrality, we should know every details of this dynamics and we must have done all possible simulations, so that we do not need to know epidemic centrality again because given any parameter pair $(\beta,\delta)$, we already have the corresponding ranking of nodes' influences.

    Evolutionary games have long been exemplary models for the emergence of cooperation in socioeconomic and biological systems \cite{Nowak2006Science3141560,Szabo2007PhysRep44697}. Till far, there are few studies considering the identification of vital nodes in evolutionary games. Simko and Csermely \cite{Simko2013PLoSONE8e67159} proposed a novel dynamic-sensitive centrality, called \emph{game centrality} (GC), which measures the ability of individually defecting nodes to convert others to their strategy. The game centrality of a node $v_i$ ($GC_i$) is defined as the proportion of defectors averaged over the last 50 simulation steps, given the initial condition with node $v_i$ being a defector and all others cooperating. The index itself is not based on a solid theory but ad hoc, while it provides a very good starting point in the study of nodes' influences in games. Similar to the question for the work by \v{S}iki\'c \emph{et al.} \cite{Sikic2013EPJB861}, the calculation of GC requires the simulation of the dynamics itself, which largely depresses the value of a centrality index that was usually thought to be a predictive tool. To this question, Simko and Csermely \cite{Simko2013PLoSONE8e67159} applied game centrality to dig out influential nodes in protein-protein interaction (PPI) networks, with an assumption that the functions in PPI networks can be to some extent described by repeated games \cite{Csermely2010TrendsBS35539}. In this application, game dynamics plays the role of a tool to find influential nodes in some specific networks just like random walk in PageRank and LeaderRank, but it is not the target dynamics and thus GC in this case is not a dynamics-sensitive centrality. Simko and Csermely \cite{Simko2013PLoSONE8e67159} briefly analyzed the correlation between GC and former centrality measures. This should be the preliminary work before further studies, and now is still far from sufficient understanding.

    Piraveenan \emph{et al.} \cite{Piraveenan2013PLoSONE8e53095} propose a novel centrality measure called \emph{percolation centrality} (PC), which takes into account the percolation states of individual nodes. They denote the percolation state of node $v_i$ by $x_i$. Specifically, $x_i=1$ indicates a fully percolated state, $x_i=0$ indicates a non-percolated state, while a partially percolated state corresponds to $0<x_i<1$. They did not consider the relationship between a real percolating process and the percolated state or discuss how to determine a node's percolated state, but showed an example that in a metapopulation epidemic model \cite{Watts2005PNAS10211157,Colizza2007NatPhys3276}, for a network of township, the percolated state of a town would be the percentage of people infected in that town. Accordingly, define the percolation centrality of a node $v_i$ as
    \begin{equation}
    PC(i)=\frac{1}{n-2}\sum_{s\neq v \neq t} \frac{\sigma_{s,r}(i)}{\sigma_{s,r}} \frac{x_s}{[\sum x_j] - x_i},
    \end{equation}
    where $\sigma_{s,t}(i)$ is the number of the shortest paths from $v_s$ to $v_t$ passing through $v_i$, and $\sigma_{s,t}$ is the total number of different shortest paths connecting $v_s$ and $v_t$. The term $w_{s,i}=\frac{x_s}{[\sum x_i] - x_i}$ is a normalization factor such that $\sum_{s\neq i} w_{s,i}=1$. If all nodes are in the same percolation level $0<\mu\leq 1$, then $x_s=\mu$ and $w_{s,i}=\frac{1}{n-1}$, so that the percolation centrality degenerates to the well-known betweenness centrality
    \begin{equation}
    PC(i)=\frac{1}{(n-1)(n-2)} \sum_{s\neq i \neq t} \frac{\sigma_{s,t}(i)}{\sigma_{s,t}} = BC(i).
    \end{equation}
    Therefore, the percolation centrality is indeed a weighted betweenness centrality with each node $v_i$ is assigned a weight $x_i$ called percolated state. In principle, this framework \cite{Piraveenan2013PLoSONE8e53095} allows us to embody dynamical features into the weights $\overrightarrow{x}$, which still asks for further exploration.

\section{Identifying a set of vital nodes}\label{Chapter6}

    The above chapters have introduced advanced techniques to find individual vital node. In many real-world applications, we are further asked to find out a small set of vital nodes that play the critical role in propagating information, maintaining network connectivity, and so on. For example, in network marketing \cite{Kim2006PhysA360493} with limited budget, the best strategy is to show the advertisements and provide discounts to a set of customers who are likely to buy the product and able to trigger many other people (including their friends, friends of friends, and so on) to buy. In the epidemic spreading with limited time and resource, we need to immunize a group of people to the best protect the whole population. In the network counterwork for military applications, one is required to destroy a few critical nodes of the enemy to the greatest reduce their communication capacity.

    A natural idea is to directly apply the centralities to find the set of vital nodes, that is to say, to rank all nodes in term of a specific centrality measure and pick up the top-$k$ nodes to form the target set. For example, the top-$k$ largest-degree nodes are usually considered as the benchmark of the set of vital nodes in many applications \cite{Albert2000Nature406378,Cohen2001PRL863682,Pastor-Satorras2002PRE65036104}. However, this method could be inefficient since the nodes of the highest degrees (or other centrality measures) may be highly clustered as indicated by the rich-club phenomenon \cite{Zhou2004IEEE8180,Colizza2006NatPhys2110}, so that to target all of them is unnecessary.

    To find a set of vital nodes subject to some specific structural or functional objectives is usually called \emph{influence maximization problem} in the literatures. Firstly, we will introduce the different descriptions of the related problems, and then show both pioneer works and the state-of-the-art progresses. Besides a few representative techniques from computer science society, this section will emphasize the chapter developed by statistical physicists.

\subsection{Influence maximization problem (IMP)}

    Given a network $G(V,E)$ with $V$ and $E$ respectively being the set of nodes and the set of links, if there exists a function $f(S)$ from a subset $S\subseteq V$ to a real number, a general description of the so-called influence maximization problem (IMP) is to find the subset $S$ with a give size $k$ (usually, $k\ll n=|V|$) that maximizes $f$. The influence function $f$ plays a critical role in this problem, which can be very simple and fully determined by the topology of $G$, or very complicated that involves mechanisms or dynamical processes with a number of parameters independent to the topology. To be convenient, we call the former as structural IMP while the latter as functional IMP.

\subsubsection{Structural IMP}

    More than 40 years ago, Corley and Chang \cite{Corley1974ManageSci21362} studied the problem to identify the $k$ most vital nodes in directed networks (an undirected network here can be considered as a very special directed network). They defined the $k$ most vital nodes in a network as those $k$ nodes whose removal, along with all directed links incident with them, would result in the greatest decrease in maximum flow between a specified node pair. Since the maximum flow is completely determined by the network structure \cite{Ford1956CJM8399}, the above problem is a typical structural IMP. Corley and Sha \cite{Corley1982ORL1157} considered a directed network with link weight representing the distance between the two ends. The $k$ most vital nodes here are those $k$ nodes whose removal will cause the greatest increase in the shortest distance between two specified nodes.

    Morone and Makse \cite{Morone2015Nature52465} argued that to find superspreaders in a spreading model or to effectively immunize a network against epidemics can be exactly mapped onto optimal percolation, where the influence maximization problem is to find the minimal set of nodes which, if removed, would break down the network into many disconnected pieces. Accordingly, a natural measure of the influence of a set of node $S$ is the size of the largest connected component after the removal of $S$ \cite{Morone2015Nature52465}. Similar issue has already been considered in the literatures \cite{Albert2000Nature406378,Cohen2001PRL863682,Holme2002PRE65056109}, but with distinct perspectives from the influence maximization problem.

    Cook \cite{Cook1971ACM151} and Karp \cite{Karp1972BOOK85} introduced the so-called feedback vertex set (FVS) problem in the early 1970s. For an undirected network, a FVS is a vertex set which contains at least one vertex of every cycle, so that after the removal of a FVS, the network is consisted of one or more trees. A feedback vertex set is also referred to as a decycling set in some references \cite{Beineke1997JGT2559,Bau2002AJC25285}, and the minimum size of all decycling sets is named the decycling number of the network. It was shown by Karp \cite{Karp1972BOOK85} that determining the decycling number of an arbitrary network is NP-complete (see some recent results on this issue in Ref. \cite{Beineke2002ENDM1181}). Instead of finding the exact decycling number, Zhou \cite{Zhou2013EPJB86455} considered a relaxed version with the size of a FVS being the objective: the smaller the better. This problem can be easily extended to weighted networks where each node is assigned a nonnegative weight and the objective function then becomes the total weight of a FVS. Although the FVS problem seems a litter bit different from the general framework of IMP, we will show later that the FVS problem is very close to the IMP and the nodes in the minimum FVS are highly overlapped with the maximum influential nodes. In fact, many classical optimization problems in graph theory are related to IMP. For example, the minimum dominating set (MDS) problem aims at constructing a node set of the smallest size such that any node of the network is either in this set or is adjacent to at least one node of this set \cite{Haynes1998BOOK,Zhao2015JSP1591154}, and the MDS or a subset of MDS can be considered as an approximate solution of the IMP.

\subsubsection{Functional IMP}

    Domingos and Richardson \cite{Domingos2001ACM,Richardson2002ACM} considered a viral market where each customer's probability of buying a specific product is a function of both the intrinsic desirability of this product (dependent on the matching between the customer's interests and the product's attributes) and the influence of other customers. With limited marketing budget, they would like to find a small group of customers to offer discounts so that to eventually maximize the total sales, resulting from not only the sales directly triggered by the discounts, but also those by the word-of-mouth effects from influential customers.

    Considering a specific product in a market of $n$ potential customers, and $X_i=1$ if the customer $i$ buys the product, and 0 otherwise. Denote $N_i$ the set of $i$'s neighbors who directly influence $i$, $Y=\{Y_1,Y_2,\cdots,Y_m\}$ the set of attributes of the product, and $M=\{M_1,M_2,\cdots,M_n\}$ the set of marketing actions where $M_i$ is the action on customer $i$. For example, $M_i$ could be a continuous variable representing the size of discount offered to the customer $i$, with $M_i=0$ indicating no discount and $M_i=1$ indicating fully free. Domingos and Richardson \cite{Domingos2001ACM,Richardson2002ACM} assumed that the probability of customer $i$ to buy the product is
    \begin{equation}
    P(X_i=1|N_i,Y,M)=\beta_iP_0(X_i=1|Y,M_i)+(1-\beta_i)P_N(X_i=1|N_i,Y,M),
    \label{TZ06}
    \end{equation}
    where $P_0$ is the customer $i$'s internal probability of purchasing the product that is affected by both the matching between $i$'s interests and the product's attributes and the marketing action on $i$, $P_N$ is the effect that $i$'s neighbors have on him, and $\beta_i$ is a scalar with $0\leq \beta_i \leq 1$ that quantifies how self-reliant $i$ is.

    Considering a simple case where $M_i$ is a Boolean variable (whether to show an advertisement to $i$, or whether to provide a discount to $i$), with limited budget $\sum_i M_i\leq k$, the optimal solution is to choose $k$ nodes to offer marketing actions to maximize the total sales $\sum_i P(X_i=1|N_i,Y,M)$. The optimal solution of the general model (see Eq. (\ref{TZ06}) and Ref. \cite{Domingos2001ACM}) is NP-hard. Richardson and Domingos \cite{Richardson2002ACM} analyzed a simple linear model with
    \begin{equation}
    P_N(X_i=1|N_i,Y,M)=\sum_{j\in N_i} w_{ij}X_j,
    \end{equation}
    where $w_{ij}$ represents how much customer $i$ is influenced by his neighbor $j$, with $w_{ij}\geq 0$ and $\sum_{j\in N_i} w_{ij}=1$. The optimal solution of this linear model can be obtained by solving a system of linear equations.

    Domingos and Richardson \cite{Domingos2001ACM,Richardson2002ACM} provided a very good starting point in the studies of the influence maximization problem, but their model is less operational. Kempe, Kleinberg and Tardos \cite{Kempe2003ACM} proposed a general framework for IMP under operational dynamical processes. They considered basic spreading models where each node can be either active or inactive, and each node's tendency to become active increases monotonically as more of its neighbors become active. If the spreading process starts with a set of initially active nodes $A$, then the influence of the set $A$, $f(A)$, is defined as the expected number of active nodes at the end of the process. Although such number usually cannot be analytically obtained, it can be accurately estimated by extensive simulations on the process. Accordingly, given the number of initially active nodes $k$, the influence maximization problem asks for a $k$-node set $A$ that maximizes $f(A)$.

    Kempe, Kleinberg and Tardos \cite{Kempe2003ACM} studied two simple yet widely applied spreading models: the \emph{linear threshold model} \cite{Granovetter1978AJS831420,Schelling1978BOOK,Watts2002PNAS995766} and \emph{independent cascade model} \cite{Goldenberg2001MarketLett12211,Goldenberg2001AMSR1}. In the linear threshold model, each node $v_i$ chooses a threshold $\theta_i$ uniformly at random from the interval $[0,1]$, and each directed link $v_i\rightarrow v_j$ is assigned a weight $w_{ij}$, satisfying that $\sum_{i \in \Gamma_j} w_{ij}\leq 1$, where $\Gamma_j$ is the set of $j$'s neighbors. Notice that, in general $w_{ij}\neq w_{ji}$. The linear threshold model starts with a number of initially active nodes, then at each time step, a node $v_i$ will become active only if $\sum_{j \in \Gamma_i^*}w_{ji}\geq \theta_i$, where $\Gamma_i^*$ is the set of $v_i$'s active neighbors. All nodes remain active if they were active in the last time step. The process unfolds in a synchronous and deterministic way, ends when no further changes of nodes' states happen. The independent cascade model also starts with a number of initially active nodes. At each time step, each active node $i$ has a single chance to activate each of $v_i$'s inactive neighbors, say $v_j$, with successful probability $p_{ij}$. Whatever $v_i$ is successful, it cannot make any further attempts to activate its inactive neighbors. Again, the process ends when no more activations are possible. The NP-complete vertex cover problem and the NP-complete set cover problem are special cases of the influence maximization problem for the linear threshold model and independent cascade model, respectively. That is to say, the influence maximization problem is NP-hard for both the linear threshold model and independent cascade model \cite{Kempe2003ACM}. After Ref. \cite{Kempe2003ACM}, the influence maximization problem becomes a mainstream challenge for more than a decade. In addition to the explicitly defined problem and the proposed greedy algorithm, a very significant contribution of \cite{Kempe2003ACM} is to show an operational way to compare the performances of different algorithms by direct simulations. In addition, Kempe, Kleinberg and Tardos \cite{Kempe2003ACM} considered a general threshold model as well as a general cascade model, and showed that these two generalizations are equivalent to each other.

    Statistical physicists are also interested in this issue, while they considered some different dynamical processes, among which the SIR model is the most popular \cite{Pei2013JSMP12002,Zhang2016arXiv160200070}. One should be aware of the fact that the SIR model is very close to the independent cascade model. If we fix the recovering rate as $\delta=1$ and at the same time let each link's successful probability be a constant $p_{ij}=\beta$, then the two models are the same. In addition, the non-progressive threshold model is very close to the SIS model. Therefore, although there exists some language and methodological barriers, to build a way of bridging computer scientists and statistical physicists and then find better solutions to the influence maximization problem can be expected. Till far, studies on influence maximization problem on other dynamics, such as synchronization, evolutionary game and transportation, are rarely reported, which is also an interesting issue for further explorations.

\subsection{Heuristic and greedy algorithms}

    As the typical information maximization problems are NP-hard \cite{Kempe2003ACM}, the most known works attempt to find approximate solutions instead of the exact solution. Heuristic algorithms are the most common among all approximate algorithms, for example, to rank all nodes according to their degree or another centrality measure and directly pick up the $k$ top-ranking nodes is a naive but widely used heuristic algorithm for $k$-IMP (i.e., to find a set of $k$ nodes that maximize the influence). Another extensively studied class of algorithms is the greedy algorithms, which add nodes one by one to the target set, ensuring that each addition brings the largest increase of influence to the previous set. Though being simple, there are many exquisite designs and elegant analyses, and this section will highlight the representatives. Notice that, the heuristic algorithms are of inexplicit borderlines and many algorithms can be considered as heuristic algorithms (e.g., percolation models), so that if some algorithms can be classified in a more specific group for their common methodologies or perspectives, we will introduce them in an independent subsection, and of course to distinguish whether an algorithm is heuristic is meaningless in this research domain. We decide not to introduce purely optimization or machine learning algorithms, while readers are encouraged to see a recent book \cite{Chen2013Book} for those information.

\subsubsection{Heuristic algorithms}
\begin{figure}[htbp]
    \centering
    \includegraphics[width=12cm]{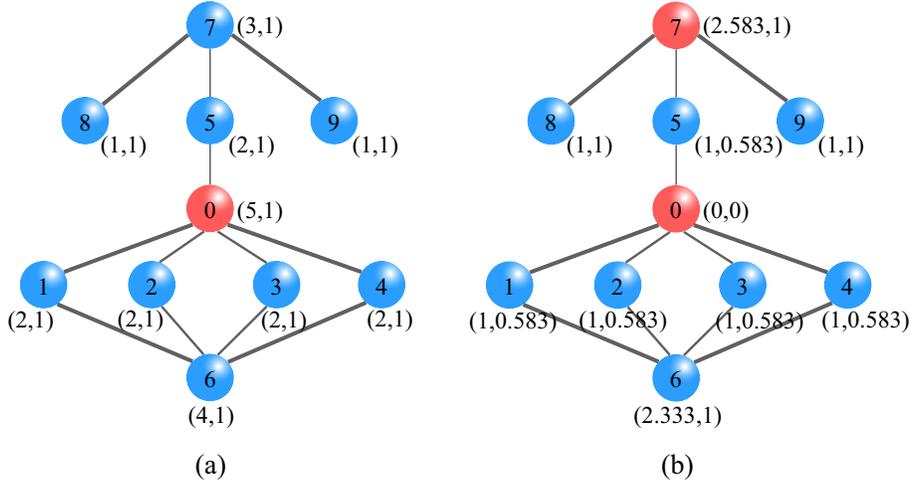}
    \caption{ An illustration of the updating rules in the VoteRank algorithm. (a) In step one, node 0 is selected as one of the vital nodes. (b) In step two, node 7 is selected as one of the vital nodes. After \cite{Zhang2016arXiv160200070}.}
    \label{Figure5}
    \end{figure}

    The most straightforward method is to directly pick up the top-$k$ nodes subject to some certain centrality measures, such as degree and betweenness \cite{Albert2000Nature406378,Cohen2001PRL863682,Pastor-Satorras2002PRE65036104}. However, as mentioned above, this method could be inefficient since the nodes of the highest degrees or betweennesses may be highly clustered. A slight improvement can be achieved by adaptive recalculation, that is, to choose the node of the largest centrality at first, and then recalculate the centralities of nodes after every step of node removal \cite{Holme2002PRE65056109}. Similar to the idea of recalculation, Chen \emph{et al.} \cite{Chen2009ACM} proposed a so-called degree discount algorithm, which performs almost the same to the greedy algorithm \cite{Kempe2003ACM} in accuracy while runs more than one-millionth faster than the fastest greedy algorithm. Considering the independent cascading model \cite{Goldenberg2001MarketLett12211,Goldenberg2001AMSR1} with constant activating rate $p_{ij}=p$ for every link. Let node $v_i$ and $v_j$ neighboring to each other, if $v_j$ has been already selected as a seed, then when considering selecting $v_i$ as a new seed based on its degree, we should discount $v_i$'s degree by one since the link $(i,j)$ contributes nothing to the spreading process. Let $s_i$ be the number of neighbors of $v_i$ that are already selected as seeds, then one should discount $v_i$'s degree as $k'_i=k_i-s_i$. To be more accurate, let's consider the expected number of additional nodes in $v_i$'s neighborhood (including $v_i$ itself) activated by selecting $v_i$ into the seed set. Firstly, the probability that $v_i$ is activated by its immediate neighbors is $1-(1-p)^{s_i}$. In this case, selecting $v_i$ as a seed contributes nothing. With the probability $(1-p)^{s_i}$, the additional nodes infected by $v_i$ include both $v_i$ itself and $(k_i-s_i)$ nodes, each of which is of probability $p$. Therefore, the total additional number should be
    \begin{equation}
    (1-p)^{s_i}\left[1+(k_i-s_i)p \right]\approx 1+\left[ k_i-2s_i-(k_i-s_i)s_ip \right]p,
    \end{equation}
    and thus one should discount $v_i$'s degree by $2s_i+(k_i-s_i)s_ip$. The degree discount algorithm works especially well when $p\ll 1$, where the approximation $(1-p)^{s_i} \approx 1-ps_i$ holds. Very recently, Zhang \emph{et al.} \cite{Zhang2016arXiv160200070} proposed another variant of adaptive recalculation named VoteRank. They assume that each node $v_i$ is characterized by a two-tuple $(S_i,V_i)$ where $V_i$ is $v_i$'s voting ability and $S_i$ is the score of $v_i$ defined as the sum of voting abilities of all $v_i$'s neighbors, as $S_i=\sum_{j\in \Gamma_i}V_i$. Initially, $V_i$ is set to $1$ for every node $v_i$ and thus $S_i$ equals the degree of $v_i$. The target set starts with an empty set, and at each time step the node $v_i$ with the largest score is selected into the target set, and then (i) the voting ability $V_i$ is set to zero; (ii) for each of $v_i$'s neighbors, its voting ability decreases by a factor $\frac{1}{\langle k \rangle}$ where $\langle k \rangle$ is the average degree, namely $V_j \leftarrow V_j - \frac{1}{\langle k \rangle}$ $(j \in \Gamma_i)$ and if $V_j<0$, we reset it as $V_j=0$. Fig.~\ref{Figure5} illustrates a simple example. The process runs for $k$ time steps and then the target set contains the required $k$ nodes. Although this method is very simple with the decaying factor $\frac{1}{\langle k \rangle}$ determined in a completely ad hoc way, it performs better than the ClusterRank \cite{Chen2013PLoSONE8e77455} and coreness \cite{Kitsak2010NatPhys6888} according to simulations in the SIR model.

    Many real networks exhibit community structures \cite{Girvan2002PNAS997821,Fortunato2010PhysRep48675}---connections within a community are dense while across communities are sparse.  This feature implies that nodes within a community are more likely to influence each other than nodes across communities. Therefore, for selecting a set of influencers, it's more efficient to choose the nodes in different communities instead of the whole network. Based on this idea, He \emph{et al.} \cite{He2015PLoSONE10e0145283} proposed a community-based method to find the top-$k$ influential spreaders located in different communities. First, the network is divided into many communities using the community detection algorithms \cite{Newman2004PRE69026113}. Then all communities are ranked in decreasing order according to their sizes. The first spreader is selected from the largest community according to a certain centrality index (e.g., to choose the node with the highest degree). Similarly, the node with the largest centrality index in the second largest community and having no edges incident to the previous communities (for the second selected spreader, there is only one previous community) is selected as the second spreader. If all communities are visited and the number of chosen spreaders is not enough, we restart the above process and choose the remaining spreaders following the same rules until $k$ spreaders are found. Clearly, the influential spreaders selected by this method are more likely to be dispersively distributed in the network. Different from above methods where the number of communities is alterable with different community detection algorithms, Zhang \emph{et al.} \cite{Zhang2013KBS4274} directly divided the network into $k$ communities based on the information transfer probability matrix and then $k$ medoids are chosen as $k$ spreaders. To obtain the information transfer probability matrix, each edge $(i,j)$ will be designated either ``open” with probability $\beta_{ij}$ or ``closed” with probability $1-\beta_{ij}$ independently, where $\beta_{ij}=1-(1-\beta)^{w_{ij}}$. The $\beta$ and $w_{ij}$ are the spreading probability and the weight of edge $(i,j)$, respectively. For unweighted networks, the process is similar to the bond percolation on network. For two nodes $v_i$ and $v_j$, if there is at least a path between them which is composed of ``open” edges, then $\omega(i,j) = 1$, otherwise 0. Then the element of information transfer probability matrix, $m_{ij}$, is defined as the average value of $\omega(i,j)$ of different trials. This method is very time consuming with time complexity in the order $O(k(n − k)^2)$ by using the most common realization of $k$-medoid clustering algorithm, \emph{Partitioning Around Medoids} \cite{Theodoridis2009BOOK}, and thus difficult to be applied to large-scale networks.

    With the similar consideration to the community-based methods, Zhao \emph{et al.} \cite{Zhao2014EPL10868005} divided the network into several independent sets, where any two nodes in one independent set are not neighboring to each other. This task is called \emph{graph coloring} in graph theory \cite{Bollobas1998BOOK}, where each node is assigned a color and two neighboring nodes cannot share the same color. To color the network, Zhao \emph{et al.} applied the well-known Welsh-Powell algorithm \cite{Welsh1967ComputJ1085}, which needs very small number of colors and is of relatively low time complexity $O(N^2)$ (for more recent algorithms, see \cite{Lu2010EJOR203241,Rossi2014SociNetAnaMin41,Hasenplaugh2014ACM166}). Given a certain centrality index, Zhao \emph{et al.} picked up the top-$k$ nodes in the independent set with the largest size (i.e., the nodes with the most popular color) to form the target set. By testing on Barab\'asi-Albert networks \cite{Barabasi1999Science286509} and two real networks subject to the contact process (a variant of SIR model, see \cite{Castellano2010PRL105218701,Yang2008PRE78026111,Yang2008PRE78066109}), Zhao \emph{et al.} showed that their method could improve the performances of many well-known centralities, including degree, betweenness, closeness, eigenvector centrality, and so on. The improvements of degree and betweenness are especially huge.

    Based on the heuristic optimal partitioning of graphs \cite{Karypis1998SIAM20359,Paul2007PRL99115701}, Chen \emph{et al.} \cite{Chen2008PRL101058701} proposed a so-called \emph{equal graph partitioning} (EGP) strategy (see also an earlier work with a similar idea \cite{Borgatti2006CMOT1221}), which performs significantly better than simply selecting the highest-degree or highest-betweenness nodes in fragmenting the network. The EGP strategy is based on the \emph{nested dissection} (ND) algorithm \cite{Lipton1979SIAM16346} that can separate a network into two equal-size clusters with a minimum number of nodes removed. Chen \emph{et al.} partitioned a network into any number of equal-size clusters by applying the ND algorithm recursively. To immunize a network with $n$ nodes so that only a fraction $F$ can be infected (i.e., the size of the giant component is no more than $Fn$), Chen \emph{et al.} separated the network into $n'\approx 1/F$ equal-size clusters. Therefore, given the target $F$, EGP will produce the minimum number of required nodes, which can be considered as the set of influential nodes required by the IMP. The EPG strategy is close to global optimization and thus more time-consuming than local algorithms. Notice that, the EPG strategy cannot be directly applied in solving the $k$-IMP since the size of the resulted set of influential nodes by EPG depends on $F$.

    Galinier \emph{et al.} \cite{Galinier2013JH19797} and Qin \emph{et al.} \cite{Qin2014EPJB87273} proposed local search algorithms to solve the FVS problem in directed and undirected networks, respectively. Considering the case of an undirected network $G$, Qin and Zhou \cite{Qin2014EPJB87273} construct a so-called legal list formed by $N$ ordered vertices as $L=(v_1,v_2,\cdots,v_N)$. For any vertex $v_i \in L$, only no more than one neighbor of $v_i$ is allowed to appear in front of $v_i$. Clearly, this constrain guarantees that the subgraph of $G$ induced by $L$ is cycle-free, so that the vertex set $G'=G-L$ is a FVS. The constrain for an directed network $D$ is similar \cite{Galinier2013JH19797}, that is, if both vertices $v_i$ and $v_j$ belong to the list $L$ and there is a directed link pointing from $i$ to $j$, $v_i$ must be in front of $v_j$ (i.e., $i<j$). Such constrain guarantees that the induced subgraph $D[L]$ does not contain any directed cycle. Since the larger size of $L$ corresponds to a smaller FVS, the energy of a legal list $L$ can be defined as
    \begin{equation}
    E(L)=n-|L|,
    \end{equation}
    where $n$ is the number of nodes in the network. Clearly, the lower $E$ corresponds to a better solution. The legal list $L$ can be constructed by using local search algorithms based on simulated annealing \cite{Kirkpatrick1983Science220671}. The algorithms are very straightforward with details can be found in Refs. \cite{Galinier2013JH19797,Qin2014EPJB87273}. The merit of such algorithms is to transfer the global constrains in FVS problem to local constrains in constructing the legal list $L$, which is also very similar to the constrains for extracting the spanning trees with maximal synchronizability \cite{Nishikawa2006PRE73065106,Nishikawa2006PhysD22477,Zhou2010NJP12043030}.

\subsubsection{Greedy algorithms}
    A natural idea to design a greedy algorithm is to ensure that each addition of a node to the target set (i.e., the set of initially infected or activated seeds) will maximize the incremental influence. Denote $f(S)$ the influence of a set of nodes $S$, which can be quantified by, for example, the number of ever infected nodes in an SIR model. Kempe, Kleinberg and Tardos \cite{Kempe2003ACM} proposed the earliest greedy algorithm for functional IMP. Their algorithm starts with an empty target set $S=\emptyset$, and at each time step, it scans all nodes to find the one $v\in V \backslash S$ that maximizes $f(S\cup \{v\})$ and then updates as $S\leftarrow S\cup \{v\}$. After $k$ time steps, one gets the target set $S$ containing $k$ influential nodes.

    To see the approximation guarantee, we first introduce the concept of \emph{submodular} \cite{Fujishige2005BOOK}. A function $f$, mapping a finite set to a non-negative real number, is a submodular function if the marginal gain from adding an element to a set $S$ is no less than the marginal gain from adding the same element to a superset of $S$. Formally, a submodular function satisfies
    \begin{equation}
    f(S\cup \{v\})-f(S) \geq f(T\cup \{v\})-f(T)
    \end{equation}
    for all elements $v$ and all pairs of sets $S\subseteq T$. If $f$ is monotone, that is, $f(S\cup \{v\}) \geq f(S)$ for all elements $v$ and sets $S$, then it has been proved that \cite{Cornuejols1977ManageSci23789,Nemhauser1978MathProg14265} the above greedy algorithm (i.e., the simplest hill-climbing algorithm) approximates to the optimum $S^*$ within a factor $1-1/e\approx 0.63$, namely
    \begin{equation}
    f(S)\geq \left(1-\frac{1}{e}\right)f(S^*).
    \end{equation}
    Kempe \emph{et al.} \cite{Kempe2003ACM} proved that for the case of the independent cascade model and linear threshold model, the objective functions on the expected number of activated nodes $f(\cdot)$ are all submodular, so the greedy hill-climbing algorithm provides a $(1-1/e)$-approximation. As shown in Ref. \cite{Kempe2003ACM}, the greedy hill-climbing algorithm performs much better than simply selecting the top-$k$ nodes of the highest degrees or the smallest closenesses.

    In the independent cascade model, the activating probability $p_{ij}$ associated with the link $(i\rightarrow j)$ is independent of the history of the dynamical process. However, information propagation in social networks exhibits memory effects \cite{Dodds2004PRL92218701,Centola2010Science3291194,Lu2011NJP13123005,Krapivsky2011JSMP12003}. Therefore, Kempe, Kleinberg and Tardos \cite{Kempe2005BOOK} further extended the independent cascade model to the so-called \emph{decreasing cascade model}, where $p_{ij}$ depends on the history. Denote $S$ the set a node $v_j$'s neighbors that have already attempted to activate $v_j$, then $v_i$'s success probability to activate $v_j$ is $p_{ij}(S)$. The decreasing cascade model contains two natural constrains: (i) \emph{order-independent}---if all nodes from a set $T$ try to activate a node $v_j$, then the order in which their attempts are made does not affect the probability of $v_j$ being active in the end; (ii) \emph{non-increasing}---the function $p_{ij}(S)$ satisfies the inequality $p_{ij}(S) \geq p_{ij}(T)$ when $S\subseteq T$. Kempe \emph{et al.} \cite{Kempe2005BOOK} proved that the objective function $f(\cdot)$ for the decreasing cascade model is also submodular, so that the greedy hill-climbing algorithm provides a $(1-1/e)$-approximation.

    An obvious and serious drawback of the original greedy algorithm is that it is very time-consuming. For the $k$-IMP on a network of $n$ nodes and $m$ links, if it requires $R$ times to run the direct simulation of a given dynamical process in order to accurately estimate the expected number of activated nodes $f(S)$ from a set of seeds $S$, the time complexity is $O(kRNM)$. In fact, it takes days to complete the algorithm for a small network with tens of thousands nodes and $k\leq 100$. Therefore, the original greedy algorithm can not be directly applied to large-scale networks in modern information society.

    Leskovec \emph{et al.} \cite{Leskovec2007ACM} noticed the submodularity property that when adding a node $v_i$ into a seed set $S$, the marginal gain $f(S\cup \{v_i\})-f(S)$ is larger (or at least equal) if $S$ is smaller. Accordingly, Leskovec \emph{et al.} \cite{Leskovec2007ACM} proposed the so-called \emph{cost-effective lazy forward} (CELF) algorithm by utilizing the fact that in each time step of finding the node for maximum marginal gain, a large number of nodes do not need to be re-evaluated because their marginal gain in the previous round are already less than that of some other nodes evaluated in the current time step. As reported in \cite{Leskovec2007ACM}, for some specific network instances, the CELF algorithm is 700 times faster than the original greedy algorithm (see also \cite{Goyal2011ACM} for a further extension of CELF, named CELF++).

    Another direction to improve the efficiency of the original greedy algorithm is to speed up the process to obtain the expected number of activated nodes $f(S)$. In fact, the direct simulation of a spreading dynamics, such as the independent cascade model and SIR model, is very inefficient. Fortunately, both the independent cascade model and SIR model are equivalent to a bond percolation \cite{Grassberger1983MathBio63157,Moore2000PRE615678,Newman2002PRE66016128,Sander2002MathBio180293,Zhou2006PNS16452} (the connections between computer science and physical models were largely ignored in the literature, for example, the well-known bootstrap percolation model \cite{Chalupa1979JPC12L31,Gao2015SciRep514662} can be considered as a special case of the linear threshold model \cite{Granovetter1978AJS831420,Schelling1978BOOK,Watts2002PNAS995766}), so that $f(S)$ can be quickly estimated by bond percolating. Chen \emph{et al.} \cite{Chen2009ACM} firstly utilized this equivalence to speed up the original greedy algorithm. Considering the independent cascade model, for each of the $R$ runs, Chen \emph{et al.} \cite{Chen2009ACM} generate a network $G'$ by removing each link $(v_i\rightarrow v_j)$ from $G$ with probability $1-p_{ij}$ (i.e., to get the percolated network). Let $R_{G'}(S)$ be the set of reachable nodes from $S$ in $G'$, then with a linear scan of $G'$ (by either depth-first search or breadth-first search with time complexity $O(M)$), one can obtain both $R_{G'}(S)$ and $R_{G'}(\{v_i\})$ for all nodes $v_i\in V$. Then for every node $v_i \in V \backslash S$, the marginal gain by adding $v_i$ to $S$ is either $|R_{G'}(\{v_i\})|$ if $v_i \notin R_{G'}(S)$ or 0 if $v_i \in R_{G'}(S)$. This algorithm is of time complexity $O(kRM)$, in principle $N$ times faster than the original greedy algorithm. It also runs remarkably faster than the CELF algorithm according to the comparison reported in Ref. \cite{Chen2009ACM}.

    Till far, there exists a huge number of extensions of the original greedy algorithm by considering the variants of the spreading dynamics \cite{Kimura2006PKDD}, by making use of the structural properties of the target networks \cite{Wang2010KDD1039}, by integrating advanced optimization methods developed for related problems \cite{Goyal2011IEEE}, and so on. This review only present the most significant progresses and tries to describe them using the language familiar to physical society and to connect them with some well-known physical models. The readers are encouraged to read the book \cite{Chen2013Book} for more examples.

\subsection{Message passing theory}
    The message-passing theory is firstly developed to deal with high-dimensional disorder systems \cite{Mezard2001EPJB20217}, which quantifies the probabilities of solutions in a dynamics that can be represented by a static constraint-satisfaction model over discrete variables. Therefore, the message-passing theory has found wide applications in combinational optimization \cite{Mezard2009BOOK} and becomes a popular tool in network analysis \cite{Karrer2010PRE82016101}. This section will introduce two examples. The former is proposed for a typical structural IMP, the FVS problem \cite{Zhou2013EPJB86455}, which can also be extended to deal with another structural IMP, the optimal percolation problem \cite{Mugisha2016arXiv160305781}. The latter is proposed for a typical functional IMP, that is, to find the most influential spreaders as initial seeds that maximize the expected number of finally active nodes in the linear threshold model \cite{Altarelli2013JSMP09011}, which can also be applied in finding a small set of nodes to immunize that best protect the whole network \cite{Altarelli2014PRX4021024}.

    Zhou \cite{Zhou2013EPJB86455} proposed an efficient message-passing algorithm, with the core idea is to turn the global cycle constraints into a set of local constraints. Let's consider an undirected simple network $G$ and define on each vertex $v_i$ a state variable $A_i$, which can take the value $A_i=0$, $A_i=i$, or $A_i=j\in \Gamma_i$, where $\Gamma_i$ is the set of $v_i$'s neighbors. If $A_i=0$ we say that vertex $v_i$ is unoccupied, if $A_i=i$ we say that vertex $v_i$ is occupied and it is a root vertex without any parent, and if $A_i=j\in \Gamma_i$ we say that vertex $v_i$ is occupied with $v_j$ being its parent vertex. Similar vertex state variables have already been used in the early studies on the Steiner tree problem \cite{Bayati2008PRL101037208,Bayati-Bechet2011PNAS108882}. Zhou \cite{Zhou2013EPJB86455} defined an edge factor $C_{ij}(A_i,A_j)$ for any edge $(i,j)$ as
    \begin{equation}
    C_{ij}(A_i,A_j)=\delta_{A_i}^0\delta_{A_j}^0+\delta_{A_i}^0(1-\delta_{A_j}^0-\delta_{A_j}^i)+\delta_{A_j}^0(1-\delta_{A_i}^0-\delta_{A_i}^j)+\delta_{A_i}^j(1-\delta_{A_j}^0-\delta_{A_j}^i)+\delta_{A_j}^i(1-\delta_{A_i}^0-\delta_{A_i}^j),
    \end{equation}
    where $\delta_i^j$ is the Kronecker symbol such that $\delta_i^j=1$ if $i=j$ and 0 otherwise. The value of the edge factor $C_{ij}(A_i,A_j)$ is either 0 or 1, and $C_{ij}(A_i,A_j)=1$ only in the following five situations: (i) both vertex $v_i$ and vertex $v_j$ are unoccupied; (ii) vertex $v_i$ is unoccupied while vertex $v_j$ is occupied, and $v_i$ is not the parent of $v_j$; (iii) vertex $v_j$ is unoccupied while vertex $v_i$ is occupied, and $v_j$ is not the parent of $v_i$; (iv) both vertex $v_i$ and vertex $v_j$ are occupied, and $v_j$ is the parent of $v_i$ but $v_i$ is not the parent of $v_j$; (iv) both vertex $v_i$ and vertex $v_j$ are occupied, and $v_i$ is the parent of $v_j$ but $v_j$ is not the parent of $v_i$. Given a microscopic configuration $\underline{A}=\{A_1,A_2,\cdots,A_N\}$, we regard each edge $(i,j)$ as a local constraint and an edge $(i,j)$ is satisfied if $C_{ij}(A_i,A_j)=1$, otherwise it is unsatisfied. If a microscopic configuration $\underline{A}$ satisfies all the edges of the network $G$, it is then referred to as a \emph{solution} of this network.

    Let's define a $c$-tree as a connected graph with a single cycle who has $n\geq 3$ vertices and $n$ edges. It can be easily proved that the occupied vertices of any solution $\underline{A}$ of a network $G$ induce a subgraph with one or more connected components, where each component is either a tree or a $c$-tree. The solutions of $G$ are closely related to the feedback vertex sets, since we can randomly remove one vertex from the cycle in each $c$-tree and turn the induced subgraph into a forest, so that the remaining nodes form a feedback vertex set. Therefore, a solution with more occupied nodes generally corresponds to a smaller FVS (this is not exactly true since the solution may contain extensive number of $c$-trees, but we do not consider such abnormal situation). Till far, as mentioned in the beginning of this section, we have turn the original FVS problem into a static constraint-satisfaction model over discrete variables. Accordingly, Zhou \cite{Zhou2013EPJB86455} defined the partition function for the system as
    \begin{equation}
    Z(x)=\sum_{\underline{A}}\exp\left[x\sum^N_{i=1}(1-\delta_{A_i}^0)w_i\right]\prod_{(i,j)\in G}C_{ij}(A_i,A_j),
    \end{equation}
    where $w_i\geq 0$ is the fixed weight of each vertex $v_i$, $x$ is a positive re-weighting parameter, and the term $\prod_{(i,j)\in G}C_{ij}(A_i,A_j)$ guarantees that only the solutions contribute to the partition function $Z(x)$.

    Denote $q_i^{A_i}$ the marginal probability that vertex $v_i$ takes the state $A_i$, which is largely influenced by $v_i$'s neighbors while at the same time $v_i$'s state will influence its neighbors' states. To avoid over-counting in computing $q_i^{A_i}$, we can first remove vertex $v_i$ from the network and then consider all the possible state combinations of the set $\Gamma_i$ in the remaining network, which is called a cavity network. Notice that, in the cavity network, the vertices in $\Gamma_i$ might still be correlated (they are uncorrected only if G is a tree), while we neglect all possible correlations and assume independence of probabilities, which is commonly known as the Bethe-Peierls approximation \cite{Bethe1935PRSLA150552,Peierls1936PRSLA154207,Peierls1936MPCPS32477} or the correlation decay assumption \cite{Mezard2009BOOK} in the statistical physics community, and works well if the network is locally like a tree (this is almost the case when the network is very sparse, and real networks are usually very sparse). According to the Bethe-Peierls approximation, the joint probability is approximately factorized as
    \begin{equation}
    P_{\backslash i}(\{A_j,j\in \Gamma_i\})\approx \prod_{j\in \Gamma_i}q_{j\rightarrow i}^{A_j},
    \label{TZ07}
    \end{equation}
    where $q_{j\rightarrow i}^{A_j}$ denotes the marginal probability of the state $A_j$ in the cavity network, where the effect of vertex $v_i$ is not considered. If all the vertices $v_j\in \Gamma_i$ are either unoccupied ($A_j=0$) or roots ($A_j=j$) in the cavity network, then $v_i$ can be a root ($A_i=i$) when it is added to the network. This is because a neighboring vertex $v_j$ can turn its state to $A_j=i$ after $v_i$ is added. Similarly, if one vertex $l\in \Gamma_i$ is occupied in the cavity network and all the other vertices in $\Gamma_i$ are either empty or roots in the cavity network, then $v_i$ can take the state $A_i=l$ when it is added to the network. These considerations, together with the Bethe-Peierls approximation (Eq. (\ref{TZ07})), result in the following expressions for $q_i^{A_i}$:
	\begin{equation}
	q_i^0=\frac{1}{z_i},
	\end{equation}
	\begin{equation}
	q_i^i=\frac{1}{z_i}\left[ e^{xw_i}\prod_{j\in \Gamma_i} \left(q_{j\rightarrow i}^0+q_{j\rightarrow i}^j\right) \right],
	\end{equation}
	\begin{equation}
	q_i^l=\frac{1}{z_i}\left[ e^{xw_i} \left( 1-q_{l\rightarrow i}^0 \right) \prod_{k\in \Gamma_i \backslash l} \left(q_{k\rightarrow i}^0+q_{k\rightarrow i}^k\right) \right],
	\end{equation}
	where $l\in \Gamma_i$ and $z_i$ is the normalization constant as
	\begin{equation}
	z_i=1+e^{xw_i}\left[\prod_{j\in \Gamma_i} \left(q_{j\rightarrow i}^0+q_{j\rightarrow i}^j\right) + \sum_{j \in \Gamma_i} \left( 1-q_{j\rightarrow i}^0 \right) \prod_{k\in \Gamma_i \backslash j} \left(q_{k\rightarrow i}^0+q_{k\rightarrow i}^k\right)\right].
	\end{equation}
    Finally, we need a set of equations for the probability $q_{i\rightarrow j}^{A_i}$, which has the same meaning as $q_i^{A_i}$ but is defined on the cavity network without vertex $v_j$. Therefore, we have
	\begin{equation}
	q_{i\rightarrow j}^0=\frac{1}{z_{i\rightarrow j}},
	\end{equation}
	\begin{equation}
	q_{i\rightarrow j}^i=\frac{1}{z_{i\rightarrow j}}\left[ e^{xw_i}\prod_{k\in \Gamma_i \backslash j} \left(q_{k\rightarrow i}^0+q_{k\rightarrow i}^k\right) \right],
	\end{equation}
	\begin{equation}
	q_{i\rightarrow j}^l=\frac{1}{z_{i\rightarrow j}}\left[ e^{xw_i} \left( 1-q_{l\rightarrow i}^0 \right) \prod_{m\in \Gamma_i \backslash \{j,l\}} \left(q_{m\rightarrow i}^0+q_{m\rightarrow i}^m\right) \right],
	\end{equation}
		where $l\in \Gamma_i \backslash j$ and $z_{i\rightarrow j}$ is the normalization constant as
	\begin{equation}
	z_{i\rightarrow j}=1+e^{xw_i}\left[ \prod_{k\in \Gamma_i \backslash j} \left(q_{k\rightarrow i}^0+q_{k\rightarrow i}^k\right) + \sum_{k\in \Gamma_i \backslash j} \left( 1-q_{k\rightarrow i}^0 \right) \prod_{m\in \Gamma_i \backslash \{j,k\}} \left(q_{m\rightarrow i}^0+q_{m\rightarrow i}^m\right)     \right].
	\end{equation}
    These self-consistent equations are obtained by the message-passing theory and commonly referred to as a set of belief propagation (BP) equations.

    The BP equations can be iteratively solved, and taking the limit of large $x$, we can in principle get the solution corresponding to the maximum total node weight in the induced subgraph and thus the minimal total weight in the FVS. However, there are still many technique details in the implementation. For example, the value of $x$ can not be real infinite but an appropriate value since the very large value will lead to the disconvergence in the iteration, and each round we only pick up a very small fraction of the nodes $v_i$ with the highest unoccupied probabilities $q_i^0$ to form the FVS and then resolve the BP equations for the new network without selected nodes. Readers who would like to implement the algorithm should read Ref. \cite{Zhou2013EPJB86455} for algorithmic details. Very recently, Mugisha and Zhou \cite{Mugisha2016arXiv160305781} argued that their algorithm can be easily extended to solve the well-known optimal percolation problem, with remarkably better performance than the algorithm in Ref. \cite{Morone2015Nature52465}, while Morone \emph{et al.} \cite{Morone2016arXiv160308273} refuted that a variant version of their original algorithm can achieve almost the same performance but runs much faster than the message-passing algorithm. It is still under debate.

    Altarelli \emph{et al.} \cite{Altarelli2013JSMP09011} considered the IMP for a progressive spreading dynamics, the linear threshold model, which is more complicated than the FVS problem. Since the essential idea of the method in Ref. \cite{Altarelli2013JSMP09011} is similar to that in Ref. \cite{Zhou2013EPJB86455}, we only highlight the  noticeable differences. Denote $x_i^t$ the state of node $v_i$ at time step $t$, the linear threshold model starts with a few active seeds $x_i^0=1$ and the updating rule reads
	\begin{eqnarray}
	x_i^{t+1}=&\left\{ \begin{matrix}
	   1~~~~~x_i^t=1\texttt{ or }\sum_{j\in \Gamma_i} w_{ji}x_j^t\geq \theta_i,  \\
	   0~~~~~~~~~~~~~~~~~~~~~~~~~~~~~\texttt{otherwise}.  \\
	\end{matrix}
	\right.
	\end{eqnarray}
    Denote $t_i$ the activation time of node $v_i$ (the initial seed is of $t_i=0$, the inactive node in the final state is of $t_i=\infty$), then the evolution of the dynamical process can be fully represented by a configuration $\underline{t}=\{t_i\}$, $v_i \in V$. In the linear threshold model, the constraint between the activation times of neighboring nodes is
	\begin{equation}
	t_i=\phi_i(\{t_j\})=\min\left\{t: \sum_{j\in \Gamma_i}w_{ji}\Theta[t_j<t] \geq \theta_i \right\},
	\end{equation}
    where $\Theta[\cdot]=1$ if the condition in $[\cdot]$ is true and 0 otherwise. Therefore, an solution $\underline{t}$ of the linear threshold model satisfies the constraints
	\begin{equation}
	\Psi_i=\Theta[t_i=0]+\Theta[t_i=\phi_i(\{t_j\})]=1
	\end{equation}
	for every node $v_i$. Accordingly, we can write down the partition function as
	\begin{equation}
	Z(\beta)=\sum_{\underline{t}}e^{-\beta E(\underline{t})} \prod_i \Psi_i(t_i,\{t_j\}_{j\in \Gamma_i}),
	\end{equation}
    where $E(\underline{t})=\sum_i E_i(t_i)$ with $E_i(t_i)$ being the cost (if positive) or revenue (if negative) incurred by activating node $v_i$ at time $t_i$. Altarelli \emph{et al.} \cite{Altarelli2013JSMP09011} set the energy function as
	\begin{equation}
	E_i(t_i)=c_i\Theta[t_i=0]-r_i\Theta[t_i<\infty],
	\end{equation}
	where $c_i$ is the cost of selecting $v_i$ as a seed and $r_i$ is the revenue generated by the activation of $v_i$.

    Different from the FVS problem, if one only consider the constraints for a single variable $t_i$, the factor network will consist of many short loops and thus the correlation decay assumption does not work. Therefore, Altarelli \emph{et al.} \cite{Altarelli2013JSMP09011} considered the pair of times $(t_i,t_j)$ for every edge $(i,j)$ instead. using the similar techniques, the marginal distribution $H_{il}(t_i,t_l)$ for a variable $(t_i,t_l)$ in the absence of node $v_j$ (a more precise description based on a bipartite factor network representation as well as the derivation can be found in \cite{Altarelli2013JSMP09011,Altarelli2013PRE87062115,Altarelli2014PRL112118701}) is
    \begin{equation}
    H_{ij}(t_i,t_j)\propto e^{-\beta E_i(t_i)} \sum_{\{t_k\}_{k\in \Gamma_i\backslash j}} \Psi_i(t_i,\{t_k\}_{k\in \Gamma_i})\prod_{k\in \Gamma_i \backslash j} H_{k,i}(t_k,t_i).
    \end{equation}
    These self-consistent equations are also solved by iteration, and by taking the limitation $\beta\rightarrow \infty$ we can get the desired results for IMP. More technique details and the extension to the network immunization can be respectively found in Ref. \cite{Altarelli2013JSMP09011} and Ref. \cite{Altarelli2014PRX4021024}.

    Notice that, the message-passing theory does not require the submodular property in objective functions. Since in some dynamical processes with phase transition, the addition of one node in the subcritical region may induce sharply change of the value of the final state, so that the submodular property may not hold. Therefore, the message-passing theory may find wider applications than the greedy algorithms. But of course, if we do not care the approximation guarantee, we can use the greedy algorithms anyway.

    \subsection{Percolation methods}
    Network percolation can be classified into bond percolation and site percolation \cite{Cohen2010BOOK}. Given an undirected network $G(V,E)$, in bond percolation, each edge is preserved (i.e., occupied) with a probability $p$ and removed with a probability $1-p$. When $p=0$, all links are removed. With $p$ increases, more links are preserved and form some small clusters. A giant connected component of size $O(|V|)$ emerges only when $p$ is larger than a critical threshold $p_c$. The process is similar in site percolation where the difference is that the preserve probability $p$ is assigned with nodes instead of edges.

    The susceptible-infected-recovered (SIR) model are widely used to describe the propagation of information. Given the spreading dynamics, the influence of a node $v_i$ can be measured by the number of the eventually infected nodes with $v_i$ being the initial seed. Newman~\cite{Newman2002PRE66016128} studied in detail the relation between the static properties of the SIR model and bond percolation and showed that the SIR model with transmissibility $p$ is equivalent to a bond percolation model with bond occupation probability $p$ on the network.

    Considering this natural relation, Ji \emph{et al.}~\cite{Ji2015ArXiv150804294} proposed a bond percolation-based method to identify the optimal combination of given number of influential spreaders. To find $W$ influential nodes, given the probability $p$, each edge will be removed with probability $p$ and then $m$ isolated clusters will appear after the link removal. Denote by $S_i (i=1,2,\cdots,m)$ the size of cluster $i$ and $L$ ($\geq W$) a tunable parameter, if $L\leq m$, the top-$L$ largest clusters will be selected and one score is assigned to the largest degree node in each cluster. If there are many nodes with the largest degree, randomly select one of them. If $m<L\leq 2m$, the largest degree node in each cluster will be firstly selected, and the rest $L-m$ nodes are chosen to be those with the second largest degree respectively from the top-$(L-m)$ largest clusters. If $L>2m$, the next largest degree nodes in each cluster will be chosen following the same rules. After $t$ times of different trials of link removal, all nodes are ranked according to their scores in a descending order and those $W$ nodes with the highest scores are suggested to be the set of initial spreaders. Clearly, this method can be parallel computed and the complexity would be $O(t|V|)$. Results on Facebook and Email-Enron networks show that compared to the un-coordinated spreaders identified by conventional methods, such as degree, betweenness, closeness, $k$-shell \emph{etc.}, the spreaders identified by the percolation method are evenly distributed within the network which greatly increases the propagation coverage and reduces its redundancy.

    Comparing with the case of finding a given number of influential spreaders, to find a minimal set of nodes that optimize a global function of influence is more complicated and was shown to be a NP-hard problem~\cite{Kempe2003KDD137}. Recently, Morone \emph{et al.} \cite{Morone2015Nature52465} pointed out that the problem of finding the minimal set of activated nodes to spread information to the whole network or to optimally immunize a network against epidemics can be exactly mapped onto optimal percolation in networks. Different from Ji's method, here the idea is based on the network site percolation and the main task is to find the minimal set of nodes which are crucial for the global connectivity of the network \cite{Albert2000Nature406378}. Consider a network of $n$ nodes and $m$ edges, let the vector $\overrightarrow{n}=(n_1,…, n_n)$ represent which node is removed ($n_i=0$, influencer) or preserved ($n_i=1$) in the network and $q=1-1/n\sum_i n_i$. The order parameter, $v_{i\rightarrow j}$, is the probability that node $v_i$ belongs to the giant component in a modified network where node $v_j$ is absent. Here $i\rightarrow j$ presents the link from $v_i$ to $v_j$. Then the optimal influence problem for a given $q(\geq q_c)$ can be rephrased as finding the optimal configuration $\overrightarrow{n}$ that minimizes the largest eigenvalue $\lambda(\overrightarrow{n};q)$ of the linear operator $\hat{\mathcal{M}}$, defined on the $2m\times2m$ directed edges as
    \begin{equation}
    \mathcal{M}_{k\rightarrow l,i\rightarrow j}\equiv\frac{\partial v_{i\rightarrow j}}{\partial v_{k\rightarrow l}}|_{\{v_{i\rightarrow j}=0\}}.
    \end{equation}
    The solution can be found by minimizing the energy of a many-body system with the extremal optimization (EO) method \cite{Boettcher2001PRL865221}. Since EO is not scalable to find the optimal configuration in large networks, a scalable algorithm, called ``Collective Influence" (CI) was proposed~\cite{Morone2015Nature52465}. Define $Ball(i,l)$ the set of nodes inside a ball of radius $l$ (defined as the shortest path) around node $v_i$, $\partial Ball(i, l)$ is the frontier of the ball. Then the CI index of node $v_i$ at level $l$ is
    \begin{equation}
    CI_l(i)=(k_i-1)\sum_{j\in \partial Ball(i,l)}(k_j-1),
    \end{equation}
    where $k_i$ is the degree of node $v_i$ and $l$ is a predefined nonnegative integer which does not exceed the network diameter for finite networks. The larger $l$ is, the better approximations to the optimal solution can be obtained. Initially, $CI_l$ is calculated based on the whole network. Then, the node with the highest $CI_l$ will be removed. Recalculate $CI_l$ for the rest nodes and remove the new top $CI_l$ node. Repeat the process until the giant component vanishes. The CI-algorithm scales as $O(N\mathrm{log}N)$ by removing a finite fraction of nodes at each step.


    As the information spreading on networks is a global process, it is commonly believed that maximizing the propagation coverage would require the whole structural information, such as the above mentioned methods. However, inspired by the famous social contagious phenomena---``Three degrees of influence” (which states that any individual's social influence ceases beyond three degrees \cite{Christakis2007NEJM357370,Christakis2013SM32556}), and by mapping the spreading dynamics (SIR family) onto bond percolation \cite{Newman2002PRE66016128} on complex networks, Hu \emph{et al.} \cite{Hu2016ArXiv150903484} found a pivotal law central to SIR family spreading---the spreading occurs in one of the two states, local phase and global phase. The former corresponds to a confined spreading which has a very limited number of active nodes, while the latter corresponds to the widespread phase whose scale is in the order of the whole network, and whose fraction is invariant with respect to initial seeds and realizations. The global and local phases are unambiguously separated by a characteristic local length scale, which could be used to predict and quantify the outcome of the spreading at the early several steps of the spreading. This revealed a very fundamental and exciting result, that is, a node's or a group of nodes' global influence can be exactly measured by using purely local network information. They gave a theoretical explanation for the presence of this local length scale which is analogous to the correlation length in a percolation transition, and obtained that the influence of a node equals the product of the giant component size and the probability that this node belongs to the giant component. For a set of $W$ nodes whose size is much smaller than the network size, the collective spreading influence equals the product of the giant component size and the probability that at least a cluster of $m$ nodes are activated by the $W$ spreaders. Here $m$ is a threshold parameter which is determined by the correlation length in critical phenomena. To find the best $W$ spreaders, they proposed a percolation-based greedy algorithm which gives a nearly optimized solution with computational time independent of the network size.

\section{On weighted networks}\label{Chapter7}
In many real systems, the interactions between nodes are usually not merely binary entities (either present or not). For example, we have both good friends and nodding acquaintances. To describe the relations we need to use weighted network representation. Denote by $G(V,E)$ an weighted undirected network, where $V$ and $E$ are the sets of nodes and weighted edges, respectively. The corresponding weighted adjacency matrix is $W$, where $W=\{w_{ij}\}$ if node $v_i$ and node $v_j$ are connected with weight $w_{ij}>0$, and $w_{ij}=0$ if node $v_i$ and node $v_j$ are not connected. Here the weight associated with a link quantifies the strength of the interaction between the two nodes. Representative weighted networks include the air transportation network where the weight between two airports is the number of available seats on the direct flight~\cite{Barrat2004PNAS1013747}, the scientific collaboration network where the weights are the numbers of coauthored papers, the neural network where the weights are the strengths of the interactions between neurons, etc. To quantify the features of weighted networks, there are many characteristic indices, such as the weighted clustering coefficient and assortativity \cite{Barrat2004PNAS1013747}, the weighted modularity \cite{Newman2004PRE70056131} and the disparity of weights \cite{Barthelemy2003PhysicaA319633}. In this section, we review some representative methods to find vital nodes in weighted networks. Many of them are natural extensions of the methods originally for unweighted networks.
    \subsection{Weighted centralities}
        \subsubsection{Node strength}
        Consider an undirected network, for an arbitrary node $v_i$, its strength is defined as the summation of the weights of the links associated to $v_i$, namely
        \begin{equation}
        s_i=\sum_{j=1}^n w_{ij}.
        \end{equation}
        The node strength integrates the information about both its connectivity and the importance of its associated links. When the weights are independent from the topology, we have $s\simeq \langle w\rangle k$ where $\langle w\rangle$ is the average weight \cite{Barthelemy2005PhysicaA34634}. While in real weighted networks, the strength is correlated with degree in a nonlinear way as $s \propto k^{\theta}$ with $\theta \ne 1$ \cite{Barrat2004PNAS1013747,Li2004PRE69046106,Wang2005PRL94188702,Ou2007PRE75021102}. For directed networks, we can also defined the in-strength and out-strength, which read
        \begin{equation}
        s_i^{in}=\sum_{j=1}^n w_{j\rightarrow i},\quad s_i^{out}=\sum_{j=1}^n w_{i\rightarrow j},
        \end{equation}
        where $w_{i\rightarrow j}$ is the weight of the directed link from node $v_i$ to node $v_j$. By normalizing the node strength, we obtain the weighted degree centrality as
        \begin{equation}
        WDC(i)=\frac{s_i}{\sum_{j=1}^n s_j}.
        \end{equation}

        \subsubsection{Weighted coreness}
\begin{figure}
		  \centering
 		  \includegraphics[width=11cm]{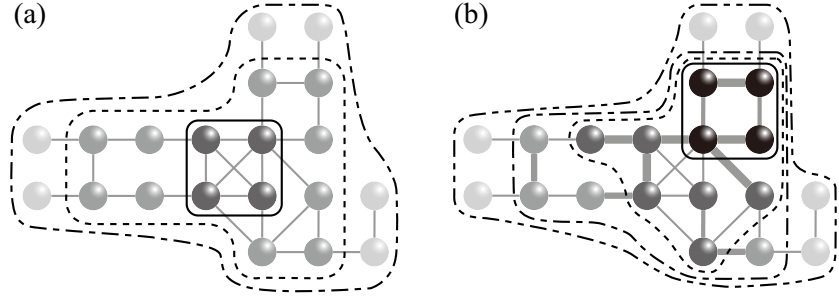}\\
		  		\caption{Illustration of network decomposition: (a) $k$-core decomposition using node degree. (b) $s$-core decomposition using node strength. All nodes inside the continuous lines belong to the same core. The link thickness indicates connection strength. Nodes in the same $k$- or $s$-shell have identical shading. From outermost towards the core: light-gray, gray, dark-gray, and black \cite{Marius2013PRE88062819}.}
		\label{Figure6}
		\end{figure}

        The classic $k$-shell decomposition can be extended to weighted network by simply replacing the node degree with weighted degree in the pruning (i.e., node removal) process. Besides node strength, Garas \emph{et al.}~\cite{Garas2012NJP1483030} defined the weighted degree of a node by considering both degree and strength, which is written as
        \begin{equation}
        k_i^W=(k_i^{\alpha}s_i^{\beta})^{\frac{1}{\alpha+\beta}},\label{w-core}
        \end{equation}
        where $k_i$ is the degree of $v_i$, $\alpha$ and $\beta$ are tunable parameters. When $\alpha=1$ and $\beta=0$, $k_i^W=k_i$ corresponding to the classic $k$-shell decomposition. When $\alpha=0$ and $\beta=1$, $k_i^W=s_i$ refers to the $s$-shell/$s$-core decomposition \cite{Marius2013PRE88062819}. When $\alpha=\beta=1$, $k_i^W=\sqrt{k_is_i}$ which means that the weight and the degree are treated equally. The $k$-core decomposition process in weighted networks is very similar to the process in unweighted networks. The only difference is that the weighted degree is usually non-integer. Denoted by $k_{min}^W=\mathrm{min} _ik_i^W$ the minimum weighted degree. The weighted $k$-core decomposition process is initiated by removing all nodes with degree $k^W=k_{min}^W$. Removing these nodes would lead to the decrease of other nodes' weighted degrees, which are named residual degree $\tilde{k}_i^W$. Then the nodes with $\tilde{k}^W\leq k_{min}^W$ also need to be removed. This iterative removals will  terminate when all the remaining nodes are with residual degree $\tilde{k}^W>k_{min}^W$. The removed nodes and their associated links form the $k_{min}^W$-shell. This pruning process is repeated for the nodes of degree $\tilde{k}^W=\tilde{k}_{min}^W$ to extract the $\tilde{k}_{min}^W$-shell, that is, in each stage the nodes with degree $\tilde{k}^W\leq\tilde{k}_{min}^W$ are removed. The process is continued until all higher-layer shells have been identified and all network nodes have been removed. Then each node $i$ is assigned a shell layer $c_i^W$, called the weighted coreness of node $i$, which means that node $v_i$ belongs to $c_i^W$-shell, but not any $c^W$-shell with $c^W \ge c_i^W$. Fig.~\ref{Figure6} gives an illustration of classic $k$-core decomposition and weighted $s$-core decomposition (i.e., $\alpha=0$ and $\beta=1$). The $s$-core structure of a network is determined by both the network topology and how the link weights are correlated with degree. Besides Eq. (\ref{w-core}), the weighted degree can be also defined in a linear way as $k_i^W=\alpha k_i+(1-\alpha)s_i$, where $\alpha\in[0,1]$ \cite{Wei2015PhysicaA420277}.

        \subsubsection{Weighted H-index}
\begin{figure}
		\centering
		\includegraphics[width=12cm]{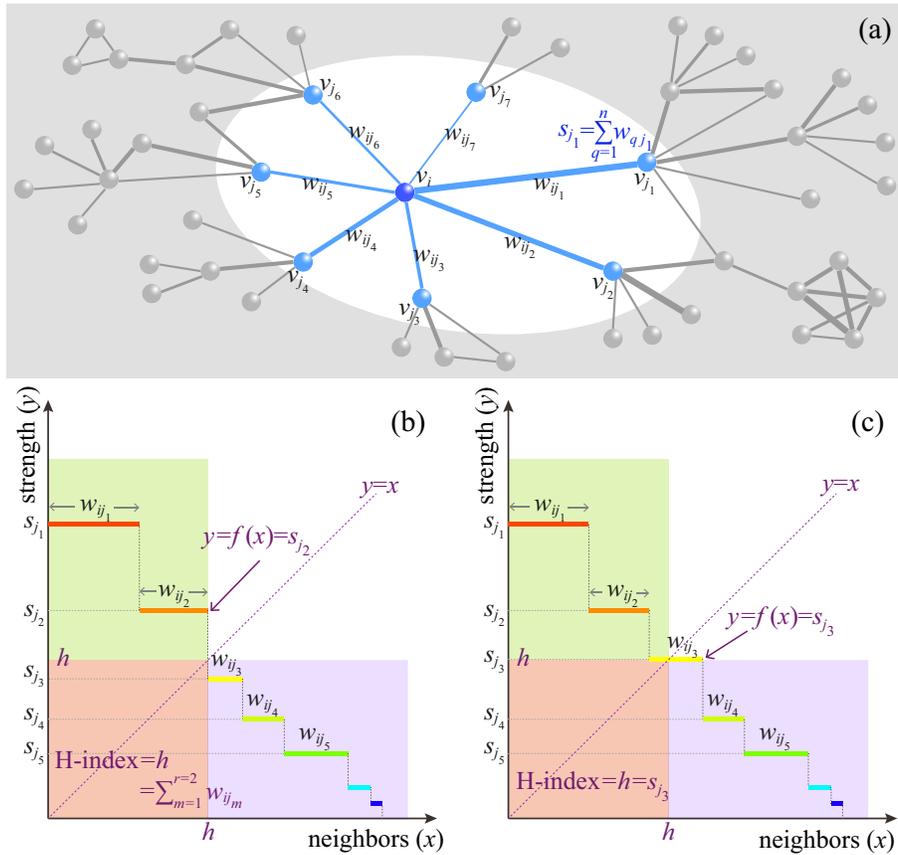}\\
		\caption{Illustration of how to calculate the weighted H-index. (a) A sample network where node $v_i$ is the target node. (b) Calculating the H-index of node $v_i$ when the curves $y=x$ and $y=f(x)$ intersect at the vertical parts of curve $y=f(x)$. (c) Calculating the H-index of node $v_i$ when the curves $y=x$ and $y=f(x)$ intersect at the horizontal parts of curve $y=f(x)$.}
		\label{Figure7}
		\end{figure}

        The extension of H-index to weighted networks is more complicated than the extension to directed networks. The weighted H-index of a node $v_i$ is calculated by the $\mathcal{H}$ function acting on the series of $v_i$'s neighbors' weighted degrees (here we take node strength as example) associated with the corresponding link weights, namely
        \begin{equation}\label{WH-index}
        h_{i}^W = \mathcal{H}[(w_{ij_1},s_{j_1}),(w_{ij_2},s_{j_2}),\cdots,(w_{ij_{k_i}},s_{j_{k_i}})],
        \end{equation}
        where $j_1,j_2,\cdots,j_{k_i}$ are the neighbours of node $v_i$, and $s_{j_1}\geq s_{j_2}\geq \cdots \geq s_{j_{k_i}}$. The $\mathcal{H}$ function would return the maximum real number $x$ which satisfies $f(x)\geq x$, where
        \begin{equation}
        f(x)=
        \begin{cases}
        s_{j_1}~~~~~~~~~\mathrm{if}~ 0<x\leq w_{ij_1};\\
        s_{j_r}~~~~~~~~~\mathrm{if}~ \sum_{m=1}^{r-1}w_{ij_m}<x\leq \sum_{m=1}^rw_{ij_m}~ \mathrm{for}~ r\geq2.
        \end{cases}
        \nonumber
        \end{equation}
        The maximum real number $x$ is our target, namely $h_{i}^W$. We borrow Fig.~\ref{Figure7} to explain how $\mathcal{H}$ works. Each horizontal line segment represents a node, and the width is proportional to the weight of the link between the node and $v_i$. In accordance with Eq. (\ref{WH-index}), the nodes with larger strength are ranked in higher order. The staircase curve is $y=f(x)$. Then the returned value $h_i^W$ equals the $x$-axis value of the intersection point of $y=f(x)$ and $y=x$. Mathematically, to obtain $h_i^W$, we need to find an integer $r^*$ ($1\leq r^* < k_i$) which meets the conditions of $\sum_{m=1}^{r^*}w_{ij_m}\leq s_{j_r^*}$ and $\sum_{m=1}^{r^*+1}w_{ij_m}\leq s_{j_{r^*+1}}$. Then there are two cases:
        (i) when $\sum_{m=1}^{r^*}w_{ij_m}\geq s_{j_{r^*+1}}$, namely the curves $y=x$ and $y=f(x)$ intersect at the vertical parts of curve $y=f(x)$, see Fig.~\ref{Figure7}(a), $h_i^W=\sum_{m=1}^{r^*}w_{ij_m}$;
        (ii) when $\sum_{m=1}^{r^*}w_{ij_m}< s_{j_{r^*+1}}$, namely the curves $y=x$ and $y=f(x)$ intersect at the horizontal parts of curve $y=f(x)$, see Fig.~\ref{Figure7}(b), $h_i^W=s_{j_{r^*+1}}$. In these two cases, $r^*$ is less than $k_i$, with which we cannot find $r^*$ if $s_{k_i}>s_i$. In such case, it is obvious that $h_i^W=s_{i}$. 

        We define the zero-order weighted H-index of node $v_i$ as $h_{i}^{W(0)}=s_{i}$ and the $n$-order weighted H-index is iteratively defined as~\cite{Lu2016NatCom710168}
        \begin{equation}\label{WH-index2}
        h_{i}^{W(n)} = \mathcal{H}[(w_{ij_1},h_{j_1}^{W(n-1)}),(w_{ij_2},h_{j_2}^{W(n-1)}),\cdots,(w_{ij_{k_i}},h_{j_{k_i}}^{W(n-1)})],
        \end{equation}
        where $j_r$ ($r=1,2,\cdots,k_i$) represents $v_i$'s neighbors, whose $(n-1)$-order weighted H-index is $h_{j_r}^{W(n-1)}$. In Eq. (\ref{WH-index2}), the $k_i$ nodes are ranked by $h_{j_1}^{W(n-1)} \geq h_{j_2}^{W(n-1)} \geq \cdots \geq h_{j_{k_i}}^{W(n-1)}$. The returned value $h_i^{W(n)}$ is the maximum real such that $f(h_i^{W(n)})\geq h_i^{W(n)}$, where $f(x)=h_{j_r}^{W(n-1)}$ if $\sum_{m=1}^{r-1}w_{ij_m}<x\leq \sum_{m=1}^rw_{ij_m}$. It can be proved that for every node $v_i\in V$ of a weighted undirected simple network $G(V, E)$, its weighted H-index sequence $h_i^{W(0)},h_i^{W(1)},h_i^{W(2)},\cdots$ will converge to the weighted coreness of node $v_i$ (i.e., in the case $k_i^W=s_i$).

        \subsubsection{Weighted closeness centrality}
        The critical point of the extension of closeness centrality to weighted network is the redefinition of shortest paths. Different from the distance of a link in unweighted network, the distance of a weighted link is related to its weight. For instance, it is faster to download files through high-bandwidth Ethernet connection than low-bandwidth connection. From the perspective of efficiency, high-bandwidth connections can shorten the distances among sites. Since the link weights in the most of the weighted networks are operationalizations of link strength and not the cost of them \cite{Opsahl2010SNet32245}, both Newman \cite{Newman2001PRE64016132} and Brandes \cite{Brandes2001JMS25163} proposed to adopt the reciprocal of weights to extend closeness centrality and betweenness centrality (more details are shown in the next part). The distance between two nodes, $v_i$ and $v_j$, is defined as
        \begin{equation}\label{EqWeightDist}
          d_{ij}^{w} = \mathrm{min}\bigg( \frac{1}{w_{ih_0}} + \frac{1}{w_{h_0h_1}} + \cdots + \frac{1}{w_{h_kj}}\bigg),
        \end{equation}
        where $v_{h_0}, v_{h_1}, \cdots, v_{h_k}$ are the intermediary nodes belonging to a path from $v_i$ to $v_j$, and the shortest path that minimizes the summation of inverse weights can be obtained by the Dijkstra's algorithm~\cite{Dijkstra1959NMath1269}. Then the weighted closeness centrality can be written as
        \begin{equation}\label{EqWeightCloseness}
          WCC(i) = \bigg[\sum_{j}^{n}d_{ij}^{w}\bigg]^{-1},
        \end{equation}

        Apparently, this definition ignores the impact of the number of intermediary nodes, i.e., $v_{h_0}, v_{h_1},\cdots, v_{h_k}$. Opsahl \cite{Opsahl2010SNet32245} claimed that this number is a significant feature and redefined the length of the shortest path, written as
        \begin{equation}\label{EqWeightDistMiddle}
          d_{ij}^{w\alpha} = \mathrm{min}\bigg( \frac{1}{(w_{ih_0})^\alpha} + \frac{1}{(w_{h_0h_1})^\alpha} + \cdots + \frac{1}{(w_{h_kj})^\alpha}\bigg),
        \end{equation}
        where $\alpha$ is a positive tunable parameter. When $\alpha=0$, it produces the same outcome as the distance in unweighed networks; whereas when $\alpha=1$, the outcome is the same as Eq. (\ref{EqWeightDist}). For $0 < \alpha < 1$, shorter paths (with fewer intermediary nodes) are favored and will be assigned with the shorter distance. Conversely, when $\alpha>1$, the impact of additional intermediary nodes is less important compared to the weights of the ties and thus longer paths are favored. Accordingly, the weighted closeness is defined as
        \begin{equation}\label{EqWeightCloseMid}
          WCC(i) = \bigg[\sum_{j}^{n}d_{ij}^{w\alpha}\bigg]^{-1}.
        \end{equation}

		\subsubsection{Weighted betweenness centrality}
        As a path-based centrality, the extension of betweenness centrality to weighted network also needs the new definitions of shortest paths. Along with the weighted closeness centrality, the weighted betweenness centrality~\cite{Brandes2001JMS25163} can be defined by employing Eq. (\ref{EqWeightDist}), as
        \begin{equation}\label{EqWeightBetweenness}
          WBC(i) = \sum_{i\neq s, i \neq t,s \neq t}\frac{g_{st}^{w}(i)} {g_{st}^{w}},
        \end{equation}
        where $g_{st}^{w}$ is the number of the shortest paths from $v_s$ to $v_t$ , and $g_{st}^{w}(i)$ is the number of the shortest paths from $v_s$ to $v_t$ that pass though node $v_i$.

        Considering the impact of intermediary nodes, the shortest path is defined as Eq. (\ref{EqWeightDistMiddle}), and the corresponding weighted betweenness centrality is given as \cite{Opsahl2010SNet32245}
        \begin{equation}\label{EqWeightBetweenness2}
          WBC(i) = \sum_{i\neq s, i \neq t,s \neq t}\frac{g_{st}^{w\alpha}(i)} {g_{st}^{w\alpha}},
        \end{equation}
        where $\alpha$ is a positive tunable parameter.

        \subsubsection{Weighted PageRank and LeaderRank}
        The extension of PageRank to weighted network is simple and clear. In each step, the PR value of a node will be distributed to its outgoing neighbors according to the link weights. That is to say, the random walk process is replaced by the weighted random walk~\cite{Ou2007PRE75021102}. Mathematically, we have
        \begin{equation}
        WPR_i^{(t)}=\sum_{j=1}^n w_{ji}\frac{WPR_j^{(t-1)}}{s_j^{out}},
        \end{equation}
        where $s_j^{out}$ is the out-strength of $v_j$. Similar extension can be applied to weighted LeaderRank algorithm. Firstly, a ground node is added to the weighted networks together with the bi-directed links between the ground node and the $n$ network nodes. Denote by $WLR_i^{(t)}$ the weighted LeaderRank score of node $v_i$ at time $t$. Initially, each network node is assigned with one unit score and the ground node with zero score. Then the weighted LeaderRank score is
        \begin{equation}\label{WLeaderRank}
        WLR_i^{(t)}=\sum_{j=1}^{n+1} w_{ji}^{\alpha}\frac{WLR_j^{(t-1)}}{b_j^{out}},
        \end{equation}
        where $\alpha$ is a tunable parameter and $b_j^{out}=\sum_i w_{ji}^{\alpha}$. When $\alpha=1$, $b_j^{out}=s_j^{out}$. When $\alpha=0$, it degenerates to the unweighted case. Note that, Eq. (\ref{WLeaderRank}) is different from Eq. (\ref{Eq_LeaderRankImproved}) that is indeed defined in an unweighted network with artificially weighting method to improve the ranking accuracy~\cite{Li2014PhysicaA40447}. After convergence, the score of the ground node will be equally distributed to the other $n$ nodes, and the final score of node $v_i$ equals the summation of its own score at steady state and the score from the ground node. Xuan \emph{et al.} \cite{Xuan2012ICSE25} applied the weighted LeaderRank algorithm to rank the contributions of software developers based on their communication network where the weight of link $i\rightarrow j$ presents the number of comments developer $v_j$ gives to $v_i$.

    \subsection{D-S evidence theory} \label{EvidenceTheory}
    The centrality of a node in weighted networks is highly related to the node's degree and strength. Intuitively, a node is considered to be more important if it has more neighbors or higher strength. These two factors are well synthesized to quantify the importance of nodes by Dempster-Shaper (D-S) evidence theory \cite{Wei2013PhysicaA3922564}, which is a general framework for reasoning with uncertainty \cite{Dempster1967AnnMathStat325,Shafer1976BOOK}. Compared with Bayesian theory of probability, D-S evidence theory needs weaker conditions (it dose not need to meet the additivity of probability); it has both the statuses of ``uncertainty'' and ``unknown''. In the case of identifying important nodes, D-S evidence theory would respectively estimate the probabilities that node $v_i$ is important and unimportant, and also allows the degree of awareness to this question is unknown \cite{Wei2013PhysicaA3922564}. Since the measurement of node importance only needs some basic theories from the complete D-S evidence theory, we directly present the calculation process.

    The importance of a node is assumed to be highly related to its degree and strength \cite{Wei2013PhysicaA3922564}. The influence of the two factors can be simply presented by two evaluation indices, namely \emph{high} and \emph{low}, which form a frame of discernment $\theta=(\emph{high},\emph{low})$. It is easy to get $k_M = \mathrm{max}\{k_1,k_2,...\}$, $k_m = \mathrm{min}\{k_1,k_2,...\}$, $s_M = \mathrm{max}\{s_1,s_2,...\}$ and $s_m = \mathrm{min}\{s_1,s_2,...\}$. Then the basic probability assignment (BPA) for the degree and strength of nodes can be respectively created. Therein, $m_{di}(h)$ represents the probability that $v_i$ is important when the influence of degree is considered, while $m_{di}(l)$ represent the probability that $v_i$ is unimportant. Considering node's strength, $m_{si}(h)$ represents the probability that $v_i$ is important while $m_{si}(l)$ represent the probability that $v_i$ is unimportant. They are given as follows:
	\begin{eqnarray}
	  m_{di}(h) = \frac{|k_i-k_m|}{k_M-k_m+2\mu}  , \quad  m_{di}(l) = \frac{|k_i-k_M|}{k_M-k_m+2\mu},  \nonumber \\
	  m_{si}(h) = \frac{|s_i-s_m|}{s_M-s_m+2\varepsilon} ,\quad  m_{si}(l) = \frac{|s_i-s_M|}{s_M-s_m+2\varepsilon}, \nonumber
	\end{eqnarray}
	where $0<\mu,\varepsilon<1$ express a kind of uncertainty of nodes' orders. Their values have no effect on nodes' orders \cite{Wei2013PhysicaA3922564}. Then the BPAs of $v_i$ with respect to the degree and strength are obtained,
	\begin{eqnarray}
	  M_d(i) &=& (m_{di}(h), m_{di}(l), m_{di}(\theta)) \nonumber, \\
	  M_s(i) &=& (m_{si}(h), m_{si}(l), m_{si}(\theta)) \nonumber,
	\end{eqnarray}
	where $m_{di}(\theta)=1-(m_{di}(h)+m_{di}(l))$, and $m_{si}(\theta)=1-(m_{si}(h)+m_{si}(l))$. The values of $m_{di}(\theta)$ and $m_{si}(\theta)$ indicate the extent that D-S evidence theory does not know whether $v_i$ is important or not. By introducing the Dempster's rule of combination, the influence value of $v_i$ can be written as
	\begin{equation}\label{EqDSEvidence1}
	  M(i)=( m_i(h), m_i(l), m_i(\theta) ),
	\end{equation}
	where
	\begin{eqnarray}
	  m_i(h)      &=& [ m_{di}(h)\cdot m_{si}(h) + m_{di}(h)\cdot m_{si}(\theta) + m_{si}(h)\cdot m_{di}(\theta)]/(1-K) \nonumber ,\\
	  m_i(l)      &=& [ m_{di}(l)\cdot m_{si}(l) + m_{di}(l)\cdot m_{si}(\theta) + m_{si}(l)\cdot m_{di}(\theta)]/(1-K) \nonumber ,\\
	  m_i(\theta) &=& [ m_{di}(\theta)\cdot m_{si}(\theta) ]/(1-K) \nonumber ,\\
	  K           &=& m_{di}(h)\cdot m_{si}(l) + m_{di}(l)\cdot m_{si}(h). \nonumber
	\end{eqnarray}
	In general, the value of $m_i(\theta)$ is distributed to $m_i(\theta)$ and $m_i(\theta)$ on average. Thus
	\begin{eqnarray}
	  M_i(h) &=& m_i(h)+\frac{1}{2m_i(\theta)} ,\nonumber \\
	  M_i(l) &=& m_i(l)+\frac{1}{2m_i(\theta)} , \nonumber
	\end{eqnarray}
	where $M_i(h)$ and $M_i(l)$ are the probabilities that $v_i$ is important and unimportant, respectively. Apparently, the higher the $M_i(h)$ is and/or the lower the $M_i(l)$ is, the more vital the $v_i$ is. Thus, the evidential centrality based on evidence theory can be defined as
	\begin{equation}\label{EqEvidenceCen}
	  \mathrm{EVC}(i)= M_i(h)-M_i(l)=m_i(h)-m_i(l).
	\end{equation}
    There is an implicit hypothesis that the degree of nodes follows uniform distribution, which restricts its effectiveness. Moreover, the EVC centrality was claimed to have ignored the global structure information of the network \cite{Gao2013PhysicaA3925490}. To tackle these problems, an improved metric, named Evidential Semi-local centrality, was proposed by combing the extension of semi-local centrality and a modified evidential centrality considering the impact of degree distribution \cite{Gao2013PhysicaA3925490}. Moreover, the topological connections among the neighbors of a node, namely local structure information, were also introduced to enhance the effectiveness of evidential centrality \cite{Ren2015IJICIC111765}.

\section{On bipartite networks}\label{Chapter8}
Different from monopartite networks, a bipartite network is constituted by two groups of nodes. Only the nodes in different groups are allowed to be connected. Given a network $G(V,E)$, where $V$ and $E$ are the sets of nodes and edges, respectively. If $V$ can be divided into two subsets $X$ and $Y$, which satisfy $X\bigcap Y=\emptyset$, and there is no edge connecting two nodes in the same subset, we call $G(V,E)$ a bipartite network, denoted by $B(X,Y,E)$. Many common networks are bipartite networks although they may not be presented by bipartite graphs where the two groups of nodes are clearly distinguished, such as tree and tetragonal lattice. Besides, many real-world networks are bipartite. For example, the heterosexual relations can be described by a bipartite networks with males as one group and females as another group \cite{Liljeros2001Nature411907}. A metabolic network is a bipartite network with chemical substance and chemical reaction as two disjoint subsets \cite{Jeong2000Nature407651}. The collaboration network is a bipartite network with participants and events as two disjoint subsets, such as the collaboration between scientists or movie actors \cite{Zhang2006PhysicaA360599}. The Internet telephone network is a bipartite network where the edge connects user ID and phone number \cite{Xuan2009Chaos19577}. Another representative example is the online e-commercial network which is presented by user-object bipartite network \cite{Shang2009EPL901303}.

\begin{figure}
  \centering
  \includegraphics[width=8cm]{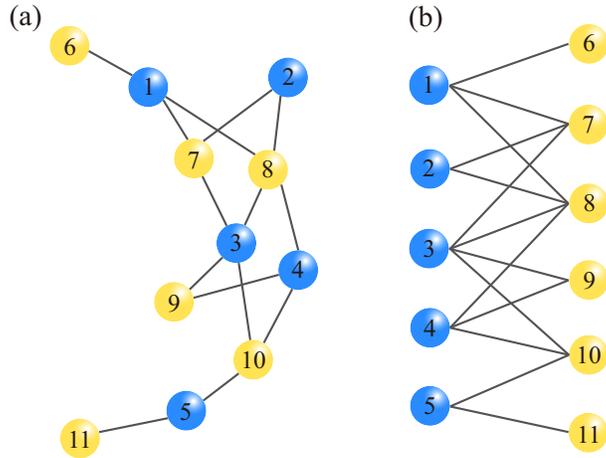}\\
  \caption{An example of a bipartite network. (a) The original network. (b) The corresponding bipartite network of (a).}
\label{Figure8}
\end{figure}

Bipartite networks have many special features: (i) they have no circuit of odd length; (ii) they can be binary coloured; (iii) they have symmetric network spectrum. Based on these features, we can judge whether an undirected simple network is bipartite with linear time complexity by breadth-first search or other methods. Firstly, we randomly select a node $v_i$ and label it with 0. Then we label $v_i$'s neighbors with 1, and further label the neighbors of $v_i$'s neighbors with 0. If there is a node which can be labeled with either 1 or 0, the network is not bipartite and the process stops. Otherwise, we continue the process. If all the nodes can be binary coloured without conflict, the network is bipartite. We can also use the depth-first search to make a judgment. Firstly, we obtain a binary coloured spanning tree by depth-first search. Then we check the edges that are not on the tree. If there is an edge between two nodes with the same colour (i.e., ``frustrated" edge), there must be an odd loop containing this edge, and thus the network is not bipartite. However, for an arbitrary monopartite network, a measure called ``bipartivity" can be used to quantify how far away from being bipartite the network is \cite{Holme2003PRE68056107,Ernesto2005PRE72046105}. Empirical results on real-world networks show that the data from the Internet community, such as email network, usually have high bipartivity, while the collaboration networks have relative lower values \cite{Holme2003PRE68056107}. Fig.~\ref{Figure8} shows an illustration of a bipartite network.

	\subsection{Reputation systems}
    In many online communities, users are freely to comment on the involved items (e.g., movie, music, book, news or scientific papers). The collection of comments is very valuable for new users. For example, in the mainstream e-commercial and entertainment websites, such as Taobao, Ebay, Amazon and Netflix, users are allowed to rate the items which they have purchased, read or watched. Some examples are shown in Table~\ref{Table1}. The average ratings is the most straightforward method to evaluate items. However, many empirical cases demonstrate that users may be misled by comments given by unreliable users. The bias can be either unconscious or intentional. Sometimes, users may be not serious about voting or not experienced in the corresponding field, and thus produce biased ratings. In addition, the intentional manipulations exist widely in these online rating systems. For example, sellers on e-commercial websits may hire a group of professional spammers to rate their items with high scores and give their competitors' items bad comments. Thus, it is very necessary to build up a reputation systems~\cite{Masum2004FM91158} to give creditable evaluations on both the user’s reputation and the item's quality aiming at reducing the damage suffered from information asymmetry between the involved parties. This is a specific problem of vital nodes (i.e., high-reputation users and high-quality items) identification in bipartite networks.

	\begin{table}
	\centering
	\caption{Some popular online rating/voting systems.}
	\label{Table1}
	\setlength\tabcolsep{16pt}
	\begin{tabular}{ccc}
	\\
	\hline
	Site & item & rating \\
	\hline
	\centering
	Taobao.com     &Products & Positive/Neutral/Negative \\
	Tmall.com      &Products & From 1 star (worst) to 5 star (best) \\
	Ebay.com       &Products & Positive/Neutral/Negative \\
	Amazon.com     &Products & From 1 star (worst) to 5 star (best) \\
	MovieLens.org  &Movies   & From 1 star (worst) to 5 star (best)\\
	Nanocrowd.com  &Movies   & From 1 star (worst) to 5 star (best)  \\
	Digg           &News     & Dig/Save/Share \\
	Zite           &News     & Yes (like) /No (dislike) \\
    Pandora.com    &Music    &  You like this\\
	Netflix.com    &DVDs, TV Shows            & From 1 star (worst) to 5 star (best) \\
	Douban.com     &Books, Movies and Music   &  From 1 star (worst) to 5 star (best)\\
	StumbleUpon    &Websites &I like it/No more like this\\
	\hline
	\end{tabular}
	\end{table}

\begin{figure}
	\centering
	\includegraphics[width=8cm]{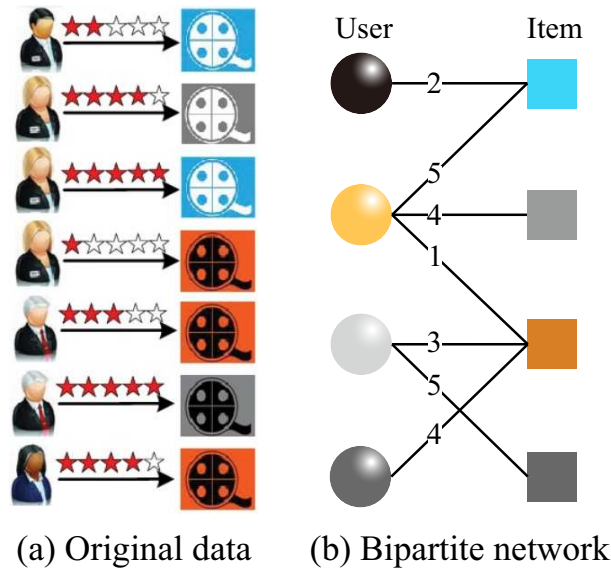}
	\caption{An examples of reputation system represented by a bipartite network. (a) is a real scene in which users see movies
	and vote them in five discrete ratings 1-5. (b) is the corresponding bipartite network,
	in which users and items are represented by circles and squares, and link weights denote the ratings.}
	\label{Figure9}
	\end{figure}

    In general, an online community is assumed to consist of $n$ users and $m$ items which can be naturally described by a bipartite network. Denoted by $G(U,I,W)$, where $U$ and $I$ are the sets of users (labeled by Latin letters, $i = 1,2,\cdots, n$) and items (labeled by Greek letters, $\alpha = 1,2,\cdots, m$), respectively. $W$ presents the set of interactions between users and items. A weighted link between user $i$ and item $\alpha$ exists if user $i$ has interacted with item $\alpha$. The weight of a link $w_{i\alpha}$ is determined by the type of interaction between the corresponding user-item pair and reflects the intensity of the interaction. $W$ can be directed or undirected. For example, in online rating systems, if user $i$ rates item $\alpha$ with score $r_{i\alpha}$, then $w_{i\alpha}=r_{i\alpha}$, otherwise $w_{i\alpha}=0$. While sometimes, the seller of the item can also rate the user, in this case we can use a directed network where $w_{i\alpha}\neq w_{\alpha i}$. Obviously, for an unweighted user-item network $G(U,I,E)$, we set $e_{i\alpha}=w_{i\alpha}=1$ if user $i$ purchases, reads, or watches item $\alpha$, otherwise $e_{i\alpha}=w_{i\alpha}=0$. We can also simplify the weighted network to its unweigthed version by setting $e_{i\alpha}=1$ if $w_{i\alpha}>w_0$. Here $w_0\in[0,w_{max})$ is the selection threshold. The larger $w_0$ is the sparser the network is. The corresponding unweighted user degree and item degree are denoted by $k_i$ and $k_{\alpha}$, respectively. Fig.~\ref{Figure9} gives an example of reputation system represented by a bipartite network.

	\subsection{Statistical methods}
	In reputation systems, the most straightforward method to quantify an item's quality $Q$ is using the average ratings (abbreviated as AR), $\bar{Q}_{\alpha}=\frac{1}{k_\alpha}\sum\limits_{i}r_{i\alpha}$, where the ratings from different users contribute equally. However, the ratings of prestigious users should be more reliable than the ratings from faithless users. Therefore, the calculation of an item's quality can be modified by
	\begin{equation}
	Q_{\alpha}=\sum\limits_{i}R_{i}r_{i\alpha},\label{WQ}
	\end{equation}
	where $R_i$ is the normalized reputation score of user $i$, which can be exogenous parameter or determined by his previous ratings on items.

	Gao \emph{et al.} proposed a group-based ranking (GR) method \cite{Gao2015EPL11028003} to quantify user's reputations. They firstly grouped users based on the patterns of their ratings, and then counted the group sizes. The basic assumption is that users who always fall into large groups are more likely to have high reputations. The GR method has five steps: (i) list the existent scores in the system, namely $\{\omega_1,\omega_2,\cdots,\omega_{n_s}\}$ where $n_s$ is the number of distinct scores; (ii) construct a score-item matrix $\Lambda$, where $\Lambda_{s\alpha}$ is the number of users who rate item $\alpha$ with score $\omega_s$; (iii) build a rating-rewarding matrix $\Lambda^*$ by $\Lambda^*_{s\alpha}=\Lambda_{s\alpha}/k_{\alpha}$; (iv) map the original rating matrix to a rewarding matrix $A′$, where $A'_{i\alpha}=\Lambda^*_{s\alpha}$ with the constrain that $r_{i\alpha}=\omega_s$. Notice that, if user $i$ does not rate item $\alpha$, the value of $A'_{i\alpha}$ is null and it will be ignored in the following calculation; (v) calculate user $i$'s reputation $R_i$ through the ration of mean value of $A'_i$ to its standard deviation, namely
	\begin{equation}
	R_i=\frac{\mu(A'_i)}{\sigma(A'_i)},
	\end{equation}
	where
	\begin{equation}
	\mu(A'_i)=\sum_{\alpha}\frac{A'_{i\alpha}}{k_i},\quad \sigma(A'_i)=\sqrt{\frac{\sum_{\alpha}(A'_{i\alpha}-\mu(A'_i))^2}{k_i}}.
	\end{equation}
    The group-based ranking method can be extended by introducing an iterative refinement process, with details presented in Ref.~\cite{Gao2015ArXiv150900594}. It has a similar frameworks as that in Chapter~\ref{Chapter8-3}.

	\subsection{Iterative methods}\label{Chapter8-3}
    Besides the statistical method based on the pattern of ratings, the user reputation and item quality can be calculated in an iterative way. In specific, items' qualities at time $t$, denoted by $Q^{(t)}$, are calculated based on users' reputations at time $t-1$, denoted by $R^{(t-1)}$; while users' reputations at time $t$ are calculated based on items' qualities at time $t-1$. The iteration starts by setting the initial value of either $Q^{(0)}$ or $R^{(0)}$ and stops when both $Q$ and $R$ converge. Following this idea, a number of methods were proposed in the literatures.

    Laureti \emph{et al.} \cite{Laureti2006EPL751006} proposed an iterative refinement (IR) method which considers users' reputation scores as inversely proportional to the mean squared error between users' rating records and the qualities of items, namely
	\begin{equation}
	IR_{i}=\frac{k_i}{\sum\limits_{\alpha\in I_i}(Q_\alpha-r_{i\alpha})^{2}},\label{IR}
	\end{equation}
    where $I_i$ is the set of items selected by user $i$. Combining this equation with Eq.~(\ref{WQ}), we can start an iterative process with $IR_i(0)=1/|I|$ to calculate $Q$ and $IR$. Note that, in each iteration $IR_i$ should be normalized.

    Zhou \emph{et al.} \cite{Zhou2011EPL9448002} proposed a correlation-based iterative method (abbreviated as CR), which assumes that a user's reputation can be reflected by the relation between his/her ratings and the corresponding items' qualities. Specifically, Pearson correlation was adopted
	\begin{equation}
	\mathrm{corr}_i=\frac{k_i\sum\limits_{\alpha\in{I_i}}r_{i\alpha}Q_{\alpha}-
	\sum\limits_{\alpha\in{I_i}}r_{i\alpha}\sum\limits_{\alpha\in{I_i}}Q_{\alpha}}{\sqrt{k_i\sum\limits_{\alpha\in{I_i}} r^2_{i\alpha}-(\sum\limits_{\alpha\in{I_i}}r_{i\alpha})^2}\sqrt{ k_i\sum\limits_{\alpha\in{I_i}}Q^2_{\alpha}-(\sum\limits_{\alpha\in{I_i}}Q_{\alpha})^2}}.\label{CR}
	\end{equation}
	If $\mathrm{corr}_i\geq0$, the user's reputation $CR_i=\mathrm{corr}_i$, otherwise $CR_i=0$. Here, we can also recall Eq.~(\ref{WQ}) to build an iterative process to calculate $Q$ and $CR$ by replacing $R_i$ with $CR_i$, and with an initial value $CR_i(0)=k_i/|I|$.

    More recently, Liao \textit{et al.} \cite{Liao2014PLoSONE9e97146} proposed the iterative algorithm with reputation redistribution (IARR) algorithm to improve the validity by enhancing the influence of prestigious users. Then in the iteration process, the user's reputation is updated according to
	\begin{equation}
	IARR_{i}=CR^\theta_i~\frac{\sum\limits_{j\in U}CR_j}{\sum\limits_{j\in U}CR^\theta_j},
	\label{IARR}
	\end{equation}
    where $\theta$ is a tunable parameter to control the influence of reputation. Obviously, when $\theta=0$, IARR is the same as the AR method. When $\theta=1$, IARR degenerates to the CR method. To further improve the reliability of the method, the authors proposed an advanced method called IARR2 by introducing a penalty factor to Eq.~(\ref{WQ}), namely
	\begin{equation}
	Q_{\alpha}=\max_{i\in U_{\alpha}}\{IARR_{i}\}\sum\limits_{i\in U_{\alpha}}IARR_{i}\cdot r_{i\alpha},
	\label{IARR2Q}
	\end{equation}
	and then $CR_{i}$ in Eq.~(\ref{IARR}) was modified as
	\begin{equation}
	CR_i=
        \begin{cases}
        \frac{\log(k_{i})}{\max\{\log(k_{i})\}}\cdot{\mathrm{corr}_i}    ~~~~~~~~~~ \mathrm{if}~ \mathrm{corr}_i>0, \\
        0 ~~~~~~~~~~~~~~~~~~~~~~~~~~~~~~ \mathrm{else}.
        \end{cases}
	\label{IARR2R}
	\end{equation}
    IARR2 emphasizes that the items rated by users with low reputations are generally of low qualities, and the users having rated only a small number of items cannot have high reputations \cite{Liao2014PLoSONE9e97146}.

	\subsection{BiHITS and variants}
    We have introduced the HITs algorithm for unipartite networks, here we introduce the HITs algorithm for bipartite networks, called ``biHITS" \cite{Liao2014PLoSONE9e112022}. Denote $R_i$ and $F_\alpha$ the reputation of user $i$ and the fitness of item $\alpha$, respectively. Fitness measures how good an item is, such as the quality of a product or the influence of a scientific paper. Consider a directed bipartite network, biHITS can be written as
	\begin{equation}
	\begin{aligned}
	&R_i=\sum_{\alpha}w_{\alpha i}F_{\alpha}, \quad \mathrm{i.e.},\ \mathbf{R}=W_{I\rightarrow U}\mathbf{F},\\
	&F_{\alpha}=\sum_iw_{i\alpha}R_i, \quad \mathrm{i.e.},\ \mathbf{F}=W_{U\rightarrow I}\mathbf{R},
	\end{aligned}
	\end{equation}
    where $W$ is the link weight matrix (i.e., the weighted adjacency matrix) of the bipartite network, and $\mathbf{R}$ and $\mathbf{F}$ are user reputation vector and item fitness vector, respectively. For undirected network, $W_{U\rightarrow I}=W_{U\rightarrow I}^T$. The solution can be found in an iterative way via the following set of equations
	\begin{equation}
	\mathbf{R}^{(t+1)}=W_{I\rightarrow U}\mathbf{F}^{(t)}, \mathbf{F}^{(t+1)}=W_{U\rightarrow I}\mathbf{R}^{(t)}. \label{biHITS}
	\end{equation}
    Initially, one can set $R_i^{(0)}=1/\sqrt{n}$ and $F_{\alpha}^{(0)}=1/\sqrt{m}$. Note that, if the network is connected, the solution is unique and independent of the selection of initial values of $R^{(0)}_i$ and $F^{(0)}_{\alpha}$ \cite{Kleinberg1999JACM46604}. In each iteration, both $R_i$ and $F_{\alpha}$ should be normalized so that the values of $\|R\|_2$ and $\|F\|_2$ are always one. The iterations will stop when the summation of absolute changes of all vector elements in $\mathbf{R}$ and $\mathbf{F}$ is less than a tiny threshold $\varepsilon$. For unweighted bipartite networks, one can replace the link weight matrix $W$ with the routine adjacency matrix.

		\subsubsection{Algorithms with content information}
        Deng \emph{et al.} proposed a generalized Co-HITS algorithm to incorporate the bipartite network with the content information from both user side and item side \cite{Deng2009ACMSIGKDD239}. They investigated two frameworks, namely iterative framework and regularization framework, from different views. The ordinary biHITS is one of the special cases under some certain parameters. Here we give brief introduction of iterative framework where the basic idea is to propagate the scores on the bipartite network via an iterative process with the constraints from both sides. To incorporate the bipartite network with the content information, the generalized Co-HITS equations can be written as
		\begin{equation}
		\begin{aligned}\label{CoHITS}
		&R_i=(1-\lambda_U)R_i^{(0)}+\lambda_U \sum_\alpha w_{\alpha i}F_{\alpha}, \\
		&F_{\alpha}=(1-\lambda_I)F_{\alpha}^{(0)}+\lambda_I \sum_i w_{i\alpha}R_i,
		\end{aligned}
		\end{equation}
        where $\lambda_U\in[0,1]$ and $\lambda_I\in[0,1]$ are the personalized parameters. The initial scores $R_i^{(0)}$ and $F_{\alpha}^{(0)}$ are calculated using a text relevance function $f$, such as the vector space model \cite{Baeza1999BOOK} and the statistical language model \cite{Ponte1998SIGIR275,Zhai2002SIGIE49}. For a given query $q$, $R^{(0)}_i=f(q,i)$ and $F_{\alpha}^{(0)}=f(q,\alpha)$. Obviously, when $\lambda_U=\lambda_I=1$, the Eq.~(\ref{CoHITS}) degenerates to the ordinary biHITS algorithm.

		\subsubsection{Algorithms with user trustiness}
        A variant of biHITS, called QTR (Quality-Trust-Reputation) was proposed by considering the users' trustiness information extracted from users' social relationships \cite{Liao2012BOOK}. In the network of users, if user $i$ trusts user $j$ or is a friend of user $j$, there will be a link from $i$ to $j$. The link weight, denoted by $T_{ij}$, represents to which extent user $i$ trusts user $j$. Consider an undirected network, the QTR method is defined as
		\begin{equation}
		\begin{aligned}
		&F_{\alpha}=\frac{1}{k_{\alpha}^{\theta_I}}\sum_i w_{i\alpha}[R_i-\rho_U\bar{R}],\\
		&R_i=\frac{1}{k_i^{\theta_U}}\sum_{\alpha} w_{i\alpha}[F_{\alpha}-\rho_I\bar{F}]+\frac{1}{f_{i}^{\theta_T}}\sum_j [R_j-\rho_U\bar{R}][T_{ji}-\rho_T\bar{T}],\\
		\end{aligned}
		\end{equation}
        where $f_i$ is the number of users who trust user $i$, $\bar{F}=\sum_{\alpha}F_{\alpha}/m$, $\bar{R}=\sum_{i}R_{i}/n$ and $\bar{T}=\sum_{ij}T_{ij}/[n(n-1)]$ are the average values of item fitness, user reputation and trust in the community, respectively. The six tunable parameters, $\theta_U$, $\theta_I$, $\theta_T$, $\rho_U$, $\rho_I$ and $\rho_T$ all belong to the range $[0,1]$. If all of these parameters are zero, QTR reduces to the ordinary biHITS. In particular, the two boundary choices of $\theta_I$ correspond to item fitness obtained by summing (when $\theta_I=0$) or averaging (when $\theta_I=1$) over reputation of neighboring users; the meaning of $\theta_U$ and $\theta_T$ are analogous. By contrast, $\rho_I$ decides whether interactions with items of low fitnesses harm user reputation (when $\rho_I>0$) or not (when $\rho_I=0$); the meaning of $\rho_U$ and $\rho_T$ are analogous. The solution can be obtained in an iterative way via setting $R_i^{(0)}=1/\sqrt{n}$ for all users and $F_{\alpha}^{(0)}=1/\sqrt{m}$ for all items. At step $t+1$, user reputation and item fitness are updated according to
		\begin{equation}
		\begin{aligned}
		&F_{\alpha}^{(t+1)}=\frac{1}{k_{\alpha}^{\theta_I}}\sum_i w_{i\alpha}[R_i^{(t)}-\rho_U\bar{R}^{(t)}],\\
		&R_i^{(t+1)}=\frac{1}{k_i^{\theta_U}}\sum_{\alpha} w_{i\alpha}[F_{\alpha}^{(t)}-\rho_I\bar{F}^{(t)}]+\frac{1}{f_{i}^{\theta_T}}\sum_j [R_j^{(t)}-\rho_U\bar{R}^{(t)}][T_{ji}-\rho_T\bar{T}].\\
		\end{aligned}
		\end{equation}
        To avoid divergence, in each step, both $R_i$ and $F_{\alpha}$ should be normalized so that the values of $\|R\|_2$ and $\|F\|_2$ are always one. The iterative process stops when the algorithm converges to a stationary state. Experimental results show that social relationships play a valuable role in improving the quality of the ranking \cite{Liao2012BOOK}.

		\subsubsection{Algorithms with the credit of item provider}
        In many rating systems, besides the information of users and items, the credits of item providers are also considered. For example, on Taobao, the largest online shopping website operated by Alibaba group, each seller has a credit score which is calculated based on the ratings that the seller has obtained. The credit is a kind of very important information. When users want to buy a new item rated by very few people, the seller's credit plays a critical role that largely impact users' decision. Generally, an item is expected to be of high quality (i.e., fitness) if it was sold by high-credit sellers; meanwhile, a high-quality item can raise the credit of its seller. This statement is also suitable to describe the relation between authors and their papers. Specifically, a paper is expected to be of high quality if it was authorized or cited by prestigious (i.e., high-credit) scientists; meanwhile, a high-quality paper can raise the prestige of its authors, see, for example, the iterative algorithm on the author-paper bipartite network proposed by Zhou \emph{et al.} \cite{Zhou2012NJP14033033}.

\begin{figure}
		  \centering
		  \includegraphics[width=7cm]{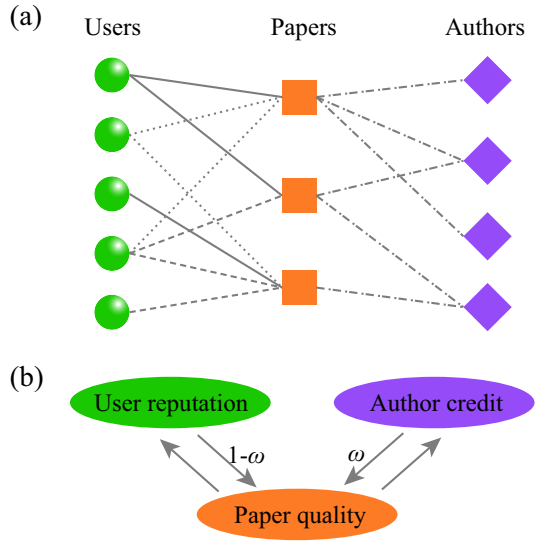}\\
		  \caption{Schematic illustration of the data and the algorithm. (a) The input data can be represented by a
		tripartite network. Different line styles indicate different interactions: paper submission, download, and
		abstract view between users and papers, and authorship between papers and authors. (b) Score flows in the
		QRC algorithm \cite{Liao2014PLoSONE9e112022}.}
		\label{Figure10}
		\end{figure}

        Based on this idea, Fujimura \emph{et al.} \cite{Fujimura2005AOA,Fujimura2005BOOK} proposed a new algorithm called `EigenRumor' that quantifies each blog entry by weighting the hub and authority scores of the bloggers based on eigenvector calculations. Under this framework, there are two bipartite networks, namely the user-item network and provider-item network. If we combine these two bipartite networks, we have a user-item-provider tripartite network. The degree of item and provider in the item-provider network are $d_{\alpha}$ and $d_m$, respectively. Denoting $\mathbf{A}$ the vector of provider credit values, the EigenRumor algorithm reads
		\begin{equation}
		\begin{aligned}\label{EigenRumor}
		\mathbf{R}=W_{I\rightarrow U}\mathbf{F}, \quad \mathbf{A}=P\mathbf{F}, \quad \mathbf{F}=\omega P^T\mathbf{A}+(1-\omega)W_{U\rightarrow I}\mathbf{R},
		\end{aligned}
		\end{equation}
       where $\omega$ is a tunable parameter in the range $[0,1]$, which determines the relative contribution of providers' credits and users' ratings to item fitness. The two matrices $W$ and $P$ can be normalized to reduce the bias towards active users and providers. Liao \emph{et al.}~\cite{Liao2013arXiv1311} applied this method to user-paper-author tripartite network and proposed a variant of EigenRumor called QRC (Quality-Reputation-Credit) by considering the negative effect of the entities with low scores, namely,
		\begin{equation}
		\begin{aligned}\label{LiaoER}
		&R_i=\frac{1}{k_i^{\theta_U}}\sum_{\alpha} w_{i\alpha}[F_{\alpha}-\rho_I\bar{F}],\\
		&A_m=\frac{1}{d_m^{\phi_O}}\sum_{\alpha} p_{m\alpha}[F_{\alpha}-\rho_O\bar{A}],\\
		&F_{\alpha}=\frac{1-\omega}{k_{\alpha}^{\theta_I}}\sum_i w_{i\alpha}[R_i-\rho_U\bar{R}]+\frac{\omega}{d_{\alpha}^{\phi_I}}\sum_m p_{m\alpha}A_m.\\
		\end{aligned}
		\end{equation}
        Note that $k_{\alpha}$ and $d_{\alpha}$ are the item's degree in user-item network and provider-item network, respectively. The meaning of $\phi_I$ and $\phi_O$ are analogous to $\theta_U$, and the meaning of $\rho_O$ is analogous to $\rho_U$. Fig.~\ref{Figure10} shows the schematic illustration of the data and the algorithm.

\section{Performance evaluation}\label{Chapter9}

In this section, we evaluate the performance of some representative algorithms on real networks. Both functional and structural importance are considered.

    \subsection{On undirected networks}\label{secPEUndirected}

    We compare the performance of eight representative algorithms on four undirected unweighted networks: (i) \emph{Amazon} \cite{Yang2012SIGKDD3} is a co-purchase network between products on Amazon website. If a product $v_i$ is frequently co-purchased with a product $v_j$, then an undirected edge exists between $v_i$ and $v_j$. (ii) \emph{Cond-mat} \cite{Leskovec2007ACM12} is a collaboration network of scientists who have posted preprints on the condensed matter archive at \emph{www.arxiv.org} from Jan. 1st, 1995 to Jun. 30th, 2003. In this network, a node represents an author and two nodes are connected if they have co-authorized at least one paper. Clearly, each preprint will form a clique where its authors are completely connected. (iii) \emph{Email-Enron}  \cite{Leskovec2009IntMath629} is a communication network that contains about half million emails. Each node is a unique email address. If an email was sent from address $v_i$ to address $v_j$, then an undirected edge exits between $v_i$ and $v_j$. (iv) \emph{Facebook}~\cite{Viswanath2009ACM37} is a friendship network extracted from \emph{facebook.com} where node represents a user and an edge $(v_i,v_j)$ indicates that users $v_i$ and $v_j$ are friends. The basic statistical features are shown in Table \ref{Table2}.
        \begin{table}
          \centering
          \caption{Basic statistical features of the four undirected networks, including the number of nodes $n$ and edges $m$, the maximum degree $k_{\max}$, the clustering coefficient $c$ and the degree heterogeneity $H$ (defined as $\frac{\langle k^2\rangle}{\langle k\rangle^2}$ \cite{Hu2008PhysicaA3873769}).}\label{Table2}
          \begin{tabular}{l llll l}
             \hline
             Networks & $n$ & $m$ & $k_{\max}$ & $c$ & $H$ \\ \hline
             Amazon  & 334863 & 925872 & 549 &0.3967 &2.0856 \\
             Cond-mat  & 27519 & 116181 & 202 &0.6546 &2.6393 \\
             Email-Enron  & 36692 & 183831 & 1383 &0.4970 &13.9796 \\
             Facebook & 63731 & 817090 & 1098 &0.2210 &3.4331 \\
             \hline
           \end{tabular}
        \end{table}

        To evaluate the performance of the ranking methods, we investigate the Kendall's tau correlation coefficient $\tau$ \cite{Kendall1938Biometrika3081} between the ranking scores of different methods and the nodes' influences obtained by the simulations of spreading processes on the networks. Higher $\tau$ indicates better performance. We consider the susceptible-infected-recovered (SIR) spreading dynamic~\cite{Anderson1992BOOK}. In the SIR model, all nodes except the infected node (i.e., the initial seed node) are susceptible initially. At each time step, each infected node will infect each of its neighbors with probability $\beta$. Then, each infected node enters the recovered state with a probability $\mu$. For simplicity, we set $\mu=1$. The spreading process ends when there is no longer any infected node. Then the spreading influence of the initial seed is defined as the number of recovered nodes \cite{Kitsak2010NatPhys6888}.

        Table \ref{Table3} shows the Kendall’s tau correlation coefficient $\tau$ between the ranking scores given by the algorithms and the real spreading influences obtained by the SIR spreading model. For each network, the infected probability is set as $\beta=1.5\beta_c$ where $\beta_c$ is the approximate epidemic threshold~\cite{Castellano2010PRL105218701}:
        \begin{equation}
          \beta_c = \langle k\rangle/(\langle k^2\rangle-\langle k\rangle).
        \end{equation}
        From the results, one can see that LocalRank \cite{Chen2012PhysicaA1777} and eigenvector centrality~\cite{Bonacich2001SNet23191} perform overall better than others. The former is the best among five local centralities, while the latter is the best among four global centralities. Especially, LocalRank performs even better than some global methods in some networks. Betweenness \cite{Freeman1977Sociometry4035, Freeman1979SNet215} is not so good, because the nodes with high betweenness values are usually the bridges connecting two communities and may not be of high spreading influences. Compared with degree, coreness and H-index perform better in all these four networks, which is in accordance with the main results in Refs. \cite{Kitsak2010NatPhys6888} and \cite{Lu2016NatCom710168}.

        \begin{table}
          \centering
          \caption{Kendall’s tau correlation coefficient $\tau$ between the ranking scores given by the algorithms and the real spreading influences obtained by the SIR spreading model. The approximate epidemic threshold $\beta_c$ of Amazon, Cond-mat, Enron eamil and Facebook are 0.095, 0.047, 0.007 and 0.011, respectively. The results are averaged over 100 independent runs. Highest value in each column is emphasized in bold.}\label{Table3}
          \begin{tabular}{l llll}
             \hline
              Networks & Amazon & Cond-mat & Email-Enron & Facebook  \\
               & $\beta=0.1425$ & $\beta=0.0705$   &$\beta=0.0105$  & $\beta=0.0165$ \\ \hline
                Degree	     &	0.2675 	       &	0.5657 	       &	0.4821 	       &	0.7348     \\
                H-index	     &	0.3219 	       &	0.6115 	       &	0.4883 	       &	0.7619    \\
                Coreness	 &	0.3289 	       &	0.6086 	       &	0.4883 	       &	0.7745       \\
                LocalRank	 &\textbf{0.6546}  &\textbf{0.8040}    &	0.5336 	       &	\textbf{0.8043}\\
                ClusterRank	 &	0.4521 	       &	0.5548 	       &	0.4001 	       &	0.7611        \\
                Closeness	 &	0.5968 	       &	0.7190 	       &	0.3271 	       &	0.7038      \\
                Betweenness	 &	0.2508 	       &	0.3277 	       &	0.4224 	       &	0.4880        \\
                Eigenvector	 &	0.3161 	       &	0.7350 	       &\textbf{0.5346}    &	0.7373     \\

             \hline
           \end{tabular}
        \end{table}

\begin{figure}
		  \centering
		  \includegraphics[width=15cm]{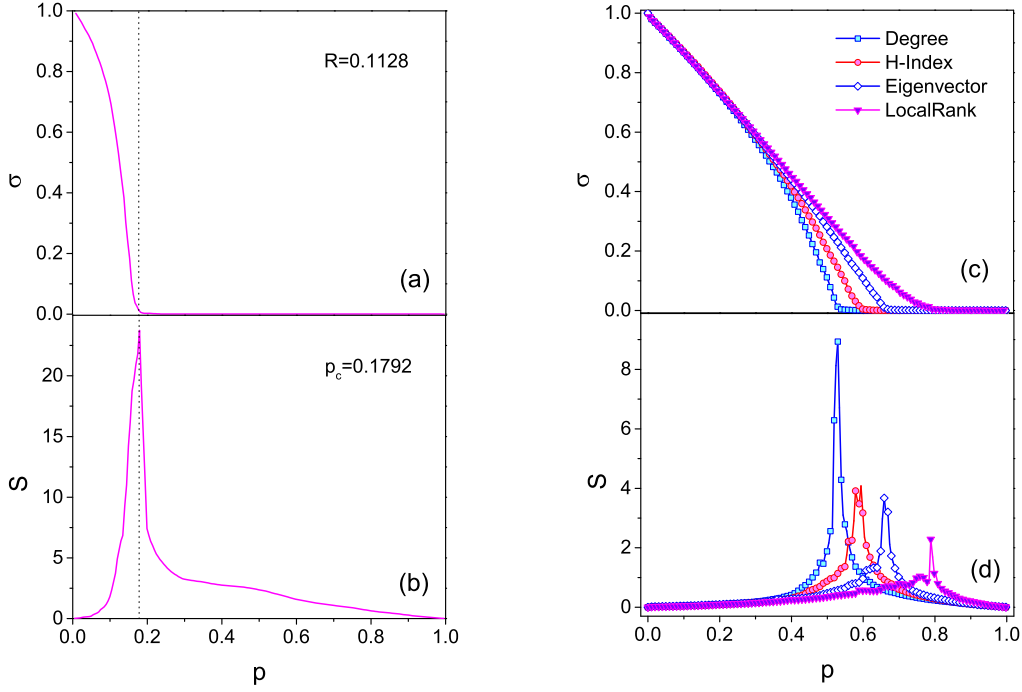}\\
		  \caption{The dependence of $\sigma$ and the corresponding susceptibility $S$ on the portion of removed nodes $p$. (a) and (b) are the schematic diagrams of $\sigma$ and $S$, where the dot line is corresponding to the collapsing point (i.e., critical portion $p_c$). (c) and (d) are the $\sigma$ and $S$ after removing the fraction $p$ of top-ranked nodes by degree, H-index, eigenvector centrality and LocalRank on Facebook.}\label{Figure11}
		\end{figure}

        Besides spreading influence, we also investigate the nodes' importance for network connectivity. Each method gives a ranking list of nodes according to their importance scores. Then we remove the nodes from the top-ranked ones and calculate the size of giant component $\sigma$ after each removal. Clearly, $\sigma$ decreases with the increasing number of removed nodes and vanishes when a critical portion ($p_c$) of nodes are removed, see a schematic diagram in Fig.~\ref{Figure11}(a) and the results of four centralities on Facebook in Fig.~\ref{Figure11}(c). To find the exact value of $p_c$, we investigate the \emph{Susceptibility} value $S$ of the network after node removal, which is defined as \cite{Cheng2010JSMP10011}:
         \begin{equation}\label{Susceptibility}
          S=\sum_{s<\sigma}\frac{{n_s}s^2}{n},
        \end{equation}
        where $n_s$ is the number of components with size $s$ and $n$ is the size of the whole network. Usually, there is a peak of $S$ at the critical portion $p_c$ at which the network collapses (i.e., the network breaks down into many disconnected pieces with small sizes), see Fig.~\ref{Figure11}(b)(d). If the network experiences multiple collapses during the node removal process, multiple peaks exist. The $p_c$ value is determined by the highest one. Obviously, according to the objective function on network connectivity, the smaller the $p_c$ is, the better the ranking algorithm is.

        \emph{Robustness} \cite{Schneider2011PNAS1083838} is another measure to quantify the performance of ranking methods. It is defined as the area under the curve of $\sigma-p$, mathematically reads
        \begin{equation}\label{robustness}
          R=\frac{1}{n}\sum_{i=1}^{n}\sigma(i/n),
        \end{equation}
        where $\sigma(i/n)$ is the size of the giant component after removing $i/n$ of nodes from the network. Clearly, the smaller $R$ is, the better the algorithm is. The robustness $R$ and $p_c$ for different methods on four real networks are listed in Table \ref{Table5} and Table \ref{Table6}, respectively. Overall speaking, degree performs the best among all the methods, indicating that the high-degree nodes are very important for the network connectivity. H-index and LocalRank methods can also give relatively better results. Betweenness is the best method among four global centralities and usually gives the second best results.

       \begin{table}
          \centering
          \caption{The robustness $R$ of the eight ranking methods on four real networks. The lowest value in each column is emphasized in bold.}\label{Table5}
          \begin{tabular}{l llll}
             \hline
             Networks & Amazon & Cond-mat & Email-Enron & Facebook  \\ \hline
              Degree      & \textbf{0.1226} &  0.1404           & \textbf{0.0404} &  0.3131  \\
              H-index     & 0.2298          &  0.2156           & 0.0605            &  0.3321  \\
              Coreness    & 0.3118          &  0.2826           & 0.0704            &  0.3383 \\
              LocalRank   & 0.3184          &  0.2787           & 0.1114            &  0.3701  \\
              ClusterRank & 0.1861          &  0.1356           & 0.0785            &  0.3404  \\
              Closeness   & 0.3778          &  0.2513           & 0.1677            &  0.3625  \\
              Betweenness & 0.1746          &  \textbf{0.1262}  & 0.0501     &  \textbf{0.2974} \\
              Eigenvector & 0.3789          &  0.3365           & 0.1113            &  0.3904  \\
             \hline
           \end{tabular}
        \end{table}

      \begin{table}
          \centering
          \caption{The critical portion of removing nodes $p_c$ for eight ranking methods on four real networks. The lowest value in each column is emphasized in bold.}\label{Table6}
          \begin{tabular}{l llll}
             \hline
             Networks & Amazon & Cond-mat & Email-Enron & Facebook  \\ \hline
              Degree      & \textbf{0.2320} &  0.2539           & \textbf{0.0948} &  \textbf{0.5289}  \\
              H-index     & 0.4649          &  0.4132           & 0.1496        &  0.5938  \\
              Coreness    & 0.6149          &  0.5277           & 0.2045        &  0.6137 \\
              LocalRank   & 0.6298          &  0.5476           & 0.4738        &  0.7884  \\
              ClusterRank & 0.3190          &  0.2240           & 0.2494        &  0.6487  \\
              Closeness   & 0.6562          &  0.5330           & 0.4252        &  0.7435  \\
              Betweenness & 0.2960          &  \textbf{0.1992}  & 0.1696        &  0.5389 \\
              Eigenvector & 0.7810          &  0.6450           & 0.4642        &  0.8136  \\
             \hline
           \end{tabular}
        \end{table}

    \subsection{On directed networks}
        Four directed networks are used to test the performance of six methods introduced in Chapter \ref{Chapter2} and Chapter \ref{Chapter3}: (i) \emph{Delicious} \cite{Lu2011PLoSONE6e21202} is a directed social network extracted from the website \emph{delicious.com}, where the primary function of users is to collect useful bookmarks with tags. Users can select other users as their ‘‘opinion leaders’’ of web browsing, in the sense that the bookmarks of the leaders are often useful and relevant. The subscriptions to leaders’ bookmarks can be made automatically. Of course users who select their leaders can in turn be the leaders of others. In that way, the users form a large-scale directed social network with information flows from leaders to followers. The link direction is always from a follower to his leader. (ii) \emph{Email-EuAll} \cite{Leskovec2007ACM12} is generated by the email data from a large European research institution from October 2003 to May 2005. Given a set of email messages, each node corresponds to an email address and a directed link exists from node $v_i$ to $v_j$ if $v_i$ received at least one email from $v_j$. (iii) \emph{Epinions} \cite{Richardson2003LNCS2870351} is a who-trust-whom online social network of a general consumer review site \emph{Epinions.com}. Members of the site can decide whether to ``trust'' each other. All the trust relationships interact and form the web of trust which is then combined with review ratings to determine which reviews are shown to the user. A directed link exists from node $v_i$ to $v_j$ if $v_i$ trusts $v_j$. (iv) The \emph{Wiki-Vote} network \cite{Leskovec2010WWW641} contains all the Wikipedia voting data from the inception of Wikipedia till January 2008. Nodes in the network represent Wikipedia users and a directed link from node $v_i$ to node $v_j$ represents that user $v_i$ voted for user $v_j$. The basic statistical features of these four networks are shown in Table \ref{Table7}.
        \begin{table}
          \centering
          \caption{The basic statistical features of four directed networks,  including the number of nodes $n$ and
edges $m$, the maximum degree $k_{max}$, the clustering coefficient $c$ and the degree heterogeneity in directed network is defined as $H=\frac{E_{fo}}{\log_2n}$, where $E_{fo}$ is the first-order entropy of the network \cite{Zhou2011SysEng290123}.}\label{Table7}
          \begin{tabular}{l llll lll}
             \hline
             Networks & $n$ & $m$ & $k_{\max}$ & $k^{in}_{\max}$ & $k^{out}_{\max}$& $c$ & $H$ \\ \hline
             Delicious & 582210 & 1679090 & 11370 & 11168 & 2767 &0.1509 &0.2760 \\
             Email-EuAll & 265009 & 418956 & 7636 & 7631 & 929 &0.0564 &0.1198 \\
             Epinions & 75879 & 508837 & 3079 & 3035 & 1801 &0.1098 &0.3471 \\
             Wiki-Vote & 7115 & 103689 & 1167 & 457 & 893 &0.0800 &0.4968 \\
             \hline
           \end{tabular}
        \end{table}

        We also consider the SIR spreading model to evaluate the algorithms' performance of identifying vital nodes with high spreading influences. In the directed networks, the information (or epidemic) spreads along the directed links. The Kendall’s tau correlation coefficient $\tau$ between the ranking scores given by the algorithms and the real spreading influences obtained by the SIR model are shown in Table \ref{Table8}. As the performances of the methods are very different on the four networks, it is difficult to say which algorithm is the best. Generally speaking, the three local methods perform better the the other three global methods. LeaderRank is better than PageRank in all four networks.

       \begin{table}
          \centering
          \caption{The Kendall’s tau correlation coefficient $\tau$ between the ranking scores given by the algorithms and the real spreading influences obtained by the SIR model. The approximate epidemic threshold $\beta_c$ of Delicious, Email-EuAll, Epinions and Wiki-Vote are 0.071, 0.0083, 0.013 and 0.023, respectively. The results are averaged over 100 independent runs. In each column, the highest value is emphasized in bold.}\label{Table8}
          \begin{tabular}{l llll}
             \hline
              Networks & Delicious & Email-EuAll & Epinions & Wiki-Vote  \\
               & $\beta=0.1065$ & $\beta=0.0125$  & $\beta=0.0195$ & $\beta=0.0345$ \\ \hline
             In-degree	        &	\textbf{0.6860}  &	0.5582 	         &	0.7867 	        &	\textbf{0.9555} 	\\
             H-index	        &	0.6766 	         &	0.6635 	         &	0.7876 	        &	0.9534 	       \\
             Coreness	        &	0.6707 	         &	\textbf{0.6877}  &	0.7878 	        &	0.9474 	      \\
             HITs (authority)   &	0.6340 	         &	0.5713 	         &	\textbf{0.7936} &	0.9524 	           \\
             PageRank$^{\dag}$	&	0.5030 	         &	0.4628 	         &	0.6591 	        &	0.9279 	  \\
             LeaderRank	        &	0.5429 	         &	0.4617 	         &	0.7182 	        &	0.9365 	    \\
             \hline
           \end{tabular}\\
            \small{$^{\dag}$ For PageRank, the random jumping probability equals 0.15. }
        \end{table}

        To evaluate the node importance for network connectivity, we investigate the weakly connected component in directed networks after node removal. The robustness $R$ and the critical portion $p_c$ of different methods on four networks are shown in Table \ref{Table10} and Table \ref{Table11}, respectively. It can be seen that the network is more vulnerable when it is attacked by in-degree centrality. Notice that the results indicated by $R$ and $p_c$ are slightly difference in some networks. For example, for Delicious, the $R$ value of in-degree is the lowest while the $p_c$ value of in-degree is larger than both PageRank and LeaderRank. Overall speaking, in-degree is a very good indicator of nodes' important in directed networks.

		\begin{table}
          \centering
          \caption{The robustness $R$ of six ranking methods on four directed networks. The lowest value in each column is emphasized in bold.}\label{Table10}
          \begin{tabular}{l llll}
             \hline
             Networks         & Delicious & Email-EuAll & Epinions & Wiki-Vote  \\ \hline
             In-degree	      &	\textbf{0.0958}  &	0.0126  &	\textbf{0.0624} &	0.1366 	\\
             H-index	      &	0.1442 	         &	0.0334 	         &	0.0722 	        &	0.1438 	       \\
             Coreness	      &	0.1677 	         &	0.0383           &	0.0758 	      &	0.1453 	      \\
             HITs (authority) &	0.1669 	         &	0.0700 	         &	0.0937        &	0.1429 	           \\
             PageRank	      &	0.1190 	         &\textbf{0.0053}    &	0.1027 	        &	0.2489 	  \\
             LeaderRank	      &	0.1463 	         &	0.0363 	         &	0.0729 	        &	\textbf{0.1330} 	    \\

             \hline
           \end{tabular}
        \end{table}

		\begin{table}
          \centering
          \caption{The critical portion of removing nodes $p_c$ for six ranking methods on four directed networks. The lowest value in each column is emphasized in bold.}\label{Table11}
          \begin{tabular}{l llll}
             \hline
              Networks & Delicious & Email-EuAll& Epinions& Wiki-Vote  \\ \hline
              In-degree     & \textbf{0.3650}   &   0.0150          & \textbf{0.1379}   & \textbf{0.2870}\\
              H-index        & 0.5300           &   0.0770          & 0.1428            &0.3052 \\
              Coreness       & 0.5300           &   0.0900          & 0.1618            &0.3008 \\
              HITs (authority) & 0.4000         &   \textbf{0.0050} & 0.2450            &0.3013 \\
              PageRank       & 0.3680            &   0.0610          & 0.3572            &0.4254 \\
              LeaderRank     & 0.7460           &   \textbf{0.0050} & 0.1978            &0.3152 \\
             \hline
           \end{tabular}
        \end{table}

   \subsection{On weighted networks}
        Four weighted networks are used to evaluate the performance of seven weighted ranking methods, including two directed networks and two undirected ones. (i) \emph{Adolescent health} \cite{Moody2001SNet23261} is a directed network which was created from a survey that took place in 1994-1995. Each student was asked to list his/her 5 best female and 5 best male friends. A node represents a student and a directed link from node $v_i$ to $v_j$ means student $v_i$ choose student $v_j$ as a friend. Higher link weights indicate more interactions. (ii) \emph{USA airports} \cite{WebPage2015konect} is a directed network of flights between USA airports in 2010. Each link represents an airline from one airport to another, and the weight of a link shows the number of flights on that connection in the given direction. (iii) \emph{King James} \cite{WebPage2015konect} is an undirected network which contains nouns – places and names – of the King James bible and information about their occurrences. A node represents one of the above noun types and an edge indicates that two nouns appeared together in the same verse. The edge weights show how often two nouns occurred together. (iv) \emph{Cond-mat} is the weighted version of the collaboration network introduced in Section \ref{secPEUndirected}. If $k$ authors co-authorized a paper, $1/k$ score is added to each of edges between any two of these $k$ authors. For each network, we normalize the edge weight with its minimal value, namely the range is from $[w_{\min},w_{\max}]$ to $[1,\frac{w_{\max}}{w_{\min}}]$, where $w_{\min}$ and $w_{\max}$ are the minimal and maximal edge weight of the original network, respectively. The basic statistical features of these four weighted networks are shown in Table \ref{Table12}. The degree heterogeneity $H$ is defined the same as we introduced in Section \ref{secPEUndirected}. 
        \begin{table}
          \centering
          \caption{Basic statistical features of four weighted networks. $s_{\max}$ is the maximal node strength. For directed networks, it is the maximal in-strength. $k_{max}$ is the largest degree of the nodes. $c$ is the clustering coefficient of the networks and $H$ is the degree heterogeneity (defined as $\frac{\langle k^2\rangle}{\langle k\rangle^2}$ \cite{Hu2008PhysicaA3873769}), ignoring the weights of the networks.}\label{Table12}
          \begin{tabular}{l llll ll}
             \hline
             Networks & $n$ & $m$ & $k_{\max}$ & $s_{\max}$& $c$ & $H$ \\ \hline
             Adolescent & 2539 & 12969 & 36 & 113 &0.1196 &0.6167 \\ 
             USA airports  & 1574 & 28236 & 596 & 86095283 &0.4885 &0.6370 \\ 
             King James & 1773 & 9131 & 364 &1585  &0.7208 &4.0115 \\
             Cond-mat & 27519 & 116181 & 202 &145&0.6546 &2.6393 \\ 
             \hline
           \end{tabular}
        \end{table}

        In weighted networks, the spreading process of SIR model is similar with that in unweighted networks. The little difference is that the infected probability is not a constant but depends on the edge weights. Yan \emph{et al.}~\cite{Yan2005CPL22510} defined the infection transmission by the spreading rate, $ \lambda_{ij}=(\frac{\omega_{ij}}{\omega_{max}})^\alpha$ in which susceptible node $v_i$ acquires the infection from its infected neighbour $v_j$, $\alpha$ is a positive constant and $\omega_{max}$ is the largest value of $w_{ij}$ in the network. In this report, we adopt another form of infection transmission. The probability that an infected node $v_i$ infects his/her susceptible neighbor $v_j$ is $1-(1-\beta)^{w_{ij}}$ \cite{Wang2014PRE90042803}, where $w_{ij}$ is the weight of edge $(v_i,v_j)$. The Kendall’s tau correlation coefficient $\tau$ between the ranking scores given by the algorithms and the real spreading influences obtained by the SIR model are shown in Table \ref{Table13}. It can be seen that H-index and coreness have advantage on identifying influential spreaders, and perform even better than some global methods. LeaderRank is better than PageRank on two directed networks.

        \begin{table}
          \centering
          \caption{Kendall’s tau correlation coefficient $\tau$ between the ranking scores given by the algorithms and the real spreading influences obtained by the SIR model. The approximate epidemic threshold $\beta_c$ of Adolescent health, USA airports, King James and Cond-mat networks are 0.0688, $7.6294\times 10^{-7}$, 0.0125 and 0.0047, respectively. We set $\beta=1.5\beta_c$ in the experiments. The results are averaged over 100 independent runs. The highest value in each column is emphasized in bold.}\label{Table13}
          \begin{tabular}{l ll|lll}
          \hline
           Networks &Adolescent & Airports  & & King James & Cond-mat  \\
           & $\beta=0.103$  & $\beta=1.144\times 10^{-6}$  && $\beta=0.019$  & $\beta=0.0071$  \\
           \hline
            In-Strength	&	0.6524 	          &	0.7692 	        & In-Strength   &	0.6468 	       &\textbf{0.6278} 	\\
            H-index		&   0.7142 	          &	0.7709 	        & H-index		&	0.6642 	       &	0.6056 	\\
            Coreness    &	0.7369 	          &	\textbf{0.7711} & Coreness      &	\textbf{0.6689}&	0.5937 	\\
            PageRank    &	0.6263 	          &	0.5693 	        & Closeness     &	0.5784 	       &	0.1033 	\\
            LeaderRank	&	\textbf{0.7470}   &	0.7652 	        & Betweenness   &	0.5141 	       &	0.3589 	\\
             \hline
           \end{tabular}
        \end{table}

        The analysis of network connectivity is the same as that in unweighted networks. The robustness $R$ of seven methods on four weighted networks are shown in Table \ref{Table15}. Different from promoting information spreading, for maintaining the network connectivity, node strength performs better than weighted H-index and weighted coreness. In fact, strength performances the best in three networks, namely Adolescent health, King James and Cond-mat, and the second best in USA airports where PageRank is the best. PageRank is better than LeaderRank on two directed networks, while betweenness is better than closeness on the two undirected networks. The critical portion of removing nodes $p_c$ is listed in Table \ref{Table16}. Similar results are obtained. Again, $p_c$ of node strength method is the smallest in most cases among all ranking methods.

		\begin{table}
          \centering
          \caption{The robustness $R$ of seven ranking methods on four weighted networks. The lowest value in each column is emphasized in bold.}\label{Table15}
          \begin{tabular}{l ll|lll}
          \hline
           Networks &Adolescent & Airports  & & King James & Cond-mat  \\ \hline
              Strength    &\textbf{0.4174}   & 0.2143           &Strength       & \textbf{0.1442} & \textbf{0.1346}  \\
              H-index     &0.4483           & 0.2284            &H-index        & 0.1630          & 0.1599   \\
              Coreness    &0.4544           &0.2284             &Coreness       &  0.1734         & 0.1730  \\
              PageRank    &0.4371           & \textbf{0.1538}   &Closeness      &  0.2222         & 0.3676   \\
              LeaderRank  & 0.4470          & 0.2149            &Betweenness    &0.1475           & 0.1512    \\
             \hline
          \end{tabular}
        \end{table}

		\begin{table}
          \centering
          \caption{The critical portion of removing nodes $p_c$ for seven ranking methods on four weighted networks. The lowest value in each column is emphasized in bold.}\label{Table16}

          \begin{tabular}{l ll|lll}
          \hline
           Networks &Adolescent & Airports  & & King James & Cond-mat  \\  \hline
              Strength    &\textbf{0.7020}  & 0.2935        &Strength       & \textbf{0.1985} & \textbf{0.2887}  \\
              H-index     &0.7872           & 0.3647        &H-index        & 0.2437        & 0.3535   \\
              Coreness    &0.7933           &0.3647         &Coreness       & 0.2798        & 0.3535   \\
              PageRank    &0.7541           & \textbf{0.2135}    &Closeness   & 0.3180      & 0.6870  \\
              LeaderRank  & 0.7906          & 0.2891        &Betweenness    & 0.2413        & 0.2556  \\
             \hline
           \end{tabular}

        \end{table}

	\clearpage
	\subsection{On bipartite networks}
    In this section, six rating-based ranking algorithms are compared in detail on two real networks and one artificial network. (i) \emph{Netflix}~\cite{Zhou2011EPL9448002} is a randomly selected subset of the famous data set released by the DVD rental company Netflix in 2006. The Netflix Prize in 2006 challenged researchers to increase accuracy based solely on the rating structure of the user-movie bipartite network. The ratings are given on the integer rating scale from $1$ to $5$. (ii) \emph{MovieLens}~\cite{Yao2015ICAI4002} is collected by GroupLens Research from the MovieLens website (\emph{http://movielens.org}). There exist three different versions of MovieLens data sets. The ratings of the data set we used here is from $0.5$ to $5$, with a step size of $0.5$. All the selected users have rated at least $20$ movies. To test the validity of the algorithms in Chapter~\ref{Chapter8}, all the movies in these two data sets which were nominated for the best picture category at the Annual Academy Awards (as known as Oscars) are adopted as the benchmark. In addition, an artificial network of 6000 users and 4000 items is generated through a preferential attachment mechanism in the evolution of the rating system. It is assumed that each item has a certain true intrinsic quality and each user has a certain magnitude of rating error~\cite{Zhou2011EPL9448002}. Items' qualities, $Q$, follow an uniform distribution on the interval $[0,1]$, and users' magnitudes of rating error, $\delta$, obey uniform distribution on the interval $[0.1, 0.5]$. The rating relationships between users and items are generated according to the preferential attachment mechanism. The rating of an item $\alpha$ given by a user $i$ is $r_{i\alpha} = Q_\alpha + \delta_{i\alpha}$ where $\delta_{i\alpha}$ is draw from the normal distribution $[0, \delta_i]$. The properties of the three data sets are summarized in Table~\ref{Table17}, including the number of users, items, ratings and the benchmark items, as well as the network sparsity.
        \begin{table}
          \centering
          \caption{Basic statistical features of bipartite networks.}\label{Table17}
          \begin{tabular}{llllll}
             \hline
             Networks &Users & Items &Ratings &Benchmark  &Sparsity \\ \hline
             Netflix & 4960 & 17238 &1462991 & 363 & 0.017\% \\
             MovieLens & 69878 & 10677 & 10000054 & 387 &0.013\% \\
             Artificial Network & 6000 & 4000 & 480000 & 200 & 0.020\% \\
             \hline
           \end{tabular}
        \end{table}

     One of the popular metric to evaluate the ranking accuracy on bipartite networks is AUC (abbreviation of the area under the receiver operating characteristic curve)~\cite{Lu2012PhysRep5191}. AUC was originally used to judge the discrimination ability of the predictive methods in the theory of signal detection~\cite{Hanley1982Radiology14329}. A simple way to compute the AUC of a ranking algorithm is by comparing its discrimination ability for good and bad items. For Netflix and Movielens data sets, two items will be selected randomly among benchmark items (i.e., the items who were nominated for Oscars) and other items, respectively. According to the quality values given by a ranking algorithm, if the benchmark item has a higher quality than the other one, AUC increases by $1$. If the two items get the same quality value, AUC increases by $0.5$. If the benchmark item obtains a lower quality than another one, then AUC keep unchanged. The final value of AUC will be normalized by the number of comparisons, mathematically reads
            \begin{equation}\label{Eq_AUC}
              AUC = \frac{n'+0.5n''}{n},
            \end{equation}
     where $n$ is the number of comparisons, $n'$ is the times when the benchmark items have higher qualities than other items, and $n''$ is the times when the benchmark items have the seam qualities with other items. For the artificial network data set, the items which get the top $5\%$ quality values and the users whose rating error belong to the lowest $5\%$ will be selected as the benchmarks.

     AUC is mainly used to test the ability of the ranking algorithm to distinguish between good and bad items, while the Pearson product moment correlation coefficient, $r$, is employed to test the ranking accuracy for all items. Pearson coefficient $r$ reflects the extent of a linear relationship between the certain true intrinsic qualities of items and their scoring qualities given by the ranking algorithm. It reads
            \begin{equation}\label{Eq_Pearson}
              r = \frac{{\sum {({x_i} - \bar x)({y_i} - \bar y)} }}{{\sqrt {\sum {{{({x_i} - \bar x)}^2}} \sum {{{({y_i} - \bar y)}^2}} } }}.
            \end{equation}
     Evidently, Pearson coefficient $r$ could also be used to test the accuracy of the ranking algorithm for users, according to their true intrinsic reputations and their scoring reputations.

     We also test the algorithms' capability to resist attack. Two kinds of attacks are considered in this section~\cite{Zhou2011EPL9448002}: (i) Random rating: the attacker will rate items with random allowable scores. (ii) Push rating: the attacker will rate items with maximum or minimum allowable scores.

     From our experiments, we found that the parameter $\theta$ in IARR and IARR2~\cite{Liao2014PLoSONE9e97146} is very sensitive and has to be selected carefully. The AUC of the ranking algorithms for Netflix and MovieLens are shown in Table~\ref{Table18}. IARR2 with $\theta=1$ performs the best for Netflix and MovieLens comparing with all the other algorithms. IR performs relatively good for Netflix, while CR and IARR perform relatively good for Movielens.
        \begin{table}
          \centering
          \caption{The AUC value of the ranking algorithms for real data sets. The highest value in each row is emphasized in bold.}\label{Table18}
          \begin{tabular}{l llllll}
             \hline
              Networks & AR    & IR       & CR      & GR    & IARR ($\theta=3$) & IARR2 ($\theta=1$) \\ \hline
              Netflix  & 0.767 & 0.808 & 0.782   & 0.767 & 0.781           & \textbf{0.809}   \\
              MovieLens& 0.850 & 0.753 & 0.851   & 0.848 & 0.851           & \textbf{0.870}   \\
             \hline
           \end{tabular}
        \end{table}

     However, it is pretty hard to say which item (e.g., movie) is the best and should be considered as the benchmark to test the validity of the ranking algorithms in particular scenarios. So the artificial networks are constructed in which every user has a real intrinsic reputation and every item has a certain true intrinsic quality. Table~\ref{Table19} comprehensively compares the discrimination abilities of the algorithms to evaluate the reputation of users and the quality of items. As indicated above, AUC reflects the accuracy of the ranking algorithms to tell good from mediocre users or items. Thus, the top 5\% users with the highest reputations and 5\% items with the highest qualities are regarded as the benchmark. Unlike the real data sets, the ratings in this artificial network are no longer limited to several fixed values. Thus, the GR \cite{Gao2015EPL11028003} method, grouping users based on their ratings, will not be analyzed here. According to the results in Table~\ref{Table19}, AR has the best performance in distinguishing the influence of different users and IARR with $\theta=3$ has the best performance of ranking items. When measuring with Pearson coefficient $r$, IARR has the best performance for ranking of both users and items.

\begin{figure}
          \centering
          \includegraphics[width=10cm]{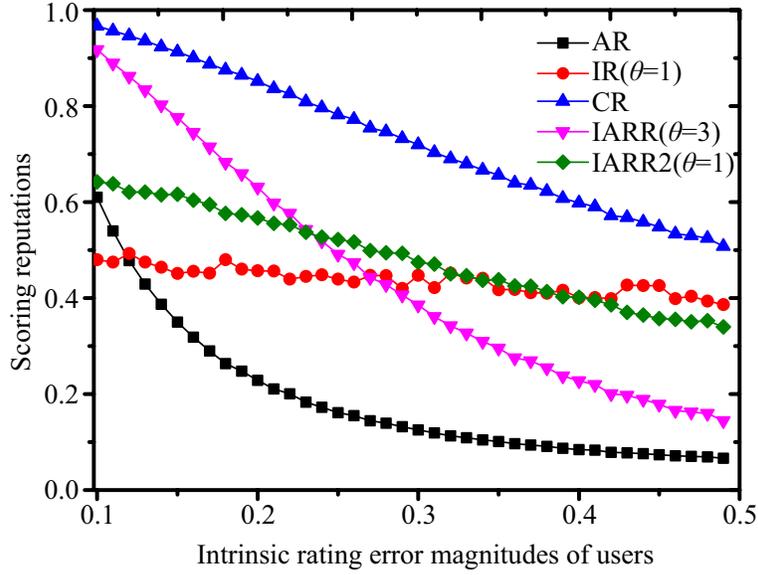}
          \caption{The relation between users' true intrinsic rating error magnitudes and their scoring reputations in different ranking algorithms. The results in this figure are averaged over 10 independent realizations. }\label{Figure12}
        \end{figure}

     For a good reputation estimation method, the users' final reputations given by the algorithm should be negatively correlated with their true intrinsic reputations. The stronger the correlation is, the better the algorithm is. We define $40$ equally distributed intervals between the minimum and maximum of the initial intrinsic reputations of all users and group the nodes whose intrinsic reputation fall in the same interval. The results of the representative methods are reported in Fig.~\ref{Figure12}.

\begin{figure}
         \centering
         \includegraphics[width=16cm]{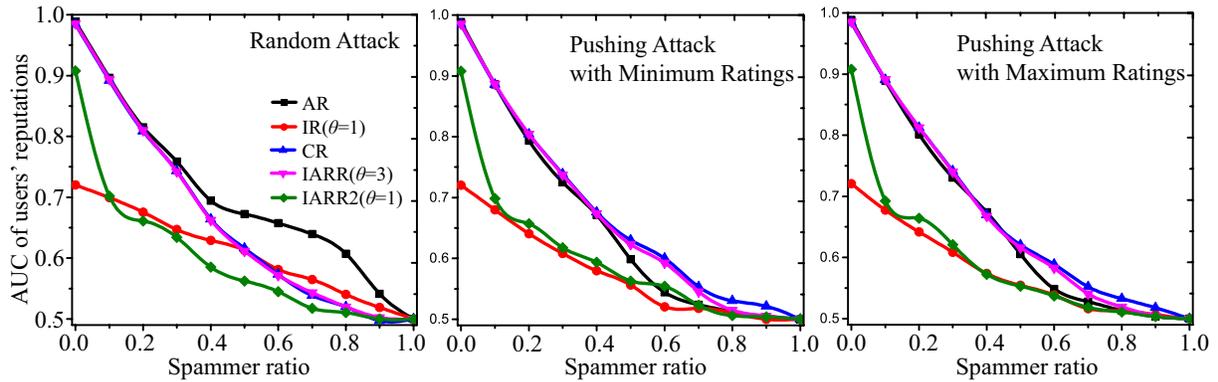}\\
         \caption{The AUC values of users' reputation scores under spammer attack. (a) Random attack. (b) Pushing attack with minimum ratings. (c) Pushing attack with maximum ratings. The results in this figure are averaged over 10 independent realizations.}\label{Figure13}
       \end{figure}

     The effectiveness of different algorithms with 20\% spammers is shown in Table~\ref{Table20}. For the random attack strategy, every item will obtain a random rating in the range [0,1]. However, for the pushing attack strategy, every item will get an extreme value, that is, $0$ for the pushing attack with minimum ratings and $1$ for the pushing attack with maximum ratings. As we have mentioned above, the top 5\% items with highest intrinsic qualities and the top 5\% users with lowest error magnitudes will be selected as the benchmark, respectively. According to the results in Table~\ref{Table20}, AR has the best performance in ranking users, and CR, as well as IARR, has the best performance in distinguishing items under random attack. CR and IARR are also the best two algorithms under pushing attack. Fig.~\ref{Figure13} gives a more clear illustration for the effectiveness of all the algorithms under spammer attack.

        \begin{table}
          \centering
          \caption{The AUC values and Pearson coefficient $r$ of the ranking algorithms for the artificial network. All the results are averaged over 10 independent runs. The results of the best performing algorithm in each row are emphasized in bold.}\label{Table19}
          \begin{tabular}{l lllll}
             \hline
              Metrics      & AR      & IR     & CR       & IARR ($\theta=3$) & IARR2 ($\theta=1$) \\ \hline
              AUC (user)  & \textbf{0.9886}   & 0.7205  & 0.9843    & 0.9855           & 0.9075  \\
              AUC (item)& 0.9860  & 0.7783  & 0.9870    & \textbf{0.9876}           & 0.9372  \\
              $r$ (user)  & -0.8398 & -0.4367 & -0.8995   & \textbf{-0.9202}          & -0.8195 \\
              $r$ (item)& 0.9923  & 0.2783  & 0.9933    & \textbf{0.9942}           & 0.9699  \\
             \hline
           \end{tabular}
        \end{table}

        \begin{table}
          \centering
          \caption{The AUC of the ranking algorithms for the artificial network with 20\% spammer attack. For the pushing attack strategy, 0 and 1 represent the pushing attack with minimum and maximum ratings, respectively. The results in this table are averaged over 10 independent realizations. The best result in each column is emphasized in bold.}\label{Table20}
          \begin{tabular}{l llll lll ll}
             \hline
             Methods & \multicolumn{2}{c}{Random Attack} && \multicolumn{2}{c}{Pushing Attack (0)} && \multicolumn{2}{c}{Pushing Attack (1)}  \\ \cline{2-3}\cline{5-6}\cline{8-9}
                          & Users & Items && Users & Items && Users & Items \\ \hline
             AR                & \textbf{0.8151} & 0.9746 &&0.7938  & 0.9617 && 0.8015 & 0.9821 \\
             IR                & 0.6755 & 0.7485 &&0.6407  & 0.7753 && 0.6417 & 0.7571 \\
             CR                & 0.8087 & \textbf{0.9840} &&0.8039  & \textbf{0.9845} && \textbf{0.8127} & \textbf{0.9842} \\
             IARR ($\theta=3$) & 0.8099 & \textbf{0.9840} &&\textbf{0.8044}  & 0.9844 && \textbf{0.8127} & 0.9840 \\
             IARR2 ($\theta=1$)& 0.6610 & 0.6878 &&0.6568  & 0.6890 && 0.6642 & 0.6906 \\
             \hline
           \end{tabular}
       \end{table}

  \clearpage
      \subsection{Finding a set of vital nodes}
      The networks used in this section are the same as those in Section \ref{secPEUndirected}, namely, Amazon, Cond-mat, Email-Enron and Facebook. To evaluate the algorithms' ability of finding a set of influential nodes, we also consider the SIR model. Initially, a set of nodes are infected as seeds, and the spreading process is the same as we have described in Section \ref{secPEUndirected}. Then the ratio of final recovered nodes is used to measure the algorithms' performance. We test seven methods including degree, betweenness, closeness, Ji's percolation method \cite{Ji2015ArXiv150804294}, Hu's percolation method \cite{Hu2016ArXiv150903484}, CI \cite{Morone2015Nature52465} and VoteRank \cite{Zhang2016arXiv160200070}. For the two parameter dependent methods, we set $L=0.05n$ for Ji's method and $l=2$ for CI. The occupation probabilities for both Ji's and Hu's methods equal to the infected probability of the spreading model.

      Table \ref{Table21} shows the average ratio of final recovered nodes triggered by initializing 5\% infected nodes selected by different methods. It can be seen that Ji's method and VoteRank have relatively higher ability to find a set of high influential nodes to maximize information coverage than other methods. Hu's method is also rather good, while closeness is the overall worst. Betweenness \cite{Freeman1977Sociometry4035} is not so bad in finding a set of vital nodes comparing with the results in finding a single node. Besides, betweenness performs better than closeness which is in contrast to the results in finding a single node. This may be caused by the fact that the nodes with high closeness values are usually closely connected, leading to high redundancy, while the nodes with high betweenness values are more likely to be dispersedly distributed.

	\begin{table}
	\centering
	\caption{The average ratio of final recovered nodes triggered by initializing 5\% infected nodes selected by different methods. The results are averaged over 100 independent runs. The highest value in each column is emphasized in bold. }\label{Table21}
	\begin{tabular}{l llll}
	\hline
	  Networks       &Amazon        & Cond-mat & Email-Enron & Facebook  \\
					 & $\beta=0.096$  & $\beta=0.048$  & $\beta=0.008$  & $\beta=0.012$ \\ \hline
	  Degree         & 16.8326      & 13.3928    &6.8292  & 11.5244 \\
	  Betweenness    & 13.9126      & 14.3589    &7.1250   & 12.8927  \\
	  Closeness      & 8.764       & 11.5714    &5.1560   & 11.8676  \\
	  CI ($l=2$)     & 17.0441      & 13.9327    &5.9509   & 12.0612  \\
	  Ji's           & \textbf{23.4133}  & \textbf{16.6341}   &7.2680   & 13.4827 \\
	  Hu's           & 12.4737      & 12.6359    &6.8030   & 11.8670 \\
	  VoteRank       & 18.0210      & 15.2614    &\textbf{7.3167}   & \textbf{13.8184} \\
	\hline
	\end{tabular}
	\end{table}

    To investigate the importance of a set of nodes on network connectivity, we remove a group of the network nodes instead of single node to calculate the robustness $R$ and the corresponding $S$. The results of $R$ and $p_c$ are shown in Table \ref{Table22} and Table \ref{Table23}, respectively. It can be seen that removing the nodes selected by Ji's and Hu's methods can hardly disintegrate the network in most cases. CI method performs the best, because the original purpose of CI is to find the most influential nodes that guarantees the global connection of the network. Degree, VoteRank and betweenness also have relatively good performance.

	\begin{table}
          \centering
          \caption{The robustness $R$ of seven ranking methods. The lowest value in each column is emphasized in bold.}\label{Table22}
          \begin{tabular}{l llll}
             \hline
              Networks    & Amazon & Cond-mat & Email-Enron & Facebook  \\ \hline
              Degree      & \textbf{0.1226} &  0.1404          & 0.0404 &  0.3131  \\
              Betweenness & 0.1746          &  0.1272          & 0.0501 &  0.2974  \\
              Closeness   & 0.3118          &  0.2826           & 0.0704            &  0.3383 \\
              CI ($l=2$)  & 0.2670          &  \textbf{0.1130} & \textbf{0.0388} &  \textbf{0.1031}  \\
              Ji's        & 0.3738          &  0.4052          & 0.3127 &  0.4252  \\
              Hu's        & 0.3748          &  0.3400          & 0.3613 &  0.4450  \\
              VoteRank    & 0.2766          &  0.1432          & 0.0594 &  0.1224 \\
             \hline
           \end{tabular}
        \end{table}

	\begin{table}
          \centering
          \caption{The critical portion of removing nodes $p_c$ for different ranking methods. The lowest in each column is emphasized in bold. }\label{Table23}
          \begin{tabular}{l llll}
             \hline
             Networks & Amazon& Cond-mat & Email-Enron & Facebook  \\ \hline
              Degree      & 0.2100      &  0.2539        & 0.0948 &  0.5289  \\
              Betweenness & 0.2600      &  0.1892        & 0.1696 &  0.5389  \\
              Closeness   & 0.6149          &  0.5277           & 0.2045        &  0.6137 \\
              CI ($l=2$)         & 0.1600      & \textbf{0.1593}  & 0.0998 &  \textbf{0.3692}  \\
              Ji's        & \textbf{0.1000}&  0.2200      & 0.6000 &  0.6200  \\
              Hu's        & 0.8000      &  0.7600         & 0.1200 &  0.7300  \\
              VoteRank    & 0.1950      &  0.1643          & \textbf{0.0698} &  0.3842 \\
             \hline
           \end{tabular}
        \end{table}

\clearpage
\section{Applications}\label{Chapter10}
     \subsection{Identifying influential spreaders in social networks}
     The development of social networks has a great impact on information society. On one hand, social networks provide a new mode of acquiring information through leveraging the power of network, e.g., collecting useful information from various experts. This strategy of collective searching has the potential to renovate the current search paradigm based on isolated queries. The key point of the strategy is to identify influential users in social communities \cite{Rabade2014AdvInteSysComp359}. On the other hand, social networks construct a new platform for information propagation, which has wide applications like the online virtual marketing \cite{Akritidis2011IEEETrans41759}. The crucial part is also to seek out influential spreaders, who are more likely to make the information spread widely.

     Who are the influential spreaders? It is determined by not only the network structure but also the dynamics being considered. The key spreaders in epidemic dynamics may be not important in information dynamics. Arruda \emph{et al.} \cite{Arruda2014PRE90032812} investigated the relations between the transmission capacities of nodes and ten centrality measures in different synthetic and real-world (both spatial and nonspatial) networks. They found that, in the epidemic dynamics in nonspatial networks, the coreness and degree centralities are most related to the capacities of nodes. Whereas in the rumor propagation in nonspatial networks, the average neighborhood degree, the closeness centrality, and accessibility have the highest performance. In the case of spatial networks, accessibility has the overall best performance in both of the two dynamic processes. The algorithms of identifying important nodes are widely applied to social networks. And the application can usually bring some appreciable social and economic values in many scenarios, such as the virtual marketing by influential spreaders~\cite{Weng2010ICWSWDM261,Akritidis2011IEEETrans41759}, and the rumor control by ``immunizing" vital people. Besides, it can also help forensic investigators identify the most influential members of a criminal group \cite{Alzaabi2015IEEETrans102196}, and monitor the exceptional events like mass mobilizations \cite{GonzalezBailon2011SciRep1197}, etc.

     Real experiments are very difficult to be launched on real online social networks, because it is not easy to find enough participants and to evaluate the performance by compare the experimental algorithm with known benchmarks. Therefore, the most of the previous research were developed through analyzing off-line data. However, we have made some large-scale experiments with the help from China Mobile company in Zhuhai City. The network under consideration is a directed short-message communication network constructed by using the message forwarding data during 31 days from Dec. 8th, 2010 to Jan. 7th, 2011. Therein, each node represents a user identified by the unique mobile phone number, and a link from user $v_i$ to user $v_j$ means that $v_i$ had sent at least one short message to $v_j$ during these 31 days. The network includes 9330493 nodes and 23208675 links, and its clustering coefficient is only 0.0043 and the largest degree is 4832. In our experiments, the task is to find a number of initial users who have higher influence. At the first step, we selected $1000$ users based on different strategies, such as selecting the users with the highest LeaderRank~\cite{Lu2011PLoSONE6e21202} scores, or with the highest out-degrees. Secondly the company sent each of them a message. Then we monitored the number of forwardings of each strategy.

\begin{figure}
          \centering
          \includegraphics[width=14cm]{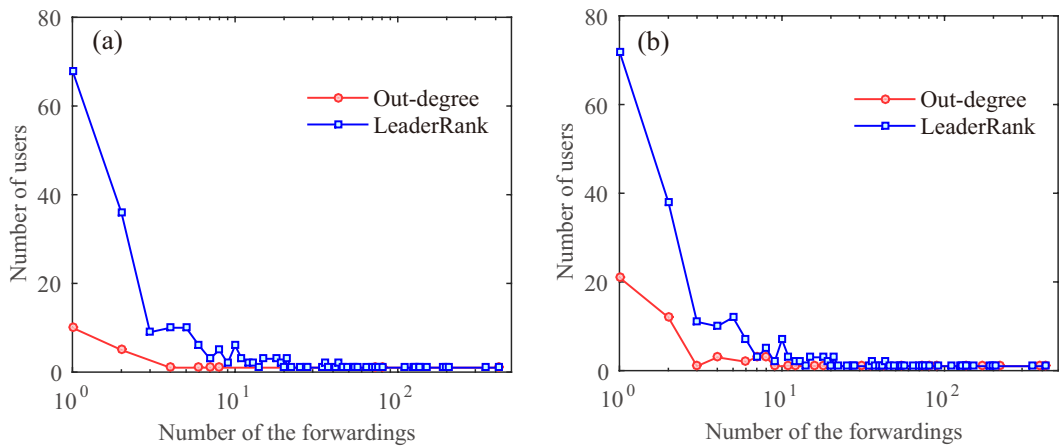}\\
          \caption{The distributions of the number of direct forwardings for out-degree centrality and LeaderRank. (a) shows the case with spammers, and (b) shows the case without spammers. The total number of forwardings achieved by LeaderRank and out-degree is 3892 and 620 for the case with spammers, and 4063 and 924 for the case without spammers.}\label{Figure14}
    \end{figure}

     To investigate the algorithm's resilience to spammers, we removed all the possible spammers whose clustering coefficient are zero, and then select the $1000$ users under each strategy. Fig.~\ref{Figure14} shows the distributions of the number of direct forwardings of the two strategies (LeaderRank and degree centrality) under two cases (with and without spammers). For the case with spammers, among the 1000 users selected by out-degree centrality there are 22 users who have forwarded the short message at least once, and the average number of forwarding is 28.18 times. However, in the case that we have removed the spammers, there are 62 users who have forwarded the short message at lease once, and the average number of forwardings rises to 31.03 times. For LeaderRank, the improvement is not obvious, indicating its great resilience to spammers. For the case with spammers, among 1000 users selected by LeaderRank there are 207 users who have forwarded the short message, and the average number of forwarding is 18.8 times. For the case without spammers, there are 221 users who forwarded the short message, and the average number of forwarding is 18.38 times. Although the average number of direct forwardings of LeaderRank is lower than that of out-degree centrality, its total number of direct forwardings is more than 6 times as much as the number for out-degree centrality for the case with spammers, and 2 times for the case without spammers.

    \subsection{Predicting essential proteins}
        Identification of essential proteins can help us in understanding the basic requirements to sustain a life form \cite{Zhang2009NuclAcidRes37455}. It also plays a significant role in the emerging field of synthetic biology that aims to create a cell with the minimal genome \cite{Glass2009MolSystBiol5330}. Additionally, previously known results suggest that essential proteins have correlations with human disease genes \cite{Furney2006BMCGeno7165}, and thus the identification of essential proteins is helpful to find drug target candidates in anti-infectious and anti-cancer therapies \cite{Csermely2013PT138333}.

        Up to now, the network-based methods play an increasing role in essential protein identification since the traditional biological experiments are expensive and time consuming. Centrality measures, such as degree, betweenness, closeness, PageRank and LeaderRank, were widely applied in the identification algorithms \cite{Wang2012IEEETrans91070}. To be more accurate, scientists usually integrated centrality measures of the target protein–protein interaction network with biological information. Representative methods include protein-protein interaction combing gene expression data centrality (PeC) \cite{Li2012BMC615}, local interaction density and protein complexes (LIDC) \cite{Luo2015PLoSONE10e0131418}, integrating orthology with PPI networks (ION) \cite{Peng2012BMCSysBiol687}, co-expression weighted by clustering coefficient (CoEWC) \cite{Zhang2013PLoSONE8e58763}, united complex centrality (UC) \cite{Li2015IEEETran1}, and so on. In PeC, a protein's essentiality is determined by the number of the protein's neighbors and the probability that the protein is co-clustered and co-expressed with its neighbors. LIDC combines local interaction density with protein complexes. ION is a multi-information fusion measures, showing the effectiveness of the orthologous information in detecting essential proteins. CoEWC captures the properties of both data hubs and party hubs despite that the two hubs have very different clustering properties. UC is a combination of the normalized alpha centrality \cite{Gosh2011PRE83066118} and in-degree of protein complexes.

\begin{figure}
          \centering
          \includegraphics[width=16cm]{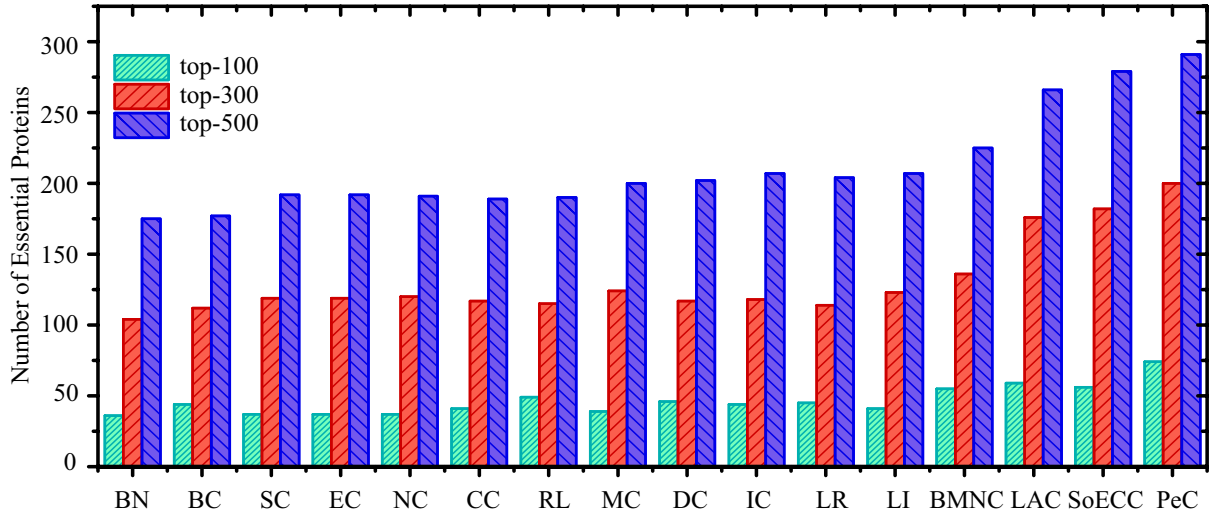}\\
          \caption{Comparison of the number of essential proteins detected by sixteen centrality measures \cite{Li2012BMC615}. For each centrality measure, a certain number of top proteins are selected as candidates for essential proteins and out of which the number of true essential proteins are determined. The number of true essential proteins detected by protein-protein interaction combing gene expression data centrality (PeC), degree centrality (DC), betweenness centrality (BC), closeness centrality (CC), subgraph centrality (SC), eigenvector centrality (EC), information centrality (IC), bottle neck (BN), density of maximum neighborhood component (DMNC), local average connectivity-based method (LAC), sum of ECC (SoECC), range-limited centrality (RL), L-index (LI), LeaderRank (LR), normalized alpha centrality (NC), and moduland-centrality (MC) from the yeast protein-protein interaction network are shown.}\label{Figure15}
     \end{figure}

\begin{figure}
          \centering
          \includegraphics[width=11cm]{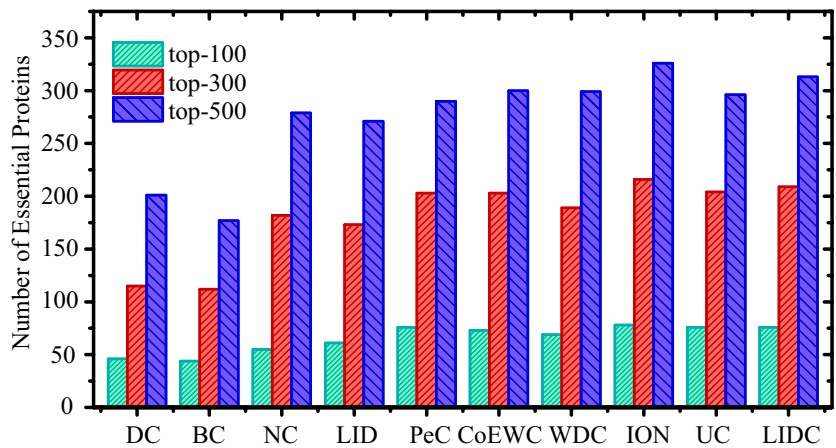}\\
          \caption{Comparison of the number of essential proteins from the top 100–600 identified by ten prediction measures in the YDIP-5093 protein interaction network, which includes 5093 proteins and 24743 interactions \cite{Luo2015PLoSONE10e0131418}.}\label{Figure16}
     \end{figure}

        As shown in Fig. \ref{Figure15}, the performance of PeC is much better than some well-known centrality measures making use of only the topological information, such as degree, betweenness, closeness, and so on. While as shown in Fig. \ref{Figure16}, the above-mentioned methods integrating both topological information and biological information exhibit similar performance.

     \subsection{Quantifying scientific influences}
     During the recent decades of scientific development, the measurement of scientific influences has experienced several significant improvements \cite{Sarli2014MisMed111399}. Therein, one intuitive and simple metric of quantifying scientist' influence, is merely the quantity of his/her publications. However, this number would underestimate the influences of authors who have published very few but highly influential publications. We can also employ the number of publications' citations to measure their influences. However, due to the abuse of self-citations or cross-citations within a small group, the absolute number of citations is not robust. This is also the disadvantage of all citation-based metrics \cite{Zhou2012NJP14033033}, including the widely applied H-index \cite{Hirsch2005PNAS10216569} and the Impact Factor of journals \cite{Garfield2007IntMicr1065}. To provide more precise measurements, some researchers borrow the idea of identifying vital nodes and then propose some new metrics by harnessing various relations among publications and authors, like quantifying the publications' influences through the citation relations among publications \cite{Chen2007JInfo18, Walker2007JSM06P06010}, ranking the scientists based on the citation relations among authors \cite{Radicchi2009PRE80056103}, and measuring the influences of both scientists and publications using multi-relations between authors and publications \cite{Zhou2012NJP14033033, Liao2014PLoSONE9e112022}.

     Through employing the citation relations, Chen \emph{et al.} \cite{Chen2007JInfo18} constructed a citation network based on all the publications in the Physical Review family of journal from 1893 to 2003. Claiming that citations could not provide a full picture of the influence of a publication, they assumed a publication is more important if it is cited by many important publications. Then they directly applied the PageRank algorithm \cite{Brin1998CNIS107}. Basing on the same data set, Walker \emph{et al.} \cite{Walker2007JSM06P06010} further considered the temporal decaying effect to depress the old citations. To harness this factor, they proposed a new ranking algorithm named CiteRank, which is inspired by the process of surfing scientific publications, where researchers usually start from a rather recent publication, and then follow random walk. The probability that a paper is visited via random walk is debased by a time-decaying function $\rho_i=\mathrm{e}^{-\mathrm{age}_i/\tau_{\mathrm{dir}}}$, where $\mathrm{age}_i$ is the age of publication $i$ and $\tau_{\mathrm{dir}}$ is a free parameter.

	\begin{figure}
	  \centering
	  \includegraphics[width=9cm]{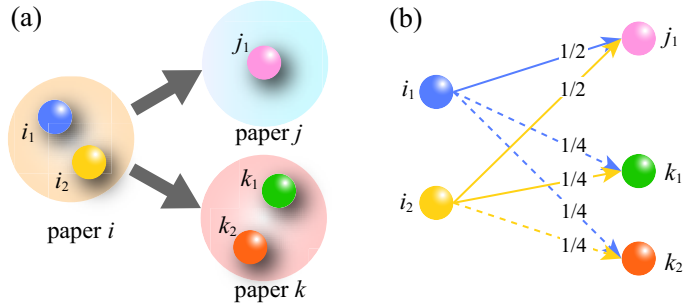}\\
	  \caption{An simple example to construct weighted author-to-author citation networks). (a) The article $i$, written by two authors $i_1$ and $i_2$, cites two papers $j$ and $k$, written by one author $j_1$ and two co-authors $k_1$ and $k_2$ respectively. (b) The weighted author-to-author citation networks is then simply generated by connecting with six directed links $(i_1 \rightarrow j_1)$, $(i_1 \rightarrow k_1)$, $(i_1 \rightarrow k_2)$, $(i_2 \rightarrow j_1)$, $(i_2 \rightarrow k_1)$ and $(i_2 \rightarrow k_2)$. The weight of links $(i_1 \rightarrow j_1)$ and $(i_2 \rightarrow j_1)$ are both defined as $\frac{1}{n_i \times n_j}=\frac{1}{2}$, where $n_i$ is the number of co-authors in paper $i$. Similarly, the weights of links $(i_1,k_1),(i_1,k_2),(i_2,k_1)$ and $(i_2,k_2)$ are all $\frac{1}{n_i \times n_k}=\frac{1}{2 \times 2}=\frac{1}{4}$. One author might have multiple citations to another author, so the weight of a directed link between two authors is calculated by the sum of all the weights over all the references. After \cite{Radicchi2009PRE80056103}.}\label{Figure17}
	\end{figure}

\begin{figure}
		  \centering
		  \includegraphics[width=8cm]{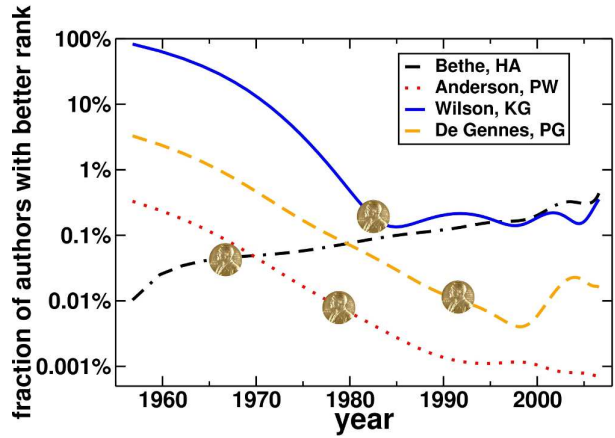}\\
		  \caption{Evolution of the relative rank expressed as top percentile of four Nobel laureates (this figure is initially drawn by Radicchi \emph{et al.} \cite{Radicchi2009PRE80056103}). ``Bethe, HA''(1967, black solid line), ``Anderson, PW'' (1977, red dotted line), ``Wilson, KG'' (1982, blue solid line), and ``De Gennes, PG'' (1992, yellow dashed line).  }\label{Figure18}
		\end{figure}

        The citations can also be used to measure the scientists' influences, but different citations should have different values depending on who is the citing scientist. From this perspective, Radicchi \emph{et al.}~\cite{Radicchi2009PRE80056103} constructed an author-to-author citation network (an simple example is shown in Fig.~\ref{Figure17}) to rank scientists, by mimicking the diffusion of scientific credits. In specific, Radicchi \emph{et al.} assigned a unit of credit to every author and assumed that the credit can be distributed to its neighbors proportionally to the weight of the directed link. In other word, the credit scores of authors depend on how much credit they can get from their neighbors. The proposed algorithm is based on an iterative process, which consists of a biased random walk and a random redistribution of the credits among all nodes. The biased random walk makes the links associated with highly ranked authors more important than those with lower-ranked authors, while the random redistribution considers the nonlocal effects of the spreading of scientific credits. More interestingly, Radicchi \emph{et al.}~\cite{Radicchi2009PRE80056103} constructed a series of networks based on the dynamical slices of database, and then got the relative rank dynamics of every author. They presented four Nobel laureates' relative rank dynamics in Fig.~\ref{Figure18}, which shows top performances are reached close to the data of the assignment of the honor.

\begin{figure}
		  \centering
		  \includegraphics[width=12cm]{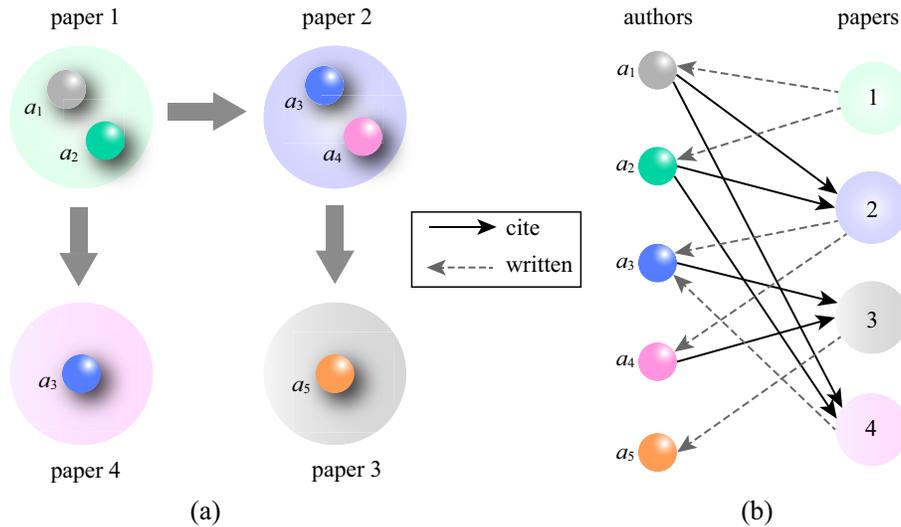}\\
          \caption{An illustration of how the author–paper bipartite network is constructed. (a) The citation relations between four papers written by five scientists. (b) The corresponding author–paper interactive network of (a), where the directed links from an author to papers indicate that the author cites these papers (named \emph{citation} relations), and the directed dash links from a paper to authors mean that this paper is written by these scientists (named \emph{written} relations). Here self-citations are not included.}\label{Figure19}
		\end{figure}

       The subjects of citation relations in the last research are either authors or publications. From a different perspective, Zhou \emph{et al.}~\cite{Zhou2012NJP14033033} considered the citation relations from authors to publications. Namely if a paper $i$, written by author $i_1$ and $i_2$, cites a paper $j$, the directed links $(i_1,j)$ and $(i_2,j)$ would be created. They also considered the \emph{written} relations between authors and publications. Thus, an author-publication bipartite network which consists of two kinds of relations can be naturally constructed, as shown in Fig.~\ref{Figure19}. Following the \emph{written} relations, the score of each publication is distributed to all its co-authors, which follows the mass diffusion process~\cite{Zhou2007PRE76046115,Zhou2010PNAS1074511}. Through the \emph{citation} relations, the calculation of papers' scores follows voting model~\cite{Liggett1999BOOK}, which adds the score of each author to all the publications he/she has cited. Thus the basic hypothesis is clearly that a paper is expected to have high quality when it is cited by prestigious scientists, while high-quality papers raise the scientists' prestige accordingly. Comparing with the Citation Counts based ranking method (named CC rank), the rank method proposed by Zhou \emph{et al.}~\cite{Zhou2012NJP14033033} (named AP rank) found some outliers, containing both scientists and publications. Some scientists with low CC ranks have higher influences, because they are appreciated by prestigious scientists; while some papers with a large number of citations are ranked lower by AP rank, namely these papers are popular but not prestigious. Therefore, this method is effective in distinguishing between prestige and popularity, and then could provide more reasonable rankings to some extent. Moreover, this method was further applied to evaluate the influence of journals by the total ranking score of their publications. The top 5 mainstream journals in econophysics were found to be ``Physica A'',``Phys. Rev. E'', ``Eur. Phys. J. B'', ``Quant. Financ.'' and ``Phys. Rev. Lett.''.

       In addition to the relations between authors and publications, the reputation of users in online scientific communities is also valuable to quantify the quality of publications, and the credit of authors \cite{Liao2014PLoSONE9e112022}. In such communities, users can submit, download and view papers. Then Liao \emph{et al.} constructed a multilayer user-paper-author network and proposed a new reputation algorithm to simultaneously measure the quality of papers, the reputation of users and the credit of authors (the detail definition of this algorithm is depicted in Chapter~\ref{Chapter8-3}). This method is able to highlight the papers that have been largely neglected by the forum users (indicated by fewer downloads), but eventually attracted considerable attention (indicated by more citations). The algorithm has been deployed at the Econophysics Forum where it helps to highlight valuable papers.

       In the above applications, the contributions of all co-authors of a publication are assumed to be equally important. However, this assumption is not very exact and has received some discussions. For example, Stallings \emph{et al.}~\cite{Stallings2013PNAS1109680} introduced an axiomatic approach to assign higher credit to the co-authors with higher order. Such method of assigning credit for one publication is named $A$-index. Then the summation of $A$-index of a particular researcher in all his publications can be used as an indicator measuring his/her scientific influence. From a quite different viewpoint, Shen \emph{et al.}~\cite{Shen2014PNAS11112325} calculated the credit of an author according to his/her contribution to a publication as perceived by the scientific community, rather than his/her actual contribution. This method assign very good credit to Nobel Prize winners, but would underestimate young scientists because they usually receive less credit than their established collaborators from co-authored publications. Above all, distinguishing co-authors' contributions is very important to properly measure authors' scientific influences, meanwhile the time-dependent factors and the whole picture of the collaboration relations also play important roles in the measurement.

   \subsection{Detecting financial risks}
       No doubt financial risks is an important problem, which can be detected through simulating the spreading of the default of an institution (institutions) \cite{Nagurney2008InBOOK273,Gai2010ProcRSocA4662401,Mistrulli2011JBankFina351114}. Suppose institution $v_i$ defaults, its creditor institution $v_j$ faces a loss. If this loss exceeds $v_j$'s equity, institution $v_j$ also defaults. This default might lead to the default of $v_j$'s creditor, and then trigger a cascade of default. This process continues until no new default occur. The system risk posed by institution $v_i$ (also called the importance of $v_i$) can be represented by the size of the cascade, which is defined in several ways. The number of institutions that fail by contagion is very intuitional \cite{Mistrulli2011JBankFina351114,Battiston2012JFinaSta8138}, and so is their total assets \cite{Mistrulli2011JBankFina351114}. Nagurney and Qiang \cite{Nagurney2008InBOOK273} introduced the network performance metric, which is proportional to the amount of financial funds that can reach the consumers through the financial intermediaries. Gai and Kapadia \cite{Gai2010ProcRSocA4662401} studied the probability and potential impact of contagion caused by the initial default. They found high connectivity could not only reduce the probability of contagion, but also increase its spread when problems occur.

       However, the default of an individual institution is typically not able to trigger a domino effect \cite{Boss2004inBOOK1070, Mistrulli2011JBankFina351114}. Inspired by this idea, Mistrulli \cite{Mistrulli2011JBankFina351114} counted the number of banks whose default causes at least one bank by contagion. But this method works poorly when measuring the importance of one bank. From a different perespective, Battiston \emph{et al.} \cite{Battiston2012SciRep2541} indicated that there exists a propagation of distress even though the default can not propagate, and the bank facing the loss would become more fragile and this also makes counterparties more fragile. Based on this, they developed DebtRank to quantify the amount of distress triggered by the initial distress of a particular bank (a set of banks). This method can be used to measure the systematic importance of the bank (the set of banks). During the propagation process of distress in DebtRank, each bank has two states variables at time-step $t$: (i) $h_i(t) \in [ 0, 1 ]$, the amount of distress of the bank $v_i$ at time $t$; (ii) $s_i(t) \in \{U, D, I \}$, the state of the bank $v_i$, which belongs to one of the three states: Undistressed, Distressed, and Inactive. A bank $v_i$ would become inactive if $h_i=1$ or it suffers distress at two successive times. After a finite number of steps $T$, the dynamics stops and all the banks in the network are either in state $U$ or $I$. The DebtRank of a default bank is determined by the number of inactive banks. Battiston \emph{et al.} \cite{Battiston2012SciRep2541} applied DebtRank in the FED (Federal Reserve Bank) data. They found that the largest borrowers were interconnected in a dense network, each of which was centrally located and could significantly impact every other bank in only one or two steps.

%

       Soon DebtRank was applied to a Japanese credit network \cite{Aoyama2013RIETI13-E-0871}, which consists of the lending/borrowing relationship between Japanese banks and Japanese firms. By introducing distress to some initial node(s) (banks or firms), they found the importance of a bank is much related to its size (total assets). The correlation is not linear, but follows power-law, written as $D^{\mathrm{(banks)}} + D^{\mathrm{(firms)}} = 6.55\times 10^{-19}S^{1.50}$ ($S$ is the size of the bank). The large exponent ($\simeq1.50$) indicates that if the size of a bank is doubled, the bank's DebtRank would increase by 1.82 times. As a result, they claimed that ``big banks are \emph{far more} important than small banks'', and merging with a partner with the same size is the optimal solution to increase the total DebtRank. Moreover, DiIasio \emph{et al.} \cite{DiIasio2013MPRA52141} applied DebtRank into a dataset covering bilateral exposures among all the Italian banks. The results showed that the systemic impact of individual banks has decreased since 2008. Tabak \emph{et al.} \cite{Tabak2013} improved DebtRank and applied it to assess the systematic risk in the Brazilian Interbank market. Puliga \emph{et al.} \cite{Puliga2014SciRep46822} proposed Group DebtRank, which is a variation of the DebtRank algorithm. In particular they focused on small shocks that hits the whole network and wanted to measure the final effect.

       Despite the process of default or distress, researchers introduced some popular centrality metrics. For instance,  compared with DebtRank, Battiston \emph{et al.} \cite{Battiston2012SciRep2541} took eigenvector centrality as the benchmark metric. Motivated by PageRank, Kaushik and Battiston \cite{Kaushik2013PLoSONE8e61815} put forward two centrality metrics. One is impact centrality, which assumes that a node is more systemically important if it impacts many systemically important nodes. Symmetrically, they proposed vulnerability centrality, supposing that a node is more vulnerable if it strongly depends on many vulnerable nodes. With the assistant of these two centrality metrics, they concluded that only a small subset of nodes need to be focused in terms of systemic impact. Basing on absorbing Markov chains (containing at least one absorbing state), Soram\"{a}ki and Cook \cite{Soramaki2013EcoOAE72013-28} proposed SinkRank to measure the systemically importance of nodes in payment networks. The absorbing Markov system reflects the fact that when a bank fails, any payments sent to this bank would be remained in this bank. Through simulating experiments, they found SinkRank can accurately rank the banks, which is helpful to estimate the potential disruption of the whole payment system. In addition, Craig and von Peter \cite{Craig2014JFI23322} tried to find the important nodes in financial networks. They named them core banks, which are defined as the intermediaries except those which do not lend to or borrow from the periphery.

       \subsection{Predicting career movements}
\begin{figure}
  \centering
  \includegraphics[width=16cm]{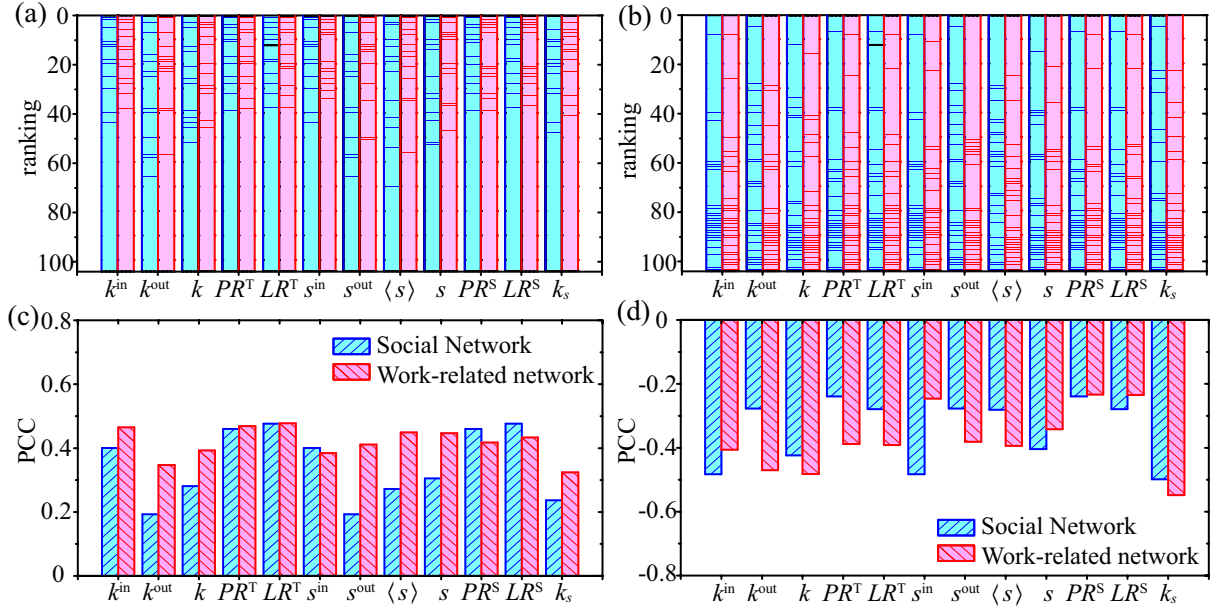}\\
  \caption{The relations between topological features and employees' career movements. (a) and (b) utilize horizontal lines to show the ranking of target employees ranked by different topological features: (a) is for promoted employees while (b) is for those who left. (c) shows the Pearson Correlated coefficient between topological features and the probabilities of being promoted, while (d) shows the correlation between topological features and the turnover risk. $k_{\mathrm{in}}$ and $k_{\mathrm{out}}$ are the in-degree and out-degree of nodes in directed networks, and $k=k_{\mathrm{in}}+k_{\mathrm{out}}$ here. $PR^{\mathrm{T}}$ and $LR^{\mathrm{T}}$ mean PageRank and LeaderRank in unweighted network, while $PR^{\mathrm{S}}$ and $LR^{\mathrm{S}}$ are for weighted network. $s_{\mathrm{in}}$ and $s_{\mathrm{out}}$ are the in-strength and out-strength of nodes in weighted directed networks (see details in Chapter \ref{Chapter7}). $\langle s\rangle$ means the average strength while $s$ means the total strength of a node (namely an employee). $k_s$ means the $k$-core centrality. }\label{Figure20}
\end{figure}

       In the domain of career movements, both the promotion and resignation can be foreseen with some clues. In 1997, Freely and Barnett~\cite{Feeley1997HumCR23370} constructed social networks based on employees' communication relationships. Through examining the relations between employees' career movements and their topological features, they showed that employees who are highly connected and in more central positions are less likely to resign. Later, Feeley \emph{et al.}~\cite{Feeley2008JACR3656} constructed another social network according to a survey during which employees were requested to report their friendships and peer relationship. At this time, they indicated that employees who reported a greater number of out-degree links with friends (including both peer relationships and friendships) were less likely to leave, however the closeness of the friendships did not play very important role. Besides, some other features were also taken into consideration, such as degree centrality, betweenness centrality, and closeness centrality~\cite{Feeley2000JACR28262, Mossholder2005AcadMJ48607, Feeley2010JACR38167, Liu2010SSRN}. The data sets adopted by those studies are all gathered from questionnaire survey. Yuan \emph{et al.}~\cite{Yuan2016PhysicaA444442} claimed that such data is probably subjective due to the psychological defense. Instead, they collected credible employees' work-related interactions and social connections from a social network platform used by a Chinese company. Accordingly, Yuan \emph{et al.} constructed two directed networks, the social network or the work-related network. Then they examined the correlations between some centrality metrics and the career movements of employees, including both turnover and promotion. In the case of turnover, they obtained similar conclusion, namely central employees in the social network or the work-related network are less likely to resign. As to promotion, the employees who get more attention (higher in-degree) in work-related network have higher chance to be promoted. Using the same dataset, Gao~\emph{et al.}~\cite{Gao2014InBook} further examined more topological features of employees' networks, and showed that these features can be used to predict the future resignation or promotion with a remarkably higher accuracy than a pure chance, as shown in Fig. \ref{Figure20}. As shown in Fig. \ref{Figure20}, the PageRank centrality and the LeaderRank centrality in social network are most positively related to promotion. The employees who have higher in-degree are also more likely to be promoted. In the case of resignation, the in-degree centrality, the in-strength centrality, and the $k$-core centrality in social network as well as the out-degree centrality, the degree centrality, and the $k$-core centrality in work-related networks are the most negatively related features.

       \subsection{Predicting failures with developer networks}
       Centrality metrics also play an important roles in predicting the failure or reopen-bugs of software. One way of constructing the network is based on the developer-module relationships, named the contribution network~\cite{Pinzger2008SIGSOFT2}, where nodes represent software modules and developers, and edges represent contributions of developers to modules. The centrality of a software module in the contribution network is used as a measure which synthetically measures the developers' contributions. Basically, central software modules are jointly contributed by various developers. Through utilizing linear and logistic regression, Pinzger \emph{et al.}~\cite{Pinzger2008SIGSOFT2} compared several centrality metrics, including degree centrality, closeness centrality and betweenness centrality in predicting failures of software. The experimental results showed that central modules are more failure-prone than modules located in surrounding areas of the network, and the closeness centrality is a significant predictor for the number of post-release failures. Furthermore, considering the dependence relations among modules, Bird~\emph{et al.}~\cite{Bird2009ISSRE109} combined it with the contribution network to construct a new network named socio-technical network. Moreover, they investigated more centrality metrics and found the predictive power could be largely increased. Another way of constructing developer network is based on collaboration relationships~\cite{Meneely2008SIGSOFT}, where the role of developers is emphasized. In such network, Meneely \emph{et al.}~\cite{Meneely2008SIGSOFT} employed the developer's closeness centrality and betweenness centrality to predict the post-release failures of softwares. Xuan \emph{et al.}~\cite{Xuan2012ICSE25} applied a variant of LeaderRank to determine the developer's priority, and found that it is an effective factor for reopened bug prediction.

\section{Outlook}\label{Chapter11}

       In this review, we have extensively reviewed the state-of-the-art progresses in identifying vital nodes in complex networks, emphasizing on the physical concepts and methods. In this section, we show some critical challenges for future studies.

       First of all, we need some benchmark experiments to evaluate different algorithms. In this research domain, the algorithms' performances depend on the objective functions under consideration (e.g., the betweenness centrality performs well in hindering epidemic spreading while it is poor in accelerating epidemic spreading), the parameters in a given dynamical process (e.g., in the SIR process, the degree centrality can better identify influential spreaders when the spreading rate is very small while the eigenvector centrality performs better when the spreading rate is close to the epidemic threshold \cite{Klemm2012SciRep2292,Liu2016SciRep621380}), as well as the network structures (see the extensive results in Chapter \ref{Chapter9}). In principle, researchers may test their algorithms for different objective functions on different networks, and only publish the results being propitious to their algorithms. Therefore, some experiments on real dynamical processes in real networks can be treated as solid benchmarks in comparing algorithms' performances. However, the design and implementation of such experiments are very challenging. Taking the information propagation in large-scale social networks as an example. If two algorithms $A$ and $B$ respectively choose $i$ and $j$ as the most influential people, we can not guarantee that $i$ and $j$ are willing to join this experiment since they are usually popular stars. Even if the algorithms are limited in some preselected volunteers, when the two volunteers try to spread the same message, there will be potential interactions that destroy the independence between two propagations since the two spreading paths could contain the same people. While if the two volunteers spreading different messages, the comparison cannot be completely fair since the prevalence of a message is very sensitive to its content \cite{Tang2009ACM,Cui2011ACM,Tsur2012ACM}. Therefore, we need a smart idea to design feasible and solid experiments. In the lack of sufficient experiments, a secondary choice is to build up an open-to-public platform that presents extensive simulations about the algorithms' performances on several representative topological and dynamical objectives for disparate real networks, with data sets and codes being freely downloadable. This platform can be considered as the benchmark in this area.

       Secondly, the methods in identifying vital nodes need further development. This review introduces tens of methods, and the methods in the literatures are even more. Taking centralities for individual vital nodes as examples. Some centralities embody closely related ideas, such as PageRank and LeaderRank, as well as betweenness and closeness, while some centralities come from far different perspectives. A valuable work is to arrange well-known centralities and classify them according to their produced rankings of nodes, the information required to produce these rankings, and other relevant features (see, for example, how to classify similarity measures in bipartite networks \cite{Liu2016SciRep618653}), so that we will have comprehensive understanding of known methods including the similarities and differences among them. Using both real networks and artificial networks with tunable topological features, we are further asked to quantify the sensitivities of each methods to the change of structural features and the dynamical properties, which help us in defining the range of applications for each method. In addition, we still have not obtained satisfied answers to some important questions. For example, how to identify the most influential node or the most influential set of nodes at a finite time $t$ instead of the steady state at $t\rightarrow \infty$ \cite{Liu2016SciRep621380}, in particular for the continuous dynamical processes, and how to identify vital nodes in incomplete networks \cite{Tan2016SciRep622916}.

       Thirdly, the scopes of the research should be largely extended. On the one hand, the topological and dynamical objectives in most previous works were limited in the size of the giant component and the spreading dynamics. In the future studies, we should consider more topological features (e.g., the network efficiency defined as the mean value of the inverses of the shortest distances of all node pairs \cite{Latora2001PRL87198701} and the coverage of a set of nodes defined as the number of nodes belonging to the set itself or neighboring to at least one node in the set \cite{Haynes1998BOOK,Zhao2015JSP1591154}) and more dynamical processes (e.g., synchronization, transportation, routing, cascading, evolutionary games, etc. See more examples in the book \cite{Barrat2013BOOK}). On the other hand, the networks considered in most previous works are of elementary types, while in the future studies, we should design efficient algorithms for some novel types of networks, which were less considered in the early studies of graph theory but attracted increasing attentions for their relevance to the real world, such as spatial networks \cite{Barthelemy2011PhysRep4991}, temporal networks \cite{Holme2012PR51997} and multilayer networks \cite{Gao2012NatPhys840,Kivela2014JCN2203}.

       Lastly, we are longing for some large-scale real applications of the mentioned algorithms in real world. Although in Chapter \ref{Chapter10}, we show many applications, where the vital nodes identification algorithms are used to solve some other research problems, most of which are themselves far from real applications. What we mean \emph{real applications} here is that the algorithms are applied in the treatments for patients, in the delivery of advertisements for real commercial companies, and so on. Real applications cannot replace experiments since doctors and business men will not take risks to systematically compare different methods. However, successful applications will largely encourage related studies, just like what Google and Amazon have contributed to the studies of information retrieval \cite{Gleich2015SIAM57321,Ermann2015RMP871261} and recommender systems \cite{Lu2012PhysRep5191,Bobadilla2013KBS46109}.

\section*{Acknowledgments}
We benefit a lot from the fruitful works collaborated with the following people, Mat\'{u}\v{s} Medo, Ming-Sheng Shang, Chi Ho Yeung, An Zeng, Hao Liao, Hao Liu , and Weiping Liu, etc. We also acknowledge the team lead by Zi-Ke Zhang (some members are the authors of the companying review) for useful exchange of research. This work is supported by the National Natural Science Foundation of China (Grant No. 61433014), the National High Technology Research and Development Program (Grant No. 2015AA7115089) and the Fundamental Research for the Central Universities (Grant No. ZYGX2014Z002), the Research Start-up Fund of Hangzhou Normal University (Grant No. PE13002004039), the Zhejiang Provincial Natural Science Foundation of China (Grant No. LR16A050001), the Swiss National Science Foundation (Grant No. 200020-143272), and the EU FP7 Grant 611272 (project GROWTHCOM).

\bibliography{Review23}

\end{document}